\documentclass[twocolumn]{aastex701}%
\usepackage{graphicx}
\usepackage{natbib}
\usepackage{amssymb,amsmath}
\usepackage{savesym}
\savesymbol{tablenum}
\usepackage{siunitx}
\restoresymbol{SIX}{tablenum}
\usepackage{color}

\usepackage{multirow}
\usepackage{subcaption}

\usepackage{booktabs}%
\usepackage{caption}

\shorttitle{}
\shortauthors{Carter, Jeffrey, Wicks, Nicolaou}

\begin{document}

\title{A new methodology for inferring the plasma conditions in solar flare energetic electron source regions from in situ electron energy spectra}

\author[0009-0003-4368-1328]{Samuel Carter}
\affiliation{School of Engineering, Physics and Mathematics, Northumbria University,
Newcastle upon Tyne, NE1 8ST, UK}
\email{samuel.carter@northumbria.ac.uk}

\author[0000-0001-6583-1989]{Natasha L. S. Jeffrey}
\affiliation{School of Engineering, Physics and Mathematics, Northumbria University,
Newcastle upon Tyne, NE1 8ST, UK}
\email{natasha.jeffrey@northumbria.ac.uk}

\author[0000-0002-0622-5302]{Robert T. Wicks}
\affiliation{School of Engineering, Physics and Mathematics, Northumbria University,
Newcastle upon Tyne, NE1 8ST, UK}
\email{robert.wicks@northumbria.ac.uk}

\author[0000-0003-3623-4928]{Georgios Nicolaou}
\affiliation{Mullard Space Science Laboratory, University College London, Holmbury Saint Mary, Dorking, Surrey, RH5 6NT, UK}
\email{g.nicolaou@ucl.ac.uk}

\begin{abstract}
The conditions within solar flares that lead to efficient electron acceleration are not well constrained. It is not clear whether the populations accelerated out into the heliosphere and inward into the chromosphere originate in the same regions. By analysing the energy distributions of heliospheric populations, modelling suggests that it should be possible to see evidence of their originating region(s), including the presence of hot, dense flare plasma. By creating and utilising a novel in situ spectral analysis package called INSPEX we have performed this analysis for flare electrons observed in situ on 09/10/2021, constructing both peak flux and fluence spectra from combined Solar Orbiter in situ electron measurements. We compare how differing methodologies for combining the datasets influence the spectral shapes and the retrieved parameters over an energy range of 0.5-80 keV. We fit different functions to the multi-component form of the energy spectra, testing combinations of thermal and/or power law components, comparing the fit statistics. We find that the spectra can be fitted with two distinct thermal curves at energies below 20 keV, corresponding to typical corona/active region and flaring material temperatures, varying between \(1.4 - 4.1\) MK and \(12.5 - 23.1\) MK depending on the rebinning window and peak flux extraction method. This study showcases how INSPEX can provide a novel and user-friendly methodology for studying electron spectra with different instrumentation, allowing investigation of multiple spectral types and signatures of acceleration and transport. This first application provides a benchmark case for the analysis of similar flares.
\end{abstract}

\section{Introduction}
Solar flares are events occurring in the solar corona, during which magnetic reconnection causes a brightening and heating of the local plasma \citep[e.g., ][]{2011-Fletcher} over a duration of minutes, or even many hours \citep[e.g., ][]{2020-French}. They are observed across the electromagnetic spectrum and can be associated with coronal mass ejections (CMEs), though not necessarily so \citep[e.g., ][]{2008-Benz}. 

Magnetic energy stored in the coronal plasma topology is converted into many different forms during a flare, of which the kinetic energy of accelerated particles makes up a significant proportion, possibly $>$30\% \citep[e.g.,][]{Emslie_2012}. As such, it is important to develop an understanding of the particle acceleration mechanisms that occur in order to develop an understanding of the energetics and dynamics of explosive reconnection processes in plasmas. The particles accelerated include both ions and electrons. Accelerated electrons from solar flares are seen in two distinct populations: those travelling outward into the heliosphere on open magnetic field lines, and those that travel sunward (i.e., trapped populations on closed magnetic field lines) and interact with the lower atmosphere. It is unclear whether these two populations are caused by the same acceleration process, or if they are accelerated in the same location \citep[e.g., ][]{2021-klein, 2004-klein}.

These two populations are observed by different techniques. Outward accelerated electrons can be observed in situ \citep[e.g., ][]{1985-Lin,2007-Krucker,2020-Dressing} and by radio emissions \citep[e.g., ][]{2008-Pick}, while inward-accelerated electrons trapped at the Sun are observed by the bremsstrahlung hard X-ray (HXRs) they produce, mainly as they interact with the dense chromosphere \citep{2011-Holman, 2011-Kontar}. Acceleration of electrons observed in the heliosphere can occur concurrently with the brightening which characterises the flare, so called prompt events, or after a short period of time, so called delayed events \citep[e.g., ][]{2007-Krucker}. Electrons accelerated during prompt events are thought to originate from the flare reconnection itself, not some subsequent  process \citep{2007-Krucker}. However, distinguishing between these scenarios is often challenging due to the complexity of multiple acceleration and transport processes \citep[e.g., ][]{2014-Jeffrey, 2014-kontar, 2016-laitinen} In this preliminary study, we focus on one suggested prompt event, in which low-energy electrons, in a range of 10-100 keV, were primarily accelerated by a flare-related process \citep{Jebaraj_2023}. Notably, the selected event is associated with a CME, though we conclude that it would have limited effect on the 10-100 keV energy range of interest \citep{Jebaraj_2023}, and thus on the modelled temperatures.

\begin{figure*}[tpbh!]
\centering
\includegraphics[width=0.49\textwidth]{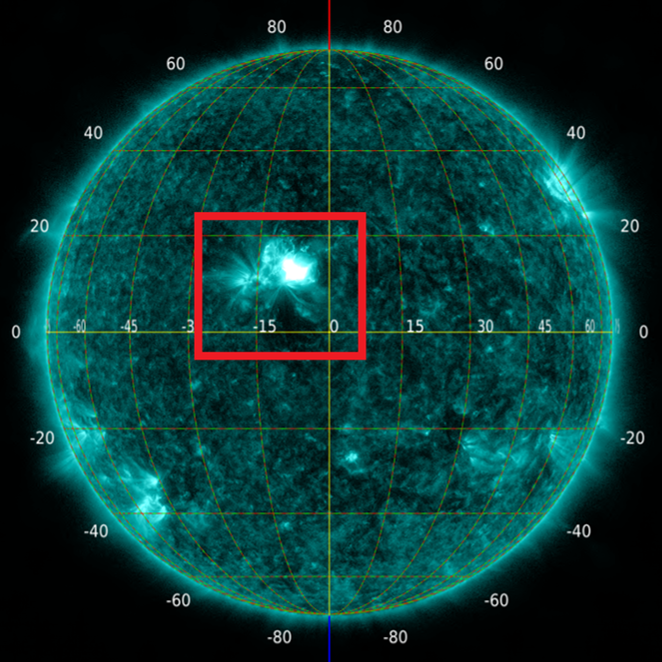}
\includegraphics[width=0.49\textwidth]{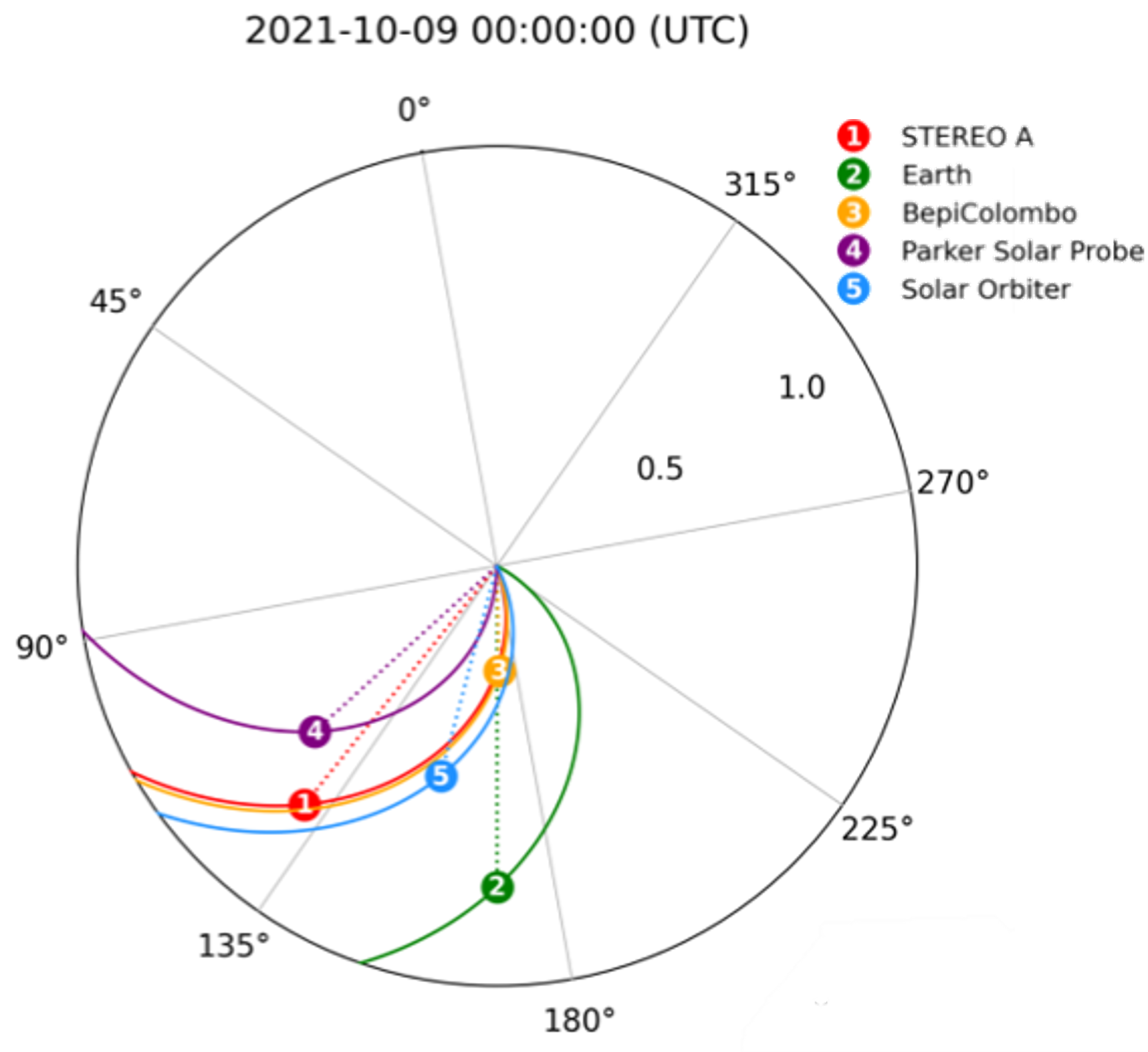}
\caption{Left: An illustration showing the SDO AIA 131~\AA~ view, from the direction of Earth, at 06:48 UTC on 09/10/2021 \citep{2012-Lemen}. The active region of interest is marked with the red box. Right: An illustration showing the spacecraft positions on the day of the flare, and their Parker spiral curves linking them to the Sun \citep{gieseler_solar-mach_2023}.}
\label{SDOAIA+solar-mach}
\end{figure*}

Through HXR spectroscopy there is a well-established methodology that allows the derivation of the inward-accelerated electron population, i.e., those in the location of acceleration \citep[e.g., ][]{1971-Brown,2019-Kontar}, as well as surrounding flare plasma conditions (temperature, number density). The Object Spectral Executive (OSPEX) \citep{ospex} package for Interactive Data Language (IDL) typifies tools for this forward-fitting method and allows for this analysis to be performed on data from e.g., Reuven Ramaty High Energy Solar Spectroscopic Imager (RHESSI)\citep{2003-Lin} and Solar Orbiter (SolO) \citep{2020-Muller} Spectrometer Telescope for Imaging X-rays (STIX) \citep{2020-Krucker}, as well as others. It generates an injected electron spectrum from the HXR spectrum, and allows the user to select physical models to fit to the spectra and to tune the initial values of the parameters of those functions, prior to optimisation. These include electron and/or X-ray transport effects such as collisions \citep[e.g., ][]{2014-Jeffrey,2015-Kontar}, return currents \citep[e.g., ][]{2006-Zharkova,2017-Alaoui} and X-ray albedo \citep[e.g., ][]{1978-Bai,2006-Kontar} that alter the properties of the emitting electron distribution. 

\begin{figure*}[hbpt]
\centering

\begin{subfigure}{0.49\textwidth}
        \centering
        \includegraphics[width=\textwidth]{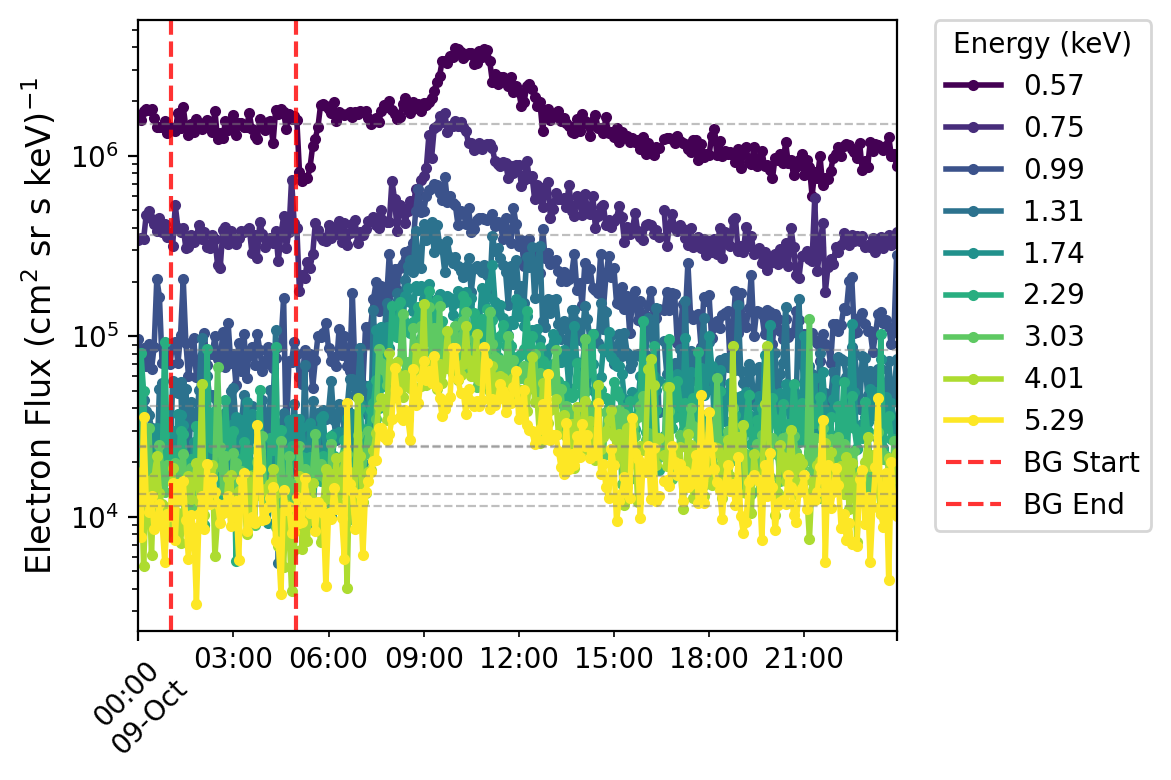}
        \caption{EAS}
        \label{EAS time series}
    \end{subfigure}
    \begin{subfigure}{0.49\textwidth}
            \centering
            \includegraphics[width=\textwidth]{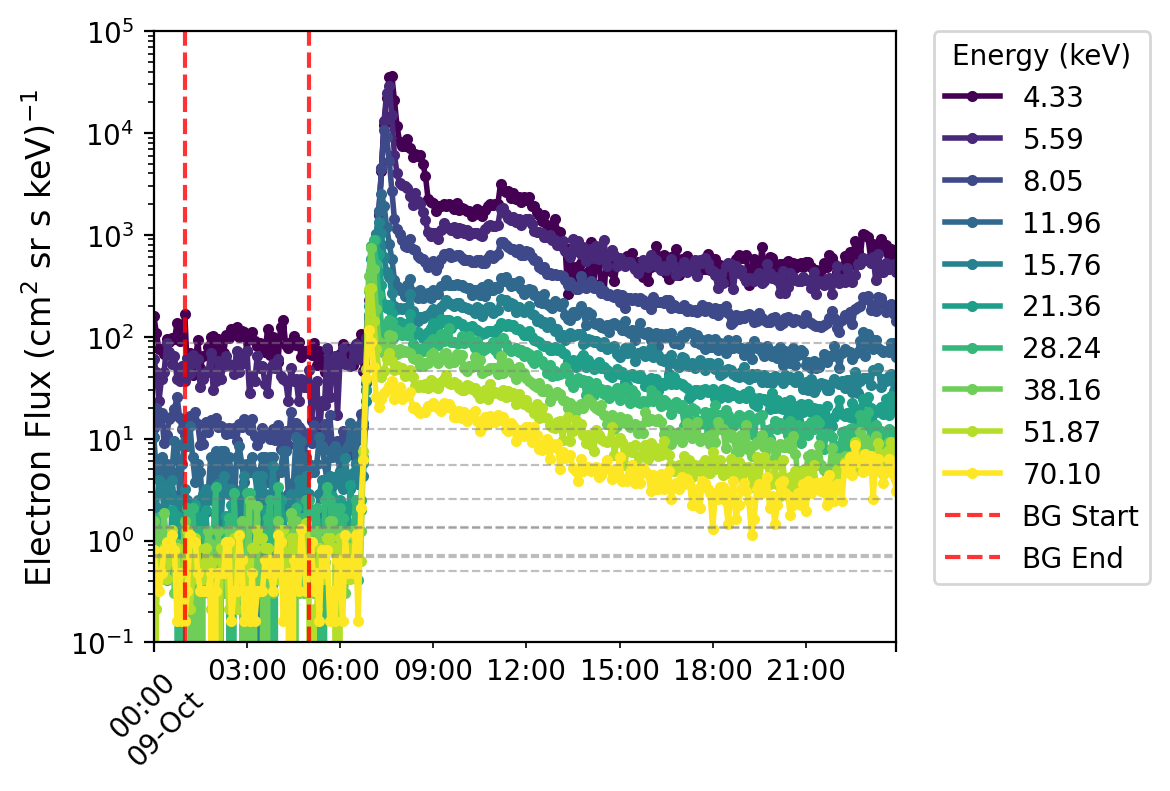}
            \caption{STEP}
            \label{STEP time series}
        \end{subfigure}

\caption{EAS (a) and STEP (b) time series resampled to 5 minute cadence, with background period shown in red and the background intensity levels calculated as an average over the full background period shown in grey.}
\label{time_series_figure}
\end{figure*}

This method has been extended in theory to in situ observations by \citet{2023-Pallister} and \citet{2025-Pallister}. This shows that electrons transported through hot, over-dense regions like those expected for the flare acceleration region should exhibit different peak flux and fluence energy spectra than those characterizing the injected populations in the origin. These changes carry information on the region they passed through. The peak locations of these spectra are moved to lower energies when passing through cold, dense regions, to a degree dependent on that region’s temperature, plasma density, and spatial extent. Therefore, these properties should be identifiable from the flare electron energy spectra, where we expect to see a Maxwellian component to the distribution \citep{2023-Pallister,2025-Pallister}. The majority of flare X-ray studies use a single temperature approximation, corresponding to the temperature of the flaring plasma and independent of the particle directivity, which we describe as an isothermal, isotropic Maxwellian distribution. The form we use for our Maxwellian neglects bulk plasma motions. One aim of this work is to extract the originating plasma temperature in particular, if possible. We evaluate the feasibility of our method by the successful determination of this key parameter.

To investigate this method, we need a tool to load in situ spectra and fit functions to them. As no tool exists to perform the in situ spectral analysis for electrons that is analogous to OSPEX for X-ray emitting electrons, we have created one. We call this new Python package In situ Spectral Executive (INSPEX) \citep{INSPEX_8_12_25_zenodo}. We have developed this tool to be able to load data from online archives, calibrate it, display it as a time series for background and integration range selection, and then fit curves to the spectrum through a user-friendly graphical user interface (GUI). In this way it allows users to go all the way from raw data to the fitted spectra. Additionally, the fitting GUI can be called without using data from the previous steps. This allows users to input their own spectral data for fitting, including modelled data. A complete step by step methodology is included in the appendix, and INSPEX, with a user guide, is available via Github\footnote{https://github.com/SamuelCarter42/INSPEX} and Zenodo \citep{INSPEX_8_12_25_zenodo}.

This work presents this new method and demonstrates its viability by showing some preliminary results, with particular focus on estimates of the temperature of flaring plasma, and potentially other physical parameters. In our methodology, we explain our choice of instruments and flares. We then introduce our new methods for generating in situ electron energy spectra from the flux data products and fitting them with physically meaningful functions. This also includes details of data processing, which proved particularly challenging. The results we present are the generated spectra and the functions we fitted to them and, in particular, the physical parameters of those functions. It is important to note that, as we analyse only a single event here, many of the conclusions we draw will be highly event-specific. This paper does not aim to conclude strongly about flares in general, but instead presents our methodology as a demonstration of the differing models that can be used. We discuss here the success of different fits in modelling the spectra of this particular event.

\section{Methodology}
\begin{figure*}[htbp]
\centering
\begin{subfigure}{0.49\textwidth}
        \centering
        \includegraphics[width=\textwidth]{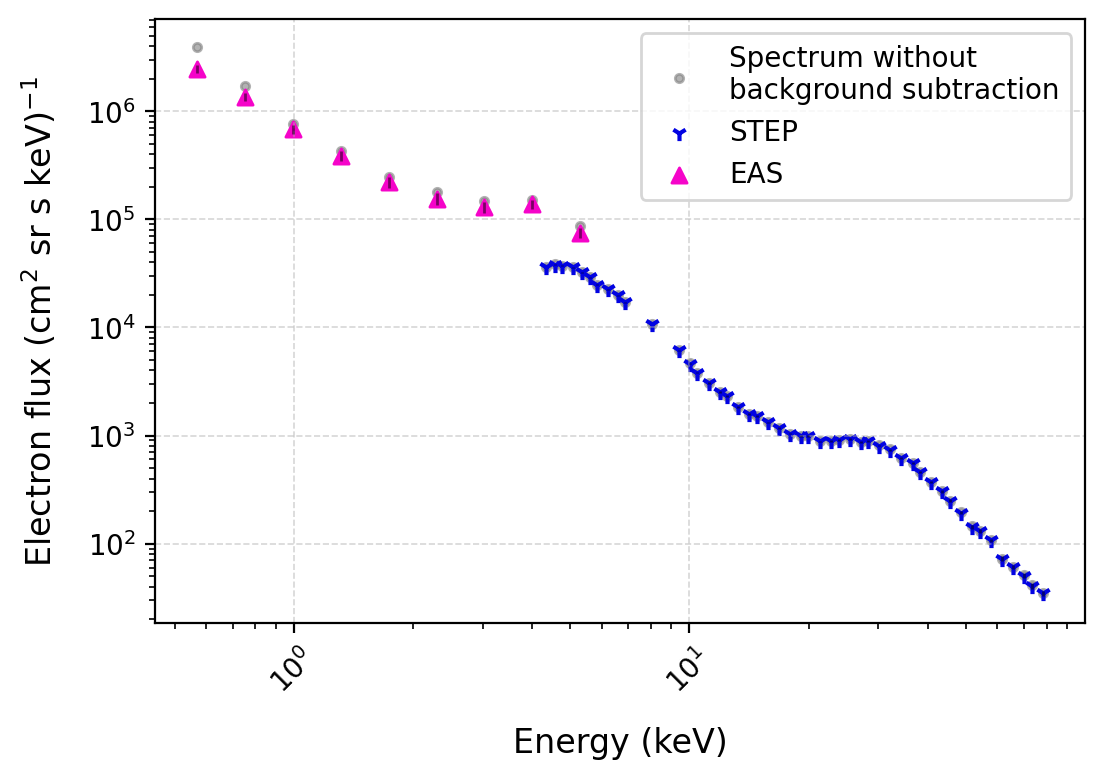}
        \caption{Peak Flux Spectrum without alignment by FAF}
        \label{Peak Flux Spectrum no FAF spec fig}
    \end{subfigure}
    \begin{subfigure}{0.49\textwidth}
        \centering
        \includegraphics[width=\textwidth]{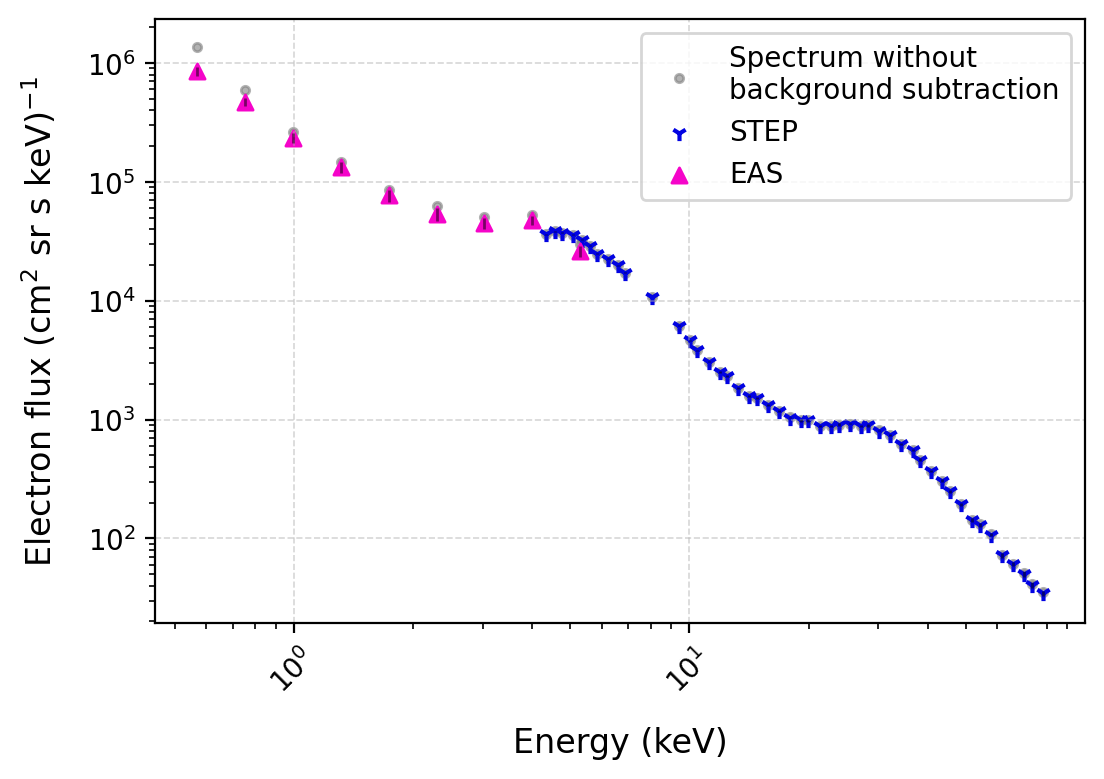}
        \caption{Peak Flux Spectrum with alignment by FAF }
        \label{Peak Flux Spectrum FAF spec fig}
    \end{subfigure}

\begin{subfigure}{0.49\textwidth}
        \centering
        \includegraphics[width=\textwidth]{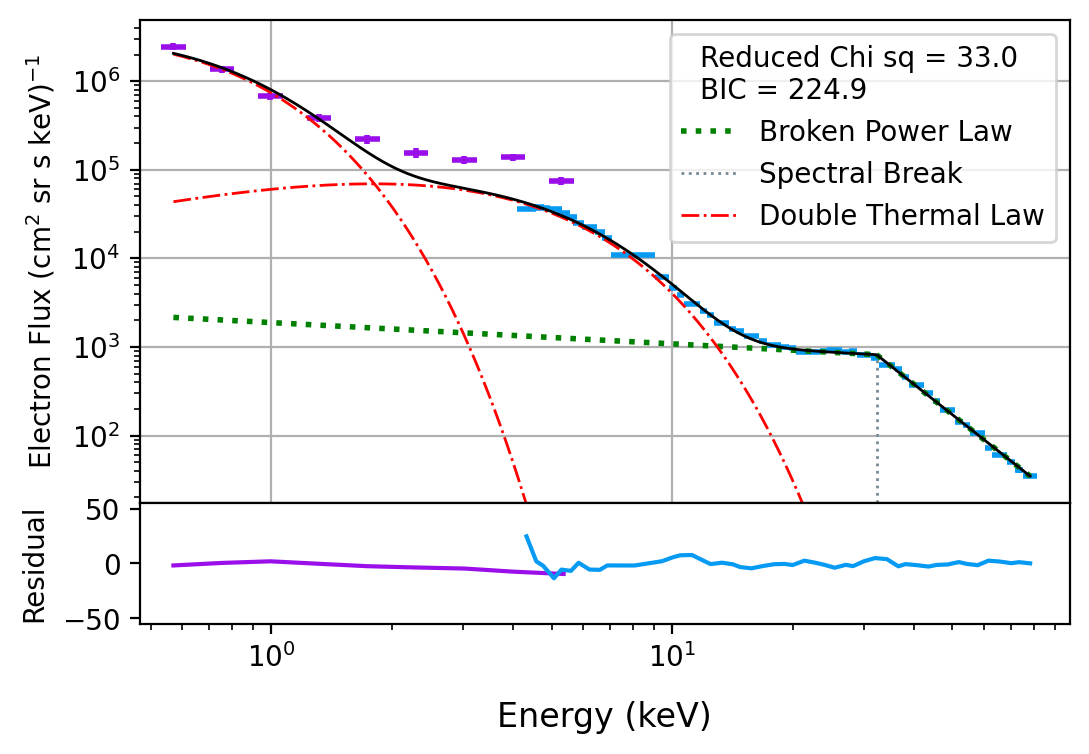}
        \caption{Function fit to the unaligned spectrum}
        \label{Peak Flux Fit no FAF spec fig}
    \end{subfigure}
    \begin{subfigure}{0.49\textwidth}
        \centering
        \includegraphics[width=\textwidth]{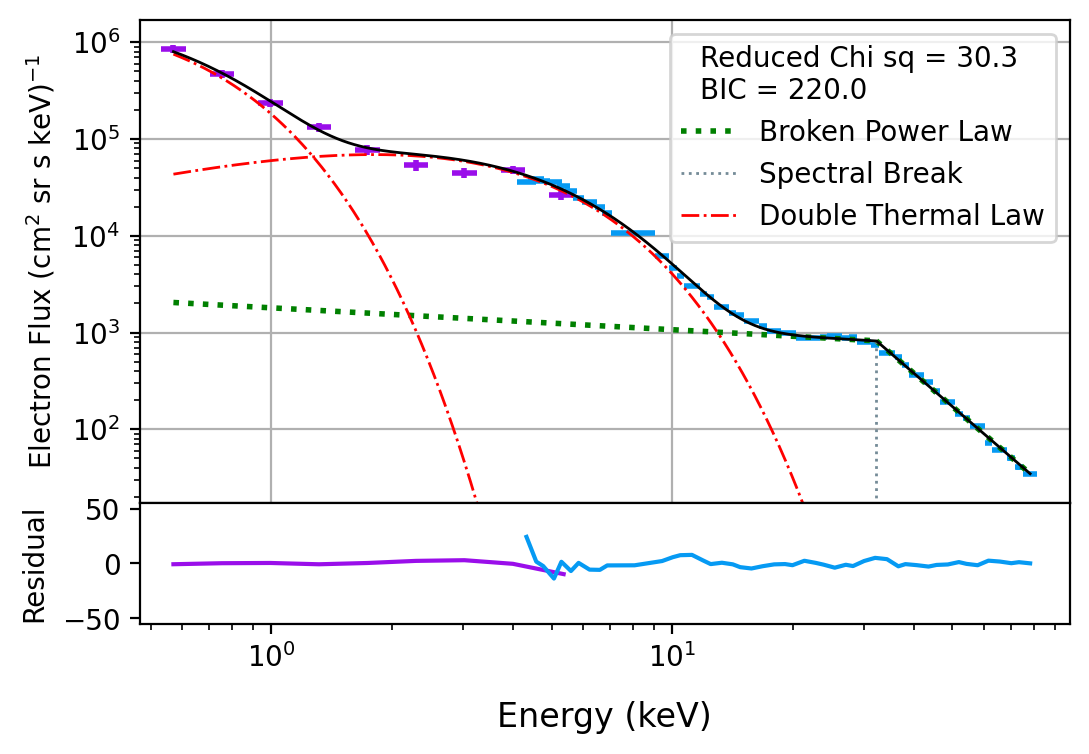}
        \caption{Function fit to the FAF aligned spectrum}
        \label{Peak Flux Fit FAF spec fig}
    \end{subfigure}
\caption{Peak flux spectra, shown without FAF on the left and with FAF on the right. a) and b) show these spectra (with the spectra generated without background subtraction shown in grey), while c) and d) show the fits to those spectra. The total fit is shown in black, with the spectral components added as shown in the legend. These fits have their respective residuals shown below to further illustrate fit quality.}
\label{peak_spectra_figure}
\end{figure*}

\begin{figure*}[hbpt]
\centering
\begin{subfigure}{0.49\textwidth}
        \centering
        \includegraphics[width=\textwidth]{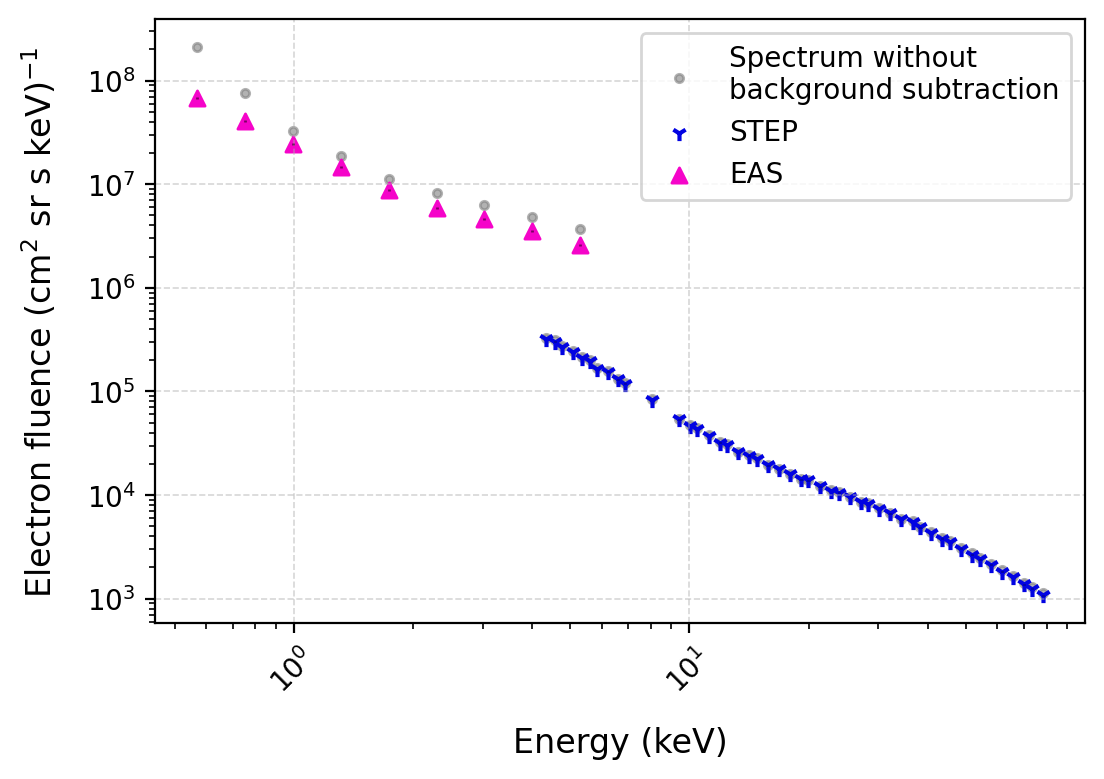}
        \caption{Fluence Spectrum without alignment by FAF}
        \label{Fluence Spectrum no FAF spec fig}
    \end{subfigure}
    \begin{subfigure}{0.49\textwidth}
        \centering
        \includegraphics[width=\textwidth]{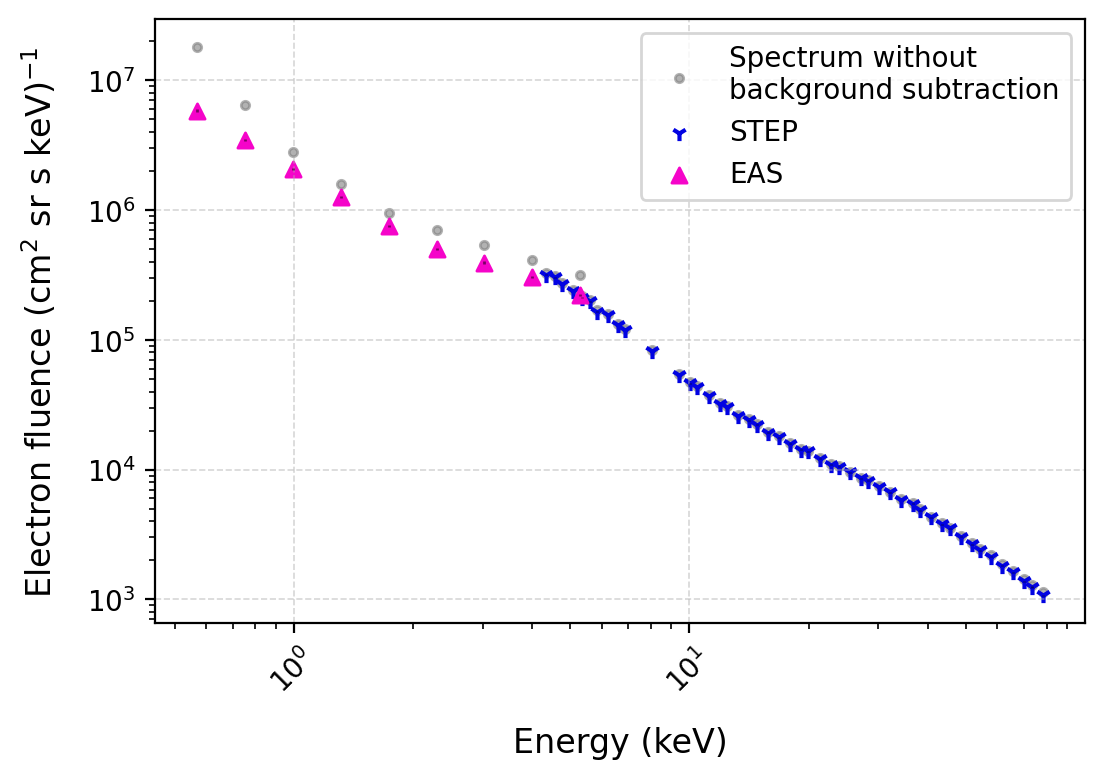}
        \caption{Fluence Spectrum with alignment by FAF }
        \label{Fluence Spectrum FAF spec fig}
    \end{subfigure}

\begin{subfigure}{0.49\textwidth}
        \centering
        \includegraphics[width=\textwidth]{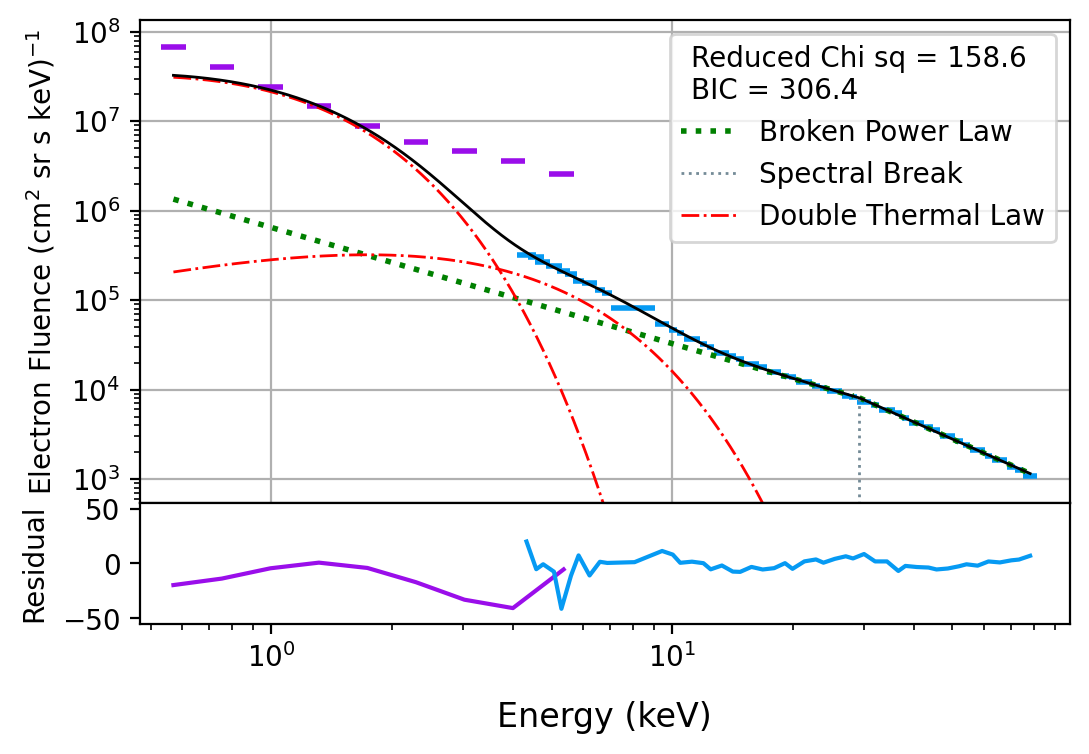}
        \caption{Function fit to the unaligned spectrum}
        \label{Fluence Fit no FAF spec fig}
    \end{subfigure}
    \begin{subfigure}{0.49\textwidth}
        \centering
        \includegraphics[width=\textwidth]{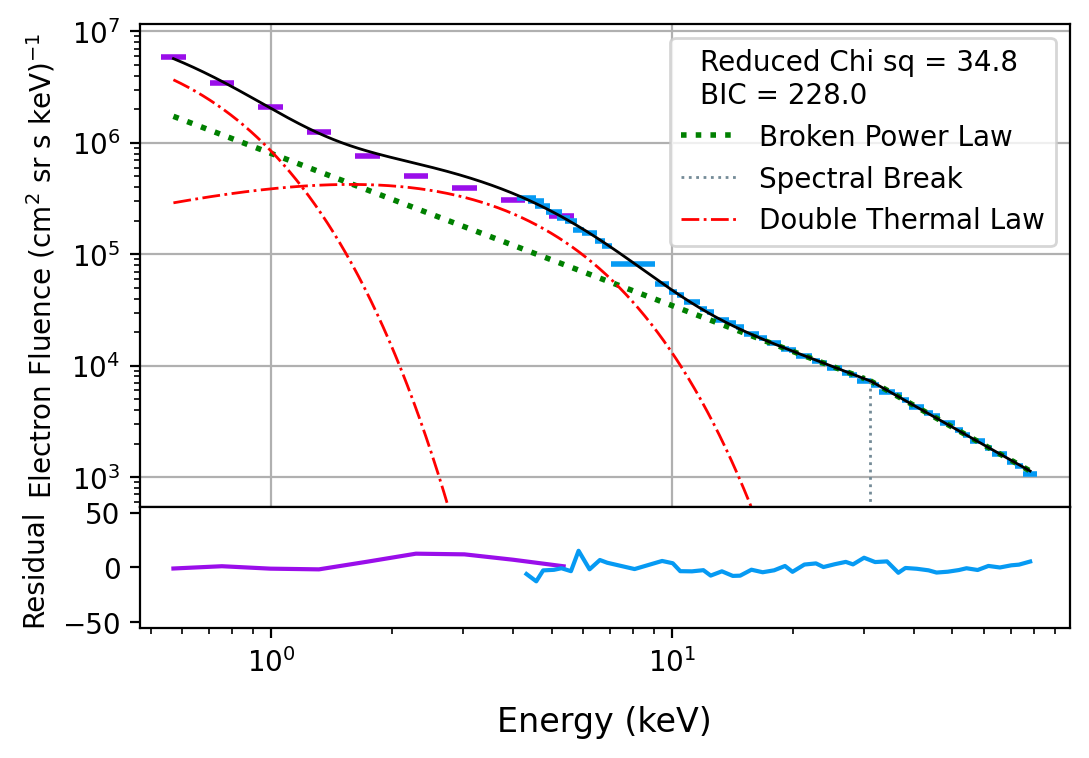}
        \caption{Function fit to the FAF aligned spectrum}
        \label{Fluence Fit FAF spec fig}
    \end{subfigure}
\caption{Fluence spectra, shown without FAF on the left and with FAF on the right. A) and B) show these spectra, with the spectra generated without background subtraction shown in grey. C) and D) show the fits to those spectra. The total fit is shown in black, with the spectral components added as shown in the legend. These fits have their respective residuals shown below to further illustrate fit quality.}
\label{flue_spectra_figure}
\end{figure*}

\subsection{Instrumentation}

The tool we have developed can be used with any electron spectrum, and includes a built-in function to load data from In situ Measurements of Particles And CME Transients (IMPACT) \citep{2007-Luhmann} Suprathermal Electron Telescope Downstream (STE-D) aboard Solar TERrestrial RElations Observatory (STEREO), and Energetic Particle Detector (EPD) \citep{2020-Rodriguez-Pacheco} SupraThermal Electrons and Protons (STEP) and Solar Wind Analyser (SWA) Electrostatic Analyser System (EAS) \citep{2020-Owen} aboard Solar Orbiter (SolO). In future, we hope to expand this built in load function to include Wind 3D Plasma Analyzer (3DP) \citep{1995-Lin} and Parker Solar Probe (PSP) Integrated Science Investigation of the Sun (ISOIS) \citep{2016-McComas}. In this study, we have chosen EPD STEP and SWA EAS aboard SolO for demonstrating our methodology. These datasets are loaded from the Solar Orbiter Archive (SOAR)\footnote{http://soar.esac.esa.int/soar/} using SunPy \citep{sunpy_community2020} and solo\_epd\_loader \citep{gieseler_solo-epd-loader_2025,2022-Palmroos}.

SolO EPD comprises two instruments: the STEP detector and the Electron Proton Telescope (EPT). We focus on fitting the multi-component spectral shapes between 0.5 and 100 keV in this work because this closely matches the energy range where signatures of hot flaring plasma and accelerated electrons are seen in HXR studies \citep[e.g.,][]{2021-Dressing, 2008-Benz}. STEP has an energy range of \(4.33\) – \(78.1\) keV at a resolution of 750 eV FWHM, which provides a better coverage of our chosen range than EPT, which covers an energy range of \(25\) – \(475\) keV at a resolution of 7 keV FWHM \citep{2020-Rodriguez-Pacheco}. The STEP detector was re-calibrated on 22/10/2021, changing the data files and the energy binning. The observations used occur before this re-calibration. However, this instrument does not measure electron flux directly. Instead, the data comprises an ``integral" particle flux channel and a ``magnetic" particle flux channel. Integral flux is the total flux of ions and electrons, while magnetic flux is just that of ions, separated out using a magnetic field. Each is binned into 48 energy channels. Electron flux is calculated by subtracting magnetic flux from integral flux, to give the flux not accounted for by the protons \citep{2020-Rodriguez-Pacheco}. The flux values are taken from the pixel averaged data to give a 1-D energy spectrum.

However, the detector does not respond to electrons in the same way as protons and therefore the calculated values must be corrected for this by the ratio of the geometric factors for each particle (Francisco Espinosa, private communication). The correction data table is available for the pre-calibration period, allowing easy conversion for spectra before this (Camille Lorffing, private communication). However, later spectra include the correction table within the data file. This data file takes a different format to the pre-calibration data, and we have added routines to INSPEX which can interpret this, allowing future work to encompass events later in solar cycle 25. Full details are given in the appendix and in the Github documentation for INSPEX \footnote{https://github.com/SamuelCarter42/INSPEX}.

SolO SWA EAS provides in situ measurements for electrons within the energy range spanning from \(1\) eV to \(5\) keV \citep{2020-Owen}. This allows us to extend our observations to lower energies. However, when using this instrument there are extra considerations we must take to ensure our results are valid. In the caveat document \footnote{Robert T. Wicks, George Nicolaou, Christopher Owen, et al. 2021, SWA EAS – Caveats to use of Science Data Products, Technical Report, SO-SWA-MSSL-IF-006, UCL} we find that we must use only the even or odd indexed energy bins, but not both together, due to the ``sawtooth issue'', in which even indexed energy bins record higher counts than the odd indexed energy bins. We elected to use only the even indexed bins. Additionally, a differing method is required for corrections between \(2\) eV and \(8\) eV, and the sweep is likely to be incorrect for points below \(2\) eV (Daniel Verscharen, private communication, \citet{owen-2022}). As our range of interest is \(\sim\) \(0.5-100\) keV, we simply discard every data point below \(0.5\) keV (\(500\) eV) to remove these issues. We do not expect significant effects from secondary and photo-electrons in the energy range being examined \citep{2021-Nicolaou, 2025-stverak, owen-2022}.

EAS provides whole-sky coverage \citep{2020-Owen}, and as we intend to use this instrument to extend down the energy range being analysed, we must ensure that we are aligning the field of view with that of STEP \citep{Lorfing-2023}, which is centred on \(35\) degrees in -y of the spacecraft x-y plane, with a field of view \(28 \times 54\) degrees \citep{2020-Rodriguez-Pacheco}. EAS consists of two detectors, EAS1 and EAS2, mounted orthogonally. Only EAS1 covers the STEP field of view, so we do not use EAS2. The data is recorded in each detector's angular bins, so must be mapped into spacecraft coordinates to allow alignment with STEP. We convert detector angles, which are given in the instrument reference frame, and energies to velocity vectors in the spacecraft reference frame \citep{2021-Nicolaou}, then select only the particles which would have hit the STEP field of view. We then average over the selected pixels to get a 1-D spectrum like that from STEP.

For both instruments aboard SolO, we have chosen to resample the time series data. By taking an average over a new cadence (a new set of fixed time bins), we eliminate spikes in the data and improve counting statistics \citep{Lorfing-2023, Nicolaou_2018}. We refer to the cadence of the observation as taken from the data files as ``raw'', and we refer to the resampled product by the length of the new cadence. For better comparison with the X-ray-emitting electrons we use units of keV. Following import of all data sets, we find that in certain places the electron flux values are negative. While this may appear not to make physical sense, it is a product of counting statistics and thus may not be excluded from the data (Lars Berger, private communication). As such, INSPEX does not automatically filter negative flux values. However, we found that whether the negative values are rounded to zero or left in the data, there is no noticeable change to either our peak flux or fluence spectra. A full step by step methodology with greater detail on these points is available in appendix section \ref{Spectral Generation Methodology}.

\subsection{Choice of Flare}

For this study we analyse a single flare, occurring on the 9th of October 2021 \citep{Jebaraj_2023,Lorfing-2023} (SOL2021-10-09T06:38). It was seen as an on-disk flare by Solar Dynamics Observatory (SDO) Atmospheric Imaging Assembly (AIA) \citep{2012-Lemen} as shown in Figure \ref{SDOAIA+solar-mach}. The peak Geostationary Operational Environmental Satellite (GOES) \citep{1994-menzel, 2019-goodman, 1994-garcia} soft X-ray flux is observed at 06:38 UTC and reaches class M1.6. As the GOES satellites are at Earth orbit, and light travel time from the Sun to the Earth is about 8.3 minutes, we can assume that the flare occurred at around 06:30 UTC. SolO STIX sees peak X-ray flux in its highest energy channel at 06:30 UTC, which corresponds to a flare time of 06:25 UTC for a light travel time of 5.6 from the Sun to SolO at 0.68 AU \citep{gieseler_solar-mach_2023}. We will use this earlier time to compare the arrival times, as this will provide an upper limit on the streaming times.

Accelerated high-energy electrons were detected in the heliosphere by both SolO STEP and EAS (Figure \ref{time_series_figure}). Figure \ref{time_series_figure} also gives an indication for the arrival times of the electrons. The rise becomes visible at around 06:35 UTC for the highest energies (100 keV) at STEP, and about 06:45 UTC for the lowest energies (0.5 keV). For EAS the highest energies rise above the averaged background (see Figure \ref{time_series_figure}) at approximately 07:05 UTC, and the lowest at 08:05 UTC. The rise in flux seen by EAS is more gradual, especially at the higher energies, making the initial rise time more difficult to distinguish than in the STEP data. The highest energy electrons arrive at SolO almost simultaneously with the peak X-ray emission seen by GOES, highlighting the promptness of this event. However, the lowest energy electrons arrive an hour and a half after the X-ray peak. 

Assuming no transport effects, we can use a simple kinetics calculation to estimate the streaming times for electrons to SolO from the Sun, by deriving their velocity from their kinetic energy and calculating their time of flight. We can retrieve the location of the spacecraft during the flare as a heliocentric distance of 0.68 AU using the Solar-MACH tool \citep{gieseler_solar-mach_2023}, and as the flare is seen by STIX close to the disk centre \citep{STIX-datacentre} we assume that this is a suitable proxy for the distance from the source region of the flare to SolO. This analysis reveals a discrepancy between the theoretical and observed arrival times, as we would expect the highest energy (100 keV) electrons to arrive at around 06:35 UTC, corresponding to a 10 minute streaming time, and the lowest energy (0.5 keV) electrons to arrive at 08:25 UTC, corresponding to a 2 hour streaming time. This suggests that the lowest electrons should have arrived later than was observed. The lowest energy electrons arrive around 20 minutes earlier than expected, while the highest energy electrons arrive at the time we calculate. This means the lowest energy electrons arrive 90 minutes after the highest energy electrons, rather than the 110 minutes expected. Because the lowest energy electrons are arriving earlier, it could be the case that they are being released earlier, or that they are in fact higher energy electrons which have been slowed during flight. Accounting for deviations in the distance travelled by electrons, such as those caused by turbulence or switchbacks, can only lengthen the path taken compared to this direct streaming route, and thus only increase the discrepancy between theory and reality.

The positions of SolO and STEREO during the flare are given in Figure \ref{SDOAIA+solar-mach}, as well as other spacecraft such as PSP and BepiColombo, and Earth, where many other instruments are located. Figure \ref{SDOAIA+solar-mach} shows that SolO and STEREO are well aligned, and connected to similar locations on the Sun by the Parker spiral. INSPEX provides an ideal tool for rapidly comparing spectra from different locations and in different data formats. Of course, these must be compared with consideration to their contexts, including the direction the instrument is facing and the spacecrafts' position, and we strongly encourage the reader to read the relevant instrument papers \citep[e.g., ][]{2020-Rodriguez-Pacheco,2020-Owen,2015-Fox,2007-Kaiser,2020-Krucker,
1995-Lin,2007-Luhmann,2016-McComas,2020-Muller,2010-Benkhoff,2012-Lemen} and develop an understanding of this when comparing different instruments and spacecraft.

A coronal mass ejection (CME) is seen on the same day as our flare originating in the same active region, N20E09 (12882). While a CME was associated with this event, which will affect the energy spectrum seen \citep{Jebaraj_2023}, we do not expect that it would have a significant effect at the lower end of the energy range we are considering \citep{Bell-1978,Jebaraj_2023}. We therefore do not attempt to characterise the effect it may have on the results we present and instead neglect it.

\subsection{In situ Spectral Types}\label{spect_types}
INSPEX can handle 3 different types of in situ spectra; peak flux, integrated fluence, and ``instantaneous" spectra. Instantaneous spectra take the flux value for each energy bin at a given time. However, since all instruments have some finite collection time these spectra are only quasi-instantaneous. These give a view of the incoming flux at a given time, and by repeating these at points over the duration of the event we can analyse the evolution of the event over time \citep{Lorfing-2023}, and the differing arrival times of the different energy bins. Peak flux spectra take the peak value of each energy bin in a given interval. Our analysis focuses on this method as it is commonly used for comparison with spectra derived from X-ray datasets \citep[e.g., ][]{2007-Krucker}. Fluence sums the flux through each energy bin over a selected time range, and is therefore useful for examining the electron population accelerated over the entire duration of the flare. The values for both peak flux and fluence can depend heavily on properly selecting the integration period, so it is important to properly consider the duration of the event and carefully choose the background range so that no electron flux from the event of interest is removed, by examining the time series over the full day and investigating the arrival times as discussed above to ensure that we do not select a background interval that coincides with the observed arrival of the electrons from the event. The background subtraction process is described in section \ref{background sub para}. A key utility of INSPEX comes from the fact both peak flux and fluence can be compared quickly and easily. The relationship between these two may help to diagnose different acceleration and transport effects.

When calculating the peak flux and fluence spectra from EAS and STEP, we found that the two instruments did not align well in flux (the vertical axis) for the overlapping energy range, as shown in Figures \ref{peak_spectra_figure} and \ref{flue_spectra_figure}. We expect that they should line up as they are sampling the same population. To account for this we implement a fitting alignment factor (FAF) which multiplies the EAS peak flux and fluence by the average difference between the last two EAS bins, and the two STEP bins closest to them. This adaptively aligns the spectra in the overlap region and ensures a continuous spectrum \footnote{This FAF may differ for other events; we intend to find the values for other events in future works.}. This assumes that the EAS counts are above noise. The reason for this mismatch is not clear, so this alignment is used in place of a more physically meaningful process. It is possible that it arises in the different detector designs, as EAS uses a pair of varying electric fields to sequentially guide electrons of different energies and elevations onto an annular ring of detector plates \citep{2020-Owen}, while STEP uses a pinhole and a segmented solid state detector to achieve angular resolution and detects the energy of the particles directly with the solid state detector segments \citep{2020-Rodriguez-Pacheco}. When we perform the viewpoint alignment, 12 pixels from EAS align with the 15 pixel field of view of STEP, as shown in appendix \ref{app_method}. Because these detector pixels are therefore different sizes, and have differing physical origins (mass spectrometry and physical optics), it is likely that the discrepancy arises in the difference between the detectors. Additionally, in the case of fluence, it should be considered that EAS measures each of the 64 energy bins sequentially, and therefore only measures 1/64 of the fluence through a given energy bin, while STEP measures each energy bin all the time so measures the full fluence through a given bin.

Each type of spectrum takes a slightly different form, with fluence often not showing a clear spectral break at high energies, and peak flux having a very clear hard-soft break in the mid-tens of keV. This means different function forms must be tested for each, as we investigate how we can extract the most information from the data. INSPEX allows for rapid testing of different function forms making it ideal for this purpose. For simplicity we assume that a thermal component can be approximated with a single Maxwellian distribution \citep{2015-jeffrey}, and that the higher energies will show a double power law, caused by non-thermal processes such as collisions or wave-particle interactions \citep{2007-Krucker,2023-Pallister,2013-reid}. Visually, Figures \ref{peak_spectra_figure} and \ref{flue_spectra_figure} would seem to correspond to two thermal distributions and a broken power law, though we additionally test other function forms to eliminate other variations and combinations of functions which may reproduce the form seen.

\subsection{In situ Background Removal and Spectral Analysis}\label{background sub para}
We take a pre-event background for each instrument, aiming to sample the conditions immediately before the arrival of the electrons from the flare.  This is taken from the fixed window resampled time series in Figure \ref{time_series_figure}. The background period is 01:00 UTC to 05:00 UTC, as shown in Figure \ref{time_series_figure}. We aim to maximise the background duration used while avoiding the event's onset. Each energy channel is averaged over the period to generate a background spectrum, which is subtracted from the chosen integration period to generate a time series of data corresponding to the flare signature alone without background solar wind electrons present. This is then used to generate our spectra. Both fluence and peak flux spectra begin at 07:00 UTC and integrate over an 8 hour window that ends at 15:00 UTC, where fluence sums the flux over that window and peak flux uses the peak value within that window. Our spectra use an energy range of 0.57 - 78.1 keV, as close to \(0.5 - 100\) keV as instrumentation permits. This is because we want to disregard background solar wind and high energy particles from other sources. The \(0.5 - 100\) keV range covers the flat, ``broken power law" components from which the acceleration properties can be retrieved, and the range in which we are investigating the presence of possible thermal components from the flare and corona/active region. Background removal is particularly important for EAS since for this particular event the electron fluxes are low in the \(0.5-5\) keV range and it is essential that we remove the high energy component of the solar wind such as the strahl which may exist in this energy range \citep[e.g., ][]{2017-Graham,2005-Maksimovic}.

\subsection{INSPEX Fitting Functions} \label{INSPEX Fitting Functions}
Currently, several functions can be fitted using INSPEX. When displaying examples of the functions here, we use $F(E)$ to denote either peak flux or fluence as a function of electron energy $E$, as both spectral types can be fitted with these functions.
In general, solar flare spectra have the form of a steeply decreasing power law. Power law forms have been fitted in previous works, \citep[e.g., ][]{2007-Krucker}, and are very typical energy distributions for accelerated particles. The breaks may be a result of differing acceleration mechanisms, changes in the environment, or various transport processes \citep[e.g., ][]{2007-Krucker,2003-conway,2005-Zharkova}. In INSPEX, we currently provide the following power law fitting functions:
\begin{itemize}
\item Single power law
\item Double power law
\item Triple power law
\item Quadruple power law
\item Quintuple power law
\end{itemize}

This choice offers flexibility when fitting spectra consisting of many different components. The double power law function is defined over two domains; above and below break energy \(E_{b}\):
\begin{equation}
F =  \left\{
\begin{array}{ll}
      A_1E^{\delta_1} &  E\leq E_{b} \\
      A_2E^{\delta_2} & E\geq E_{b} \\
\end{array}
\right.
\end{equation}
where \(A_1\) and \(A_2\) are the amplitudes for each component, and \(\delta_1\) and \(\delta_2\) are the spectral indices of the components. The amplitudes are fixed so that the function is continuous at \(E = E_{b}\).
\par
We also define triple, quadruple and quintuple component power law functions in a similar manner, and those definitions are given in full in the GitHub documentation for INSPEX. These functions may be truncated at the low energy end over a finite decay width to provide the user a way to transition from one regime to another. However, in our analysis we have chosen to simply consider them as continuing to below the bottom of our range of interest.

Since one aim of this paper is to perform a preliminary investigation of the presence of hot flaring plasma signatures in the in situ data, we add thermal functions to INSPEX. These include a single isothermal Maxwellian function:

\begin{equation}
  F = A E e^{\frac{-E}{k_B T}}
\end{equation}
where $A$ is an amplitude, $T$ is the temperature of the originating plasma and $k_{B}$ is the Boltzmann constant. 
\par
A pair of isothermal functions summed together is also included as an additional function for considering the contributions of hot, flaring plasma and the cooler corona/active region to the spectrum. It is formulated simply as
\begin{equation}
  F = A_1 E e^{\frac{-E}{k_B T_1}} + A_2 E e^{\frac{-E}{k_B T_2}}
\end{equation}
with the variables taking the same meanings as in the single isothermal case.
\par

\par
The double power law and thermal component were combined to create a single function, as a tool to rapidly test a complete, multi-component function form which covered the expected shape. The parameters were defined in such a way that the equation should be continuous between the two components, giving only a single amplitude for the whole combined function which then generates the other amplitudes according to their gradients under the condition that the components are equal at the point where they meet. This proved useful in the iterative development process of the function forms, as it was often faster to set parameters which result in continuous functions using this form. However, due to the more detailed control required to obtain the final results and the potential for the added continuity constraint to introduce bias, this was not used beyond that initial stage.

\par
In addition to the functions specifically chosen for their physical relevance, some general functions are also included in the INSPEX package for utility. First, there is a Gaussian normal distribution. This has utility for fitting any regular perturbations present in the spectral shape, particularly when the true shape is unknown and an approximate shape is needed. Thus, it has no exact physical meaning, though that does not mean that physical information cannot be extracted if the user has context to the origin of the perturbation. It has the form:

\begin{equation}
F = A e^{\frac{-(E-E_0)^2}{2 \sigma^2}}  
\end{equation}
where $A$ is an amplitude parameter, \(E_0\) is the energy at which the Gaussian is centered, and \(\sigma\) is the standard deviation.

\par
While not used in this work, we have also considered other physical functions such as the kappa function. This function is often used to describe the distribution of electron energies in hot flaring plasma, providing a smooth transition between thermal and non-thermal populations \citep{2013_oka}. We plan to include it in future publications considering the forms given in \citet{pierrard_2010,jeffrey_non-Gaussian_2017,2009-Liavdiotis,2013-livadiotis,2016-nicolaou, 2015-Battaglia}, and implement it into INSPEX along with other physical functions as the tool grows.

Using INSPEX's functionality to combine the above functions, both the electron peak flux spectra and fluence spectra are tested with the following combinations of fit functions:
\begin{itemize}
\item Double thermal and double power law
\item Quadruple power law
\item Thermal and Quadruple power law
\item Triple Thermal
\item Quintuple power law
\item Double thermal and triple power law
\end{itemize}

We are choosing to use both the reduced chi-squared ($\chi^{2}_{\nu}$) and the Bayesian information criterion (BIC) to evaluate our fits, given by the lmfit code \citep{newville_2015} used in INSPEX.  BIC is better for comparing the fits to each other than for directly measuring how well the curve fits the data, while $\chi^{2}_{\nu}$ is suitable for both applications. When considering BIC, it is important to note that for one model to be described as better than another, a difference in BIC of greater than 10 is required \citep{kass-1995}.  

Reduced chi-squared is given by
\begin{equation}
\chi^2_{\nu} = \frac{\chi^2}{\nu} = \frac{1}{n - k} \sum_{i=1}^{n} \left( \frac{y_i - f_i}{\sigma_i} \right)^2
\end{equation}
where \(\chi^2\) is the total chi-squared, \(\nu = n - k\) is the degrees of freedom, n is the number of data points, k is the number of fitted parameters, \(y_i\) are the observed data, \(f_i\) are the model predictions, \(\sigma_i\) are the uncertainties on the data points taken from the data files.
BIC is given by 
\begin{equation}
\text{BIC} = n \cdot \ln\left(\frac{\chi^2}{n}\right) + k \cdot \ln(n)
\end{equation}
where n is the number of data points, k is the number of free parameters, and \(\chi^2\) is the chi-squared value as described above.

\subsection{Instrumental, Data, and Fitting Issues} \label{Instrumental, Data and Fitting Issues}
The Python fitting package used within INSPEX, lmfit, encounters issues when generating the uncertainties on the fitted parameters, when those parameters do not change from their initial guess values, hit their maximum or minimum limits, or the minimiser cannot generate a non-singular covariance matrix for any other reason. An alternative was therefore required in cases where the built in processes fail. We use a Markov Chain Monte Carlo (MCMC) \citep{1992-press-numerical} method built in to the lmfit package to generate the Bayesian posterior probability for the parameters, which we then use to estimate the standard errors of the parameters at the maximum likelihood solution. This is then only used in the cases where the parameters are generated but the built-in uncertainty calculations do not run  for the reasons described. 

An additional contributing factor to large reduced chi-squared and BIC in our fits is the low uncertainty values given with the data products, as can be seen in Figures \ref{peak_spectra_figure} and \ref{flue_spectra_figure}. In Figures \ref{peak_spectra_figure} and \ref{flue_spectra_figure}, the uncertainties in the data files are in fact so small that the lines failed to render on some of the points, a fact also noted by \citet{Lorfing-2023}. These values include the combination of uncertainties during the calculation of electron flux for STEP from the integral and magnetic channels which should result in an increase in uncertainty. Despite this, the uncertainties are still not visible. The consequences of this issue are particularly clear in Figure \ref{5min flue 4pl spec fit fig}, where the goodness of fit, $\chi^{2}_{\nu}$, takes a value in the hundreds, despite this fit visually matching the curve well. The data product uncertainties are a Poisson statistics calculation, which is quite simplistic and does not account for any other systematic or statistical errors. Therefore, the fit is over-constrained because the error is under-estimated. This is the cause of our large chi-squared values, rather than our fits being poor. A better method could be maximum likelihood fitting, which produces better results for low count data  \citep{2024-Nicolaou}. Future implementations of INSPEX will use this method to improve the accuracy of the retrieved parameters.

As the instruments record flux in energy bins, representing a range of energies together, we treat the bins as centered around the middle value of the bin. We do not include this bin width as an uncertainty in our fitting calculations and statistics, as this significantly increases the required computing power for each fit. This is an under representation of the energy uncertainty in the data points, and may cause an under-estimation of uncertainty on the retrieved parameters. However, the binning is shown in the spectrum plots for clarity. The process of accounting for energy binning as an uncertainty in STEP data, using orthogonal distance regression \citep{boggs_1992}, is outlined in \cite{fedeli_2026}, and readers should refer there for an understanding of that alternative treatment.

Both datasets (EAS and STEP) overlap in the \(4-5\) keV range. This energy range is important to this work as it is in this range that the hotter of the isothermal components possibly relating to flaring plasma which we hope to find, at around 20 MK, is expected to dominate \citep[e.g.,][]{2014-Jeffrey, 2015-Kontar}.

\section{Results}
\subsection{Time Series and Spectra}
The time series generated for SolO STEP and EAS are shown in Figure \ref{time_series_figure}. Figures \ref{peak_spectra_figure} and \ref{flue_spectra_figure} plot examples of the electron peak flux spectrum and fluence spectrum using combined EAS and STEP datasets between 0.57 - 78.1 keV. Here, 5-minute resampling is used, following the methodology of \cite{Lorfing-2023} and the fluence spectrum is calculated over an eight-hour period. Figures \ref{peak_spectra_figure} and \ref{flue_spectra_figure} also compare the spectra with and without background subtraction. For this particular event, we can see that background subtraction only produces a small difference to the STEP data but is more important for the EAS data. It is also more important in the fluence case, where the background subtraction from each value compounds over the integration time to give a larger difference. Figures \ref{peak_spectra_figure} and \ref{flue_spectra_figure} show that both spectra are comprised of multiple components. This in turn shows the importance of using a tool like INSPEX to extract the wealth of information contained within this spectral region.

In addition to the concerns outlined in section \ref{Instrumental, Data and Fitting Issues}, Figure \ref{time_series_figure} shows that the two instruments are not completely consistent in the overlapping energy range, with STEP finding background values in this energy range of the order of \(10^4\) (cm$^2$\ s\ sr\ keV)$^{-1}$, while EAS find backgrounds of the order of \(10^6\) (cm$^2$\ s\ sr\ keV)$^{-1}$. Conversely, the peak fluxes and fluences through the detectors sees a reverse pattern, with EAS lower than STEP, though by not so great a degree. Panel a) of Figures \ref{peak_spectra_figure} and \ref{flue_spectra_figure} illustrate this best, though it can be seen in Figure \ref{time_series_figure} as well. This implies that we see a higher pre-event background flux in EAS, but not as strong a signal above that flux. Thus, the difference between the background subtracted and raw spectrum
is greater for EAS than STEP, as can be seen in Figures \ref{peak_spectra_figure} and \ref{flue_spectra_figure}. The peaks in STEP are larger than those in EAS not just in comparison to background, but also in absolute terms.

As already discussed in \ref{spect_types}, the STEP and EAS spectra do not align well (for peak flux or fluence measurements) and we applied FAF to ensure a continuous spectra. In Figures \ref{peak_spectra_figure} and \ref{flue_spectra_figure}, we show two versions of the same peak flux (Figure \ref{peak_spectra_figure}) and fluence (Figure \ref{flue_spectra_figure}) spectra, one without the FAF correction (left) and one with the FAF correction (right). In Figures \ref{peak_spectra_figure} and \ref{flue_spectra_figure}, the peak flux and fluence spectra are also fitted using the chosen fitting function combination of double isothermal and double power law. How well different combinations of functions fit the data is discussed in detail in the next subsection \ref{ss_sf}. Both spectra are discontinuous without the FAF, and as such are poorly fitted by our functions when compared to the spectra where FAF was applied. All spectra in subsequent sections use the FAF correction so that different combinations of fits can be compared consistently.

\subsection{Spectral Fits}\label{ss_sf}
\begin{figure*}[!hptb]
    \centering
        \begin{subfigure}{0.49\textwidth}
        \centering
        \includegraphics[width=\textwidth]{5min_flux_FAF_2p2t.png}
        \caption{Double thermal law and double power law (Table \ref{tab:thermal_fits})}
        \label{5min flux 2 therm 2 power spec fit fig}
    \end{subfigure}
    \begin{subfigure}{0.49\textwidth}
        \centering
        \includegraphics[width=\textwidth]{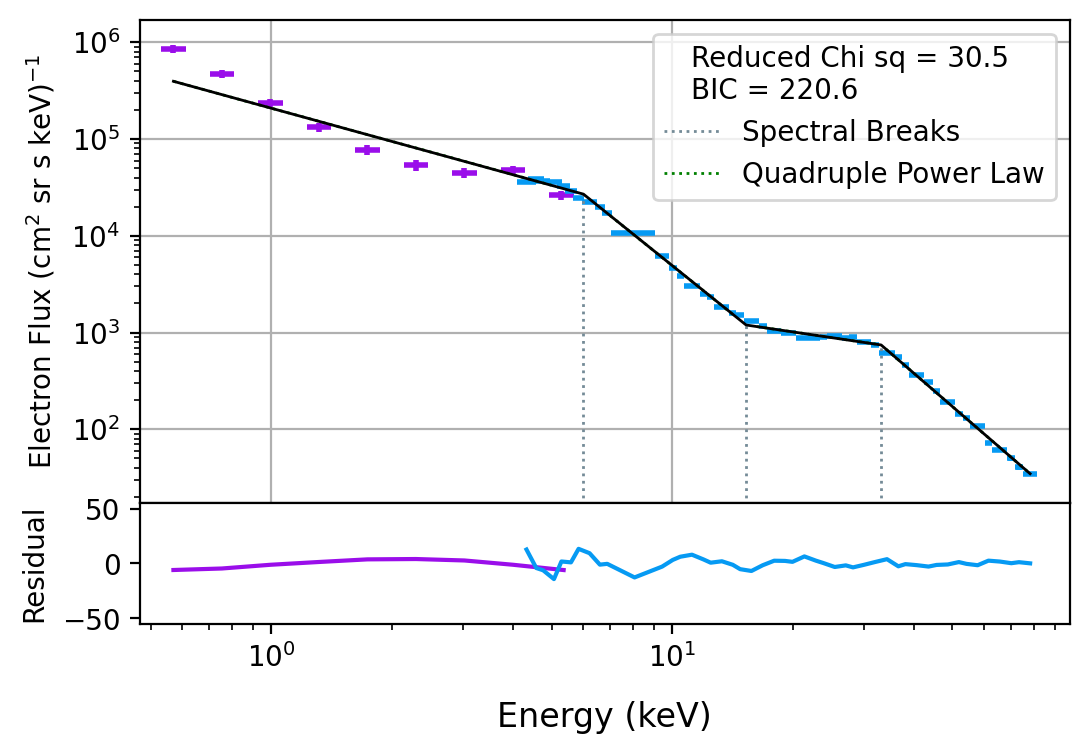}
        \caption{Quadruple power law (Table \ref{tab:quad_fits})}
        \label{5min flux 4pl spec fit fig}
    \end{subfigure}
    %\vspace{0.5cm} % Adds some vertical space
    
    \begin{subfigure}{0.49\textwidth}
        \centering
        \includegraphics[width=\textwidth]{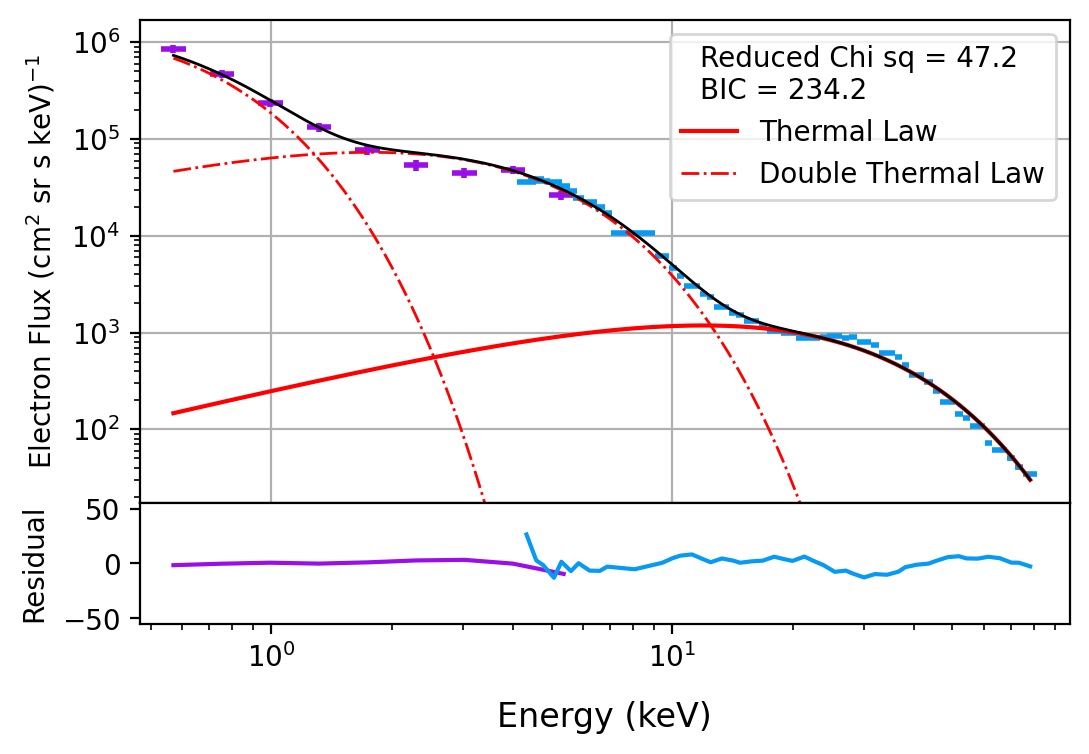}
        \caption{Triple thermal law (Table \ref{tab:3_thermal_fits})}
        \label{5min flux 3 thermal spec fit fig}
    \end{subfigure} 
    \begin{subfigure}{0.49\textwidth}
        \centering
        \includegraphics[width=\textwidth]{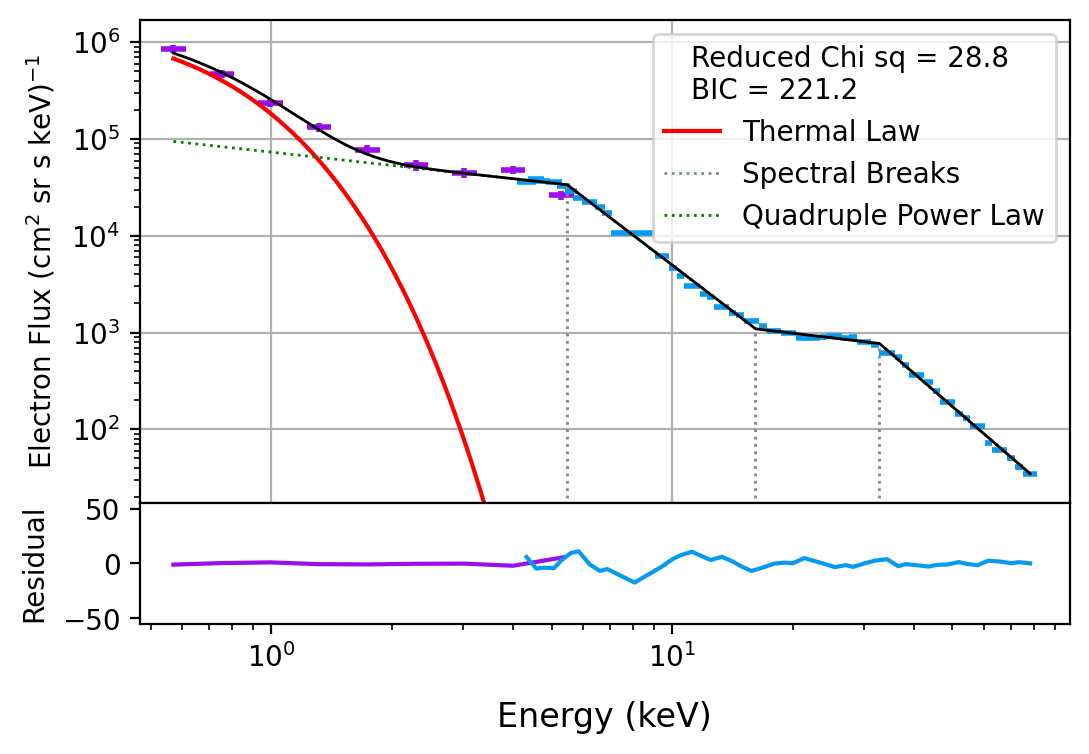}
        \caption{Thermal law and Quadruple power law (Table \ref{tab:q+thermal_fits})}
        \label{5min flux Thermal and 4pl spec fit fig}
    \end{subfigure}

    \begin{subfigure}{0.49\textwidth}
        \centering
        \includegraphics[width=\textwidth]{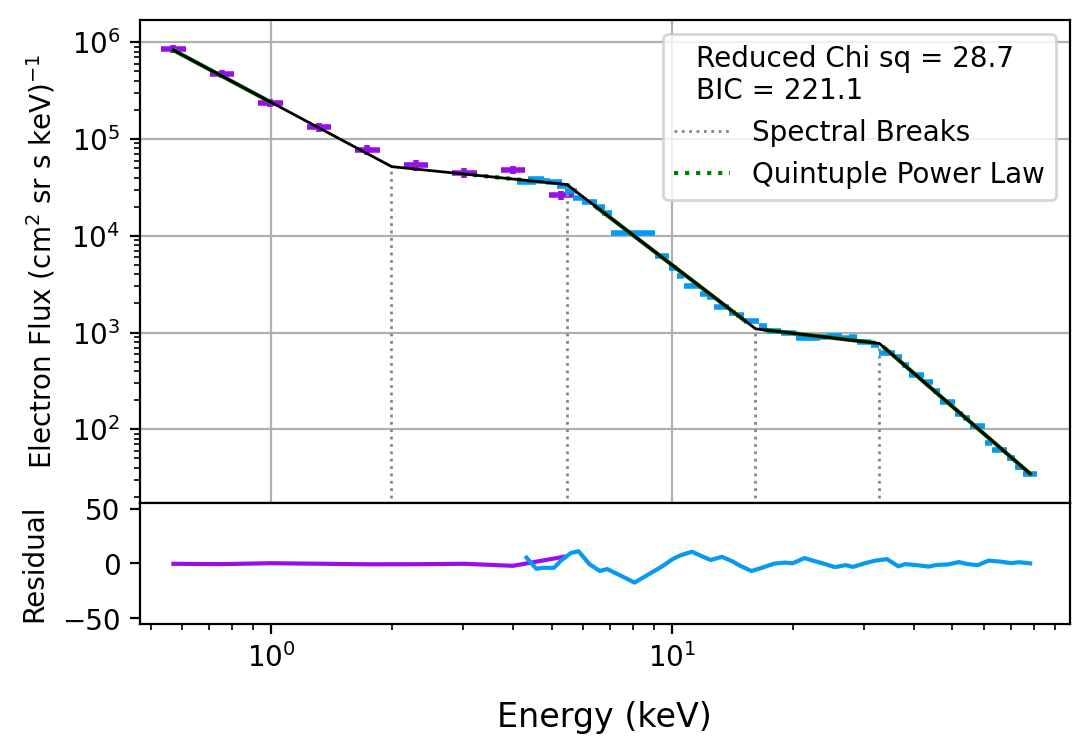}
        \caption{Quintuple power law (Table \ref{tab:quint_fits})}
        \label{5min flux 5pl spec fit fig}
    \end{subfigure}    
    \begin{subfigure}{0.49\textwidth}
        \centering
        \includegraphics[width=\textwidth]{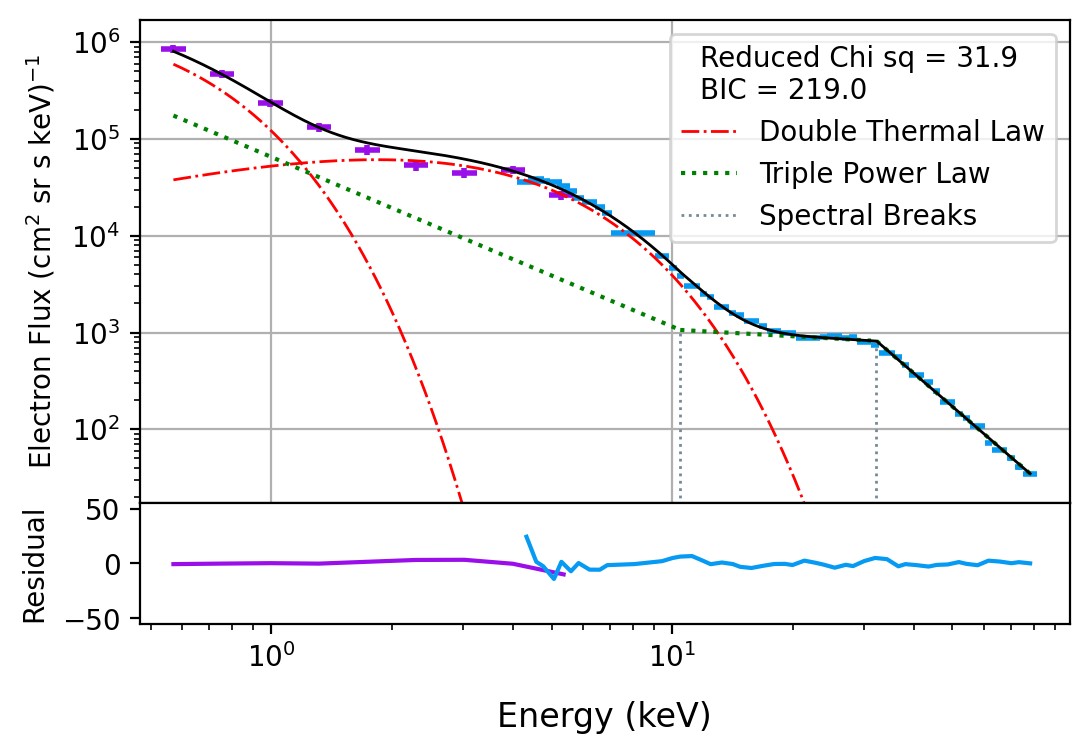}
        \caption{Double thermal law and triple power law (Table \ref{tab:2t3p fits})}
        \label{5min flux 2 therm 3 power spec fit fig}
    \end{subfigure}
    \caption{Examples of fits to the peak flux spectrum generated using time series resampled to 5 minute cadence. The total fit is shown in black, with the spectral components added as shown in the legend. These fits have their respective residuals shown below to further illustrate fit quality. Panel \ref{5min flux 2 therm 2 power spec fit fig} showing the 2 thermal component, 2 power law fit is replicated from Figure \ref{peak_spectra_figure}. The parameters of these fits are given in the tables referenced in their labels.}
    \label{flux_spectra_fits}
\end{figure*}

\begin{figure*}[!h]
    \centering
        \begin{subfigure}{0.49\textwidth}
        \centering
        \includegraphics[width=\textwidth]{5min_flue_FAF_2p2t.png}
        \caption{Double thermal law and double power law (Table \ref{tab:thermal_fits})}
        \label{5min flue 2 therm 2 power spec fit fig}
    \end{subfigure}
        \begin{subfigure}{0.49\textwidth}
        \centering
        \includegraphics[width=\textwidth]{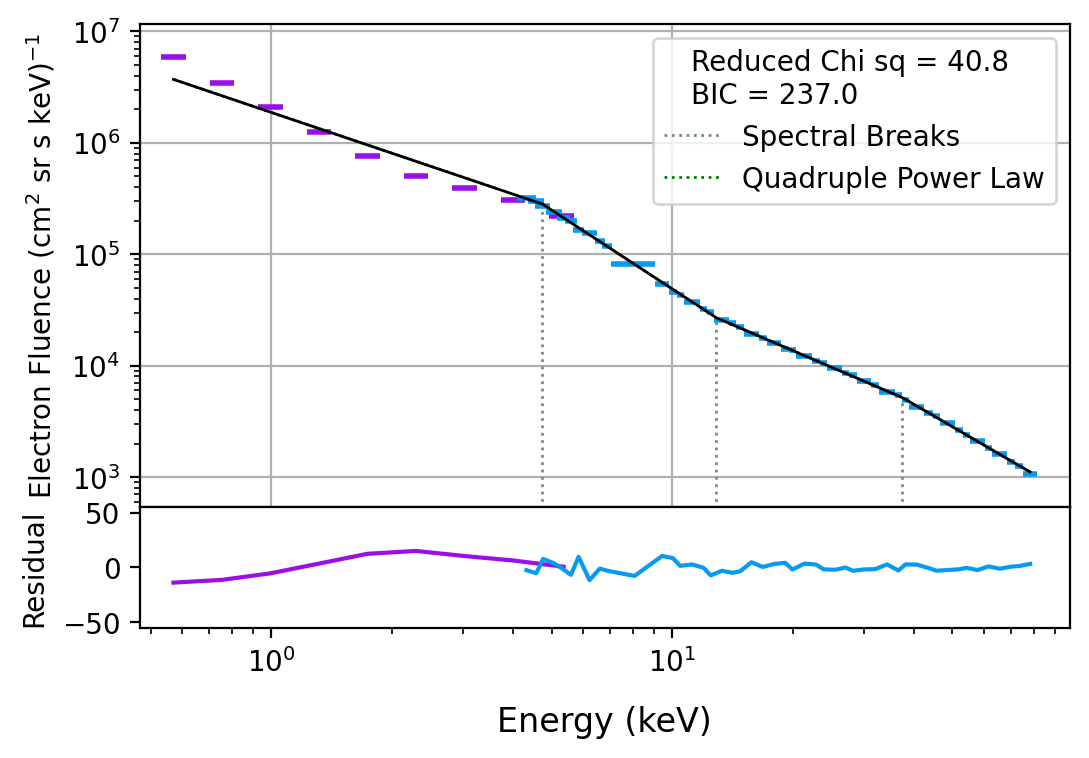}
        \caption{Quadruple power law (Table \ref{tab:quad_fits})}
        \label{5min flue 4pl spec fit fig}
    \end{subfigure}

        \begin{subfigure}{0.49\textwidth}
        \centering
        \includegraphics[width=\textwidth]{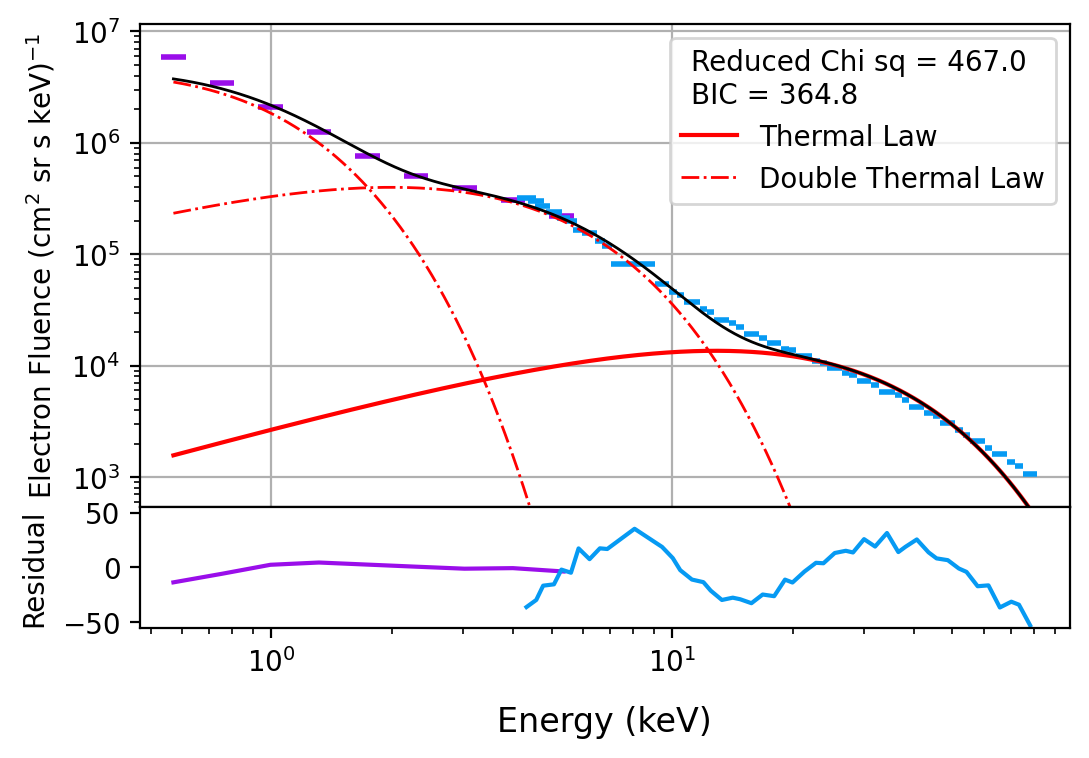}
        \caption{Triple thermal law (Table \ref{tab:3_thermal_fits})}
        \label{5min flue 3 thermal spec fit fig}
    \end{subfigure}
    \begin{subfigure}{0.49\textwidth}
        \centering
        \includegraphics[width=\textwidth]{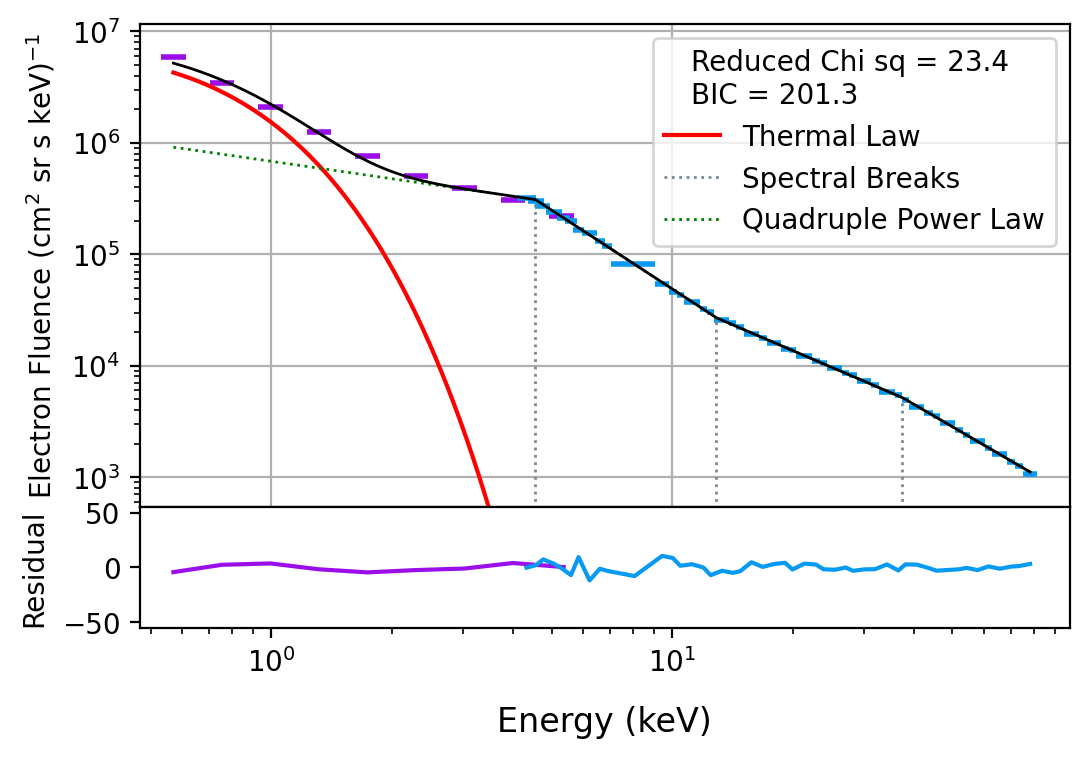}
        \caption{Thermal function and quadruple power law (Table \ref{tab:q+thermal_fits})}
        \label{5min flue  therm 4 power spec fit fig}
    \end{subfigure}

    %\vspace{0.5cm} % Adds some vertical space

    \begin{subfigure}{0.49\textwidth}
        \centering
        \includegraphics[width=\textwidth]{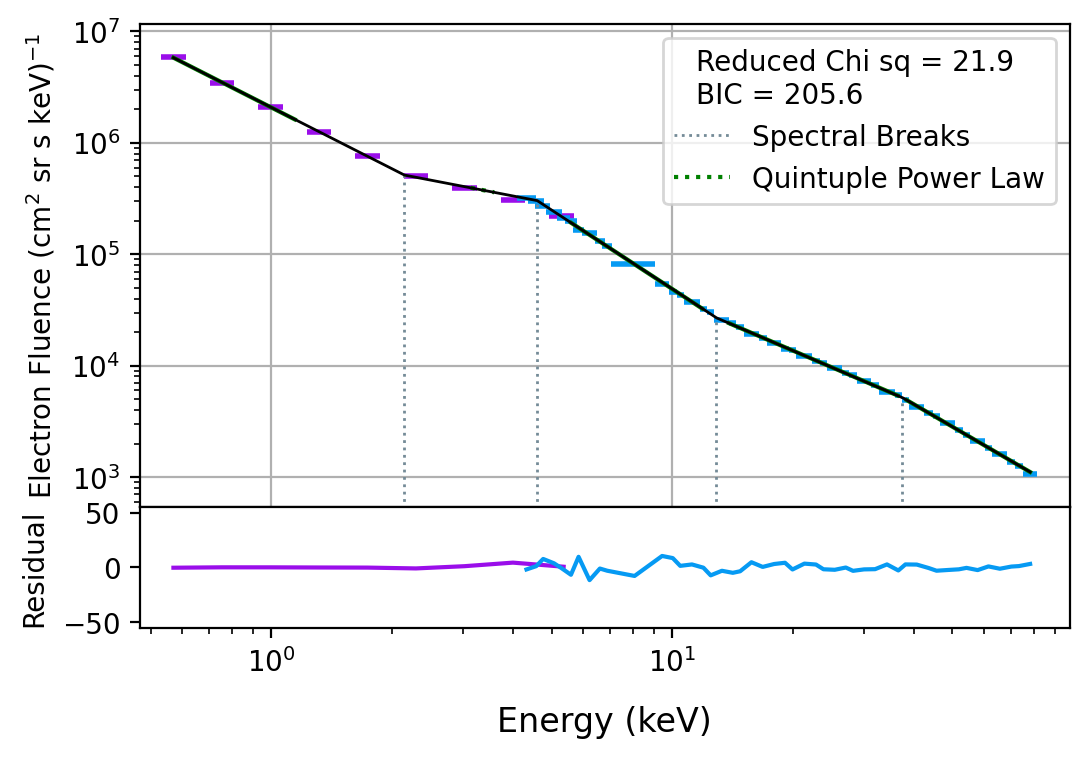}
        \caption{Quintuple power law (Table \ref{tab:quint_fits})}
        \label{5min flue 5pl spec fit fig}
    \end{subfigure}    
    \begin{subfigure}{0.49\textwidth}
        \centering
        \includegraphics[width=\textwidth]{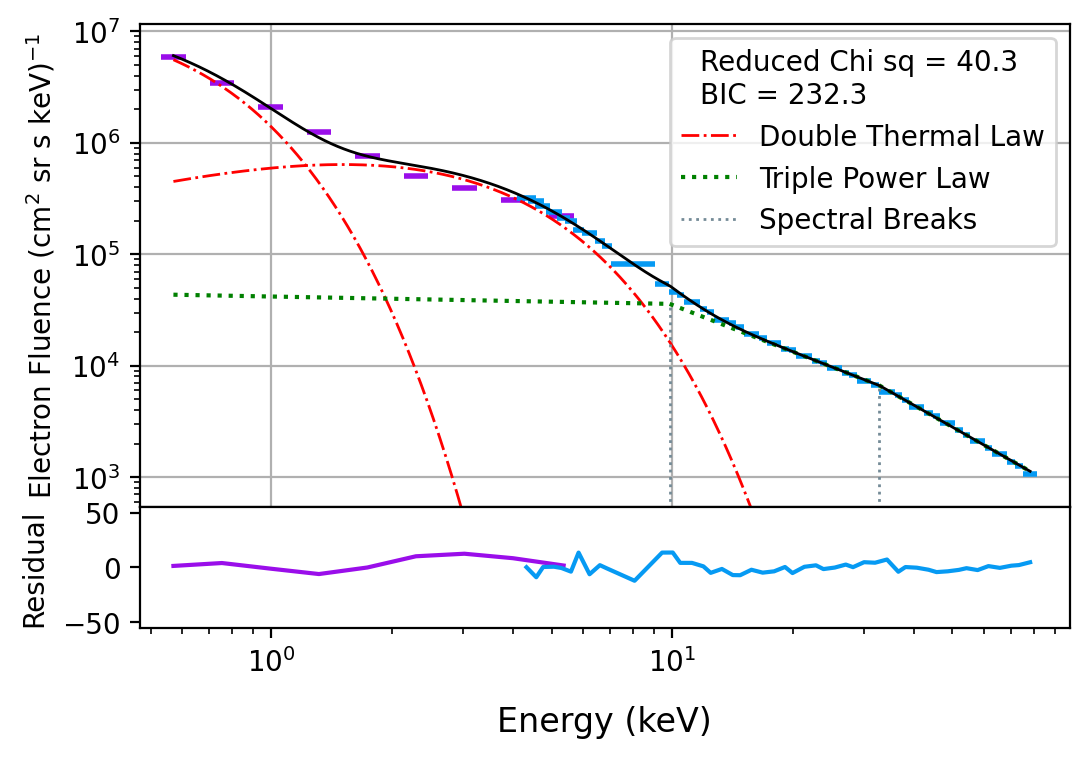}
        \caption{Double thermal law and triple power law (Table \ref{tab:2t3p fits})}
        \label{5min flue 2 therm 3 power spec fit fig}
    \end{subfigure}
    \caption{Examples of fits to the 07:00 to 15:00 UTC fluence spectrum generated using time series resampled to 5 minute cadence. The total fit is shown in black, with the spectral components added as shown in the legend. These fits have their respective residuals shown below to further illustrate fit quality. Panel \ref{5min flue 2 therm 2 power spec fit fig} showing the 2 thermal component, 2 power law fit is replicated from Figure \ref{flue_spectra_figure}. The parameters of these fits are given in the tables referenced in their labels.}
    \label{fluence_spectra_fits}
\end{figure*}

\begin{table*}[!hbpt]
    \begin{center}
    \renewcommand{\arraystretch}{1.2}  % Adjust row spacing
    \setlength{\tabcolsep}{10pt}       % Adjust column spacing
    \begin{tabular}{c c c c c c}
        \toprule
        & \(T_1\) (MK) & \(T_2\) (MK) & \(\delta_1\) & \(\delta_2\) & \(E_{b_1}\) (keV) \\ 
        \midrule
        Peak Flux   & 2.51 \(\pm\) 0.15 & 20.89 \(\pm\) 0.29 & -0.23 \(\pm\) 0.13 & -3.58 \(\pm\) 0.05 & 32.29 \(\pm\) 0.62 \\ 
        Fluence & 2.45 \(\pm\) 0.68 & 18.33 \(\pm\) 0.38 & -1.37 \(\pm\) 0.02 & -2.03 \(\pm\) 0.01 & 31.18 \(\pm\) 0.71 \\ 
        \bottomrule
    \end{tabular}
        \end{center}
    \caption{The parameters of the two thermal component, two power law fits shown in Figures \ref{Peak Flux Fit FAF spec fig} and \ref{Fluence Fit FAF spec fig}, and again in Figures \ref{5min flux 2 therm 2 power spec fit fig} and \ref{5min flue 2 therm 2 power spec fit fig}. \(T_1\) and \(T_2\) are the temperatures of the two thermal distributions, \(\delta_1\) and \(\delta_2\) are the two spectral indices, and \(E_{b_1}\) is the break energy of the double power law.}
    \label{tab:thermal_fits}
    \centering
    \renewcommand{\arraystretch}{1.2}  % Adjust row spacing
    \setlength{\tabcolsep}{5pt}       % Adjust column spacing
    \begin{tabular}{c c c c c c c c}
        \toprule
        & \(\delta_1\) & \(E_{b_1}\) (keV)& \(\delta_2\) & \(E_{b_2}\) (keV)& \(\delta_3\) & \(E_{b_3}\) (keV)& \(\delta_4\) \\ 
        \midrule
        Peak Flux   & -1.14 \(\pm\) 0.02 & 6.00 \(\pm\) 0.02 & -3.34 \(\pm\) 0.01 & 15.27 \(\pm\) 0.08 & -0.61 \(\pm\) 0.03 & 33.16 \(\pm\) 0.18& -3.58 \(\pm\) 0.03\\ 
        Fluence & -1.21 \(\pm\) 0.007 & 4.74 \(\pm\) 0.01 & -2.36 \(\pm\) 0.004 & 12.86 \(\pm\) 0.06 & -1.54 \(\pm\) 0.005 & 37.46 \(\pm\) 0.37 & -2.10 \(\pm\) 0.01 \\ 
        \bottomrule
    \end{tabular}
    \caption{The parameters of the quadruple power law fits shown in Figures \ref{5min flux 4pl spec fit fig} and \ref{5min flue 4pl spec fit fig}. The \(\delta\)s are the spectral indices, and the \(E_{b_n}\) are the break energies between them.}
    \label{tab:quad_fits}
    \begin{center}
    \renewcommand{\arraystretch}{1.2}  % Adjust row spacing
    \setlength{\tabcolsep}{10pt}       % Adjust column spacing
    \begin{tabular}{c c c c }
        \toprule
        & \(T_1\) (MK) & \(T_2\) (MK) &  \(T_3\) (MK)\\ 
        \midrule
        Peak Flux   & 2.66 \(\pm\) 0.14 & 20.47 \(\pm\) 0.05 & 138.29 \(\pm\) 0.50 \\ 
        Fluence & 4.12\(\pm\) 0.05 & 23.07 \(\pm\) 0.03 & 149.18 \(\pm\) 0.25 \\ 
        \bottomrule
    \end{tabular}
        \end{center}
    \caption{The parameters of the 3 thermal component fits shown in Figures \ref{5min flux 3 thermal spec fit fig} and \ref{5min flue 3 thermal spec fit fig}.}
    \label{tab:3_thermal_fits}
    \begin{center}
    \renewcommand{\arraystretch}{1.2}  % Adjust row spacing
    \setlength{\tabcolsep}{5pt}       % Adjust column spacing
    \begin{tabular}{c c c c c c c c c}
        \toprule
        & T (MK)& \(\delta_1\) & \(E_{b_1}\) (keV)& \(\delta_2\) & \(E_{b_2}\) (keV)& \(\delta_3\) & \(E_{b_3}\) (keV)& \(\delta_4\) \\ 
        \midrule
        Peak Flux   & 2.65 \(\pm\) 0.18& -0.45 \(\pm\) 0.04 & 5.47 \(\pm\) 0.02 & -3.17 \(\pm\) 0.01 & 16.12 \(\pm\) 0.09 & -0.49 \(\pm\) 0.03 & 32.84 \(\pm\) 0.18& -3.58 \(\pm\)0.03 \\  
        Fluence & 3.14 \(\pm\) 0.08 & -0.52 \(\pm\) 0.04 & 4.55 \(\pm\) 0.01 & -2.35 \(\pm\) 0.004 & 12.89 \(\pm\) 0.07 & -1.54 \(\pm\) 0.005 & 37.46 \(\pm\) 0.38& -2.10 \(\pm\)0.01 \\ 
        \bottomrule
    \end{tabular}
        \end{center}
    \caption{The parameters of the single thermal component, quadruple power law fits shown in  Figures \ref{5min flux Thermal and 4pl spec fit fig} and \ref{5min flue  therm 4 power spec fit fig}. \(T_1\) is the temperature of the thermal distribution, the \(\delta\)s are the spectral indices, and the \(E_{b_n}\) are the break energies between them.}
    \label{tab:q+thermal_fits}
    \centering
    \renewcommand{\arraystretch}{1.2}  % Adjust row spacing
    \setlength{\tabcolsep}{5pt}       % Adjust column spacing
    \footnotesize
    \begin{tabular}{c c c c c c c c c c}
        \toprule
        & \(\delta_1\) & \(E_{b_1}\) (keV)& \(\delta_2\) & \(E_{b_2}\) (keV)& \(\delta_3\) & \(E_{b_3}\) (keV)& \(\delta_4\)& \(E_{b_4}\) (keV)& \(\delta_5\) \\ 
        \midrule
        Peak Flux   & -2.21 \(\pm\) 0.14 & 1.99 \(\pm\) 0.13 & -0.42 \(\pm\) 0.04 & 5.46 \(\pm\) 0.02 & -3.17 \(\pm\) 0.01 & 16.12 \(\pm\) 0.09& -0.49 \(\pm\)0.03&32.84 \(\pm\) 0.17&-3.58 \(\pm\)0.03\\ 
        Fluence & -1.83 \(\pm\) 0.03 & 2.14 \(\pm\) 0.06 & -0.69 \(\pm\) 0.04 & 4.60 \(\pm\) 0.01 & -2.36 \(\pm\) 0.004 &12.86 \(\pm\) 0.06& -1.54 \(\pm\) 0.005&37.46 \(\pm\) 0.38&-2.10 \(\pm\) 0.01\\ 
        \bottomrule
    \end{tabular}
    \caption{The parameters of the quintuple power law fits shown in Figures \ref{5min flux 5pl spec fit fig} and \ref{5min flue 5pl spec fit fig}. The \(\delta\)s are the spectral indices, and the \(E_{b_n}\) are the break energies between them.}
    \label{tab:quint_fits}
    \begin{center}
    \renewcommand{\arraystretch}{1.2}  % Adjust row spacing
    \setlength{\tabcolsep}{5pt}       % Adjust column spacing
    \begin{tabular}{c c c c c c c c}
        \toprule
        & \(T_1\) (MK)& \(T_2\) (MK)&\(\delta_1\) & \(E_{b_1}\) (keV)& \(\delta_2\) & \(E_{b_2}\) (keV)& \(\delta_3\) \\ 
        \midrule
        Peak Flux   & 2.32 \(\pm\) 0.13& 21.30 \(\pm\) 0.25 & -1.75 \(\pm\) 0.06 & 10.47 \(\pm\) 0.07 & -0.23 \(\pm\) 0.12 & 32.31 \(\pm\) 0.78 & -3.58 \(\pm\) 0.17  \\  
        Fluence & 2.56 \(\pm\) 0.08 & 17.47 \(\pm\) 0.26 & -0.07 \(\pm\) 0.27 & 9.85 \(\pm\) 0.18 & -1.40 \(\pm\) 0.03 & 32.78 \(\pm\) 1.49 & -2.05 \(\pm\) 0.03 \\ 
        \bottomrule
    \end{tabular}
        \end{center}
    \caption{The parameters of the double thermal component, triple power law fits shown in Figures \ref{5min flux 2 therm 3 power spec fit fig} and \ref{5min flue 2 therm 3 power spec fit fig}. \(T_1\) and \(T_2\) are the temperatures of the thermal distributions, the \(\delta\)s are the spectral indices, and the \(E_{b_n}\) are the break energies between them.}
    \label{tab:2t3p fits}
    
\end{table*}

The fits shown in Figure \ref{flue_spectra_figure} show larger values of BIC and reduced chi-squared than the corresponding fits in Figure \ref{peak_spectra_figure}, as the integration performed to calculate the fluence inherently reduces the uncertainties of the flux values, thus increasing the calculated residuals. Table \ref{tab:thermal_fits} shows the fitted parameters to the fits where FAF has been applied, and shows that peak flux and fluence fits produce temperatures lying within each others error bars. Visually, the fluence spectrum does not show as clear spectral components as the peak flux spectrum, and this is reflected in the parameters fitted. The fluence spectrum is fitted with thermal components of \(2.45 \pm 0.68\) MK and \(18.33 \pm 0.38\) MK, a difference of 15.88 MK, while for peak flux the temperatures are \(2.51 \pm 0.15\) MK and \(20.89 \pm 0.29\) MK, a greater difference of 18.38 MK. The same trend is seen for the spectral indices, with fluence having indices of \(-1.37 \pm 0.02\) and \(-2.03 \pm 0.01\), a difference of 0.66, while for peak flux the indices are \(-0.23 \pm 0.13\) and \(-3.58 \pm 0.05\), a difference of 3.35. The more distinct differences seen in peak flux are representative of the stronger signal we expect to see from this spectral type. The less distinct spectrum for fluence however comes from the hardening of the second spectral index compared to peak flux and lower temperature of the second, higher temperature thermal component, \(T_2\). The spectral break energy also moves lower for the fluence spectrum which indicates a hardening of the spectrum down to lower energies compared to peak flux.

In Figure \ref{flux_spectra_fits}, we compare combinations of different fitting functions. We see that according to $\chi^{2}_{\nu}$, a quintuple power law function (fit \ref{5min flux 5pl spec fit fig}) best fits the peak flux spectrum, while three thermal curves (fit \ref{5min flux 3 thermal spec fit fig}) fits worst. We expected this, as the high energy component of the distribution ought to be dominated by non-thermal process that would not be well modelled with a thermal function. However, all the values of BIC are within 10 of each other, with the exception of the three thermal law function. This suggests that those fits are all equally good \citep{kass-1995}. This comes in part from the very low measured errors on the STEP data, which weight the fit strongly to that instrument's points. As the greatest differences between fits can be seen visually within the EAS data range, these differences are being masked by this weighting. This is visible most clearly in Figure \ref{5min flux 2 therm 2 power spec fit fig}. The fit is visually very poor for data points below ~10 keV, but has a BIC close to the other fitted functions. It is worth considering what each function represents physically. The functions which are simply power laws, \ref{5min flux 5pl spec fit fig} and \ref{5min flux 4pl spec fit fig}, represent a model including only non-thermal acceleration. The triple thermal function, \ref{5min flux 3 thermal spec fit fig}, represents a model where we are considering only heating. The remaining functions all represent some combination of the two processes.

The final spectral break, which is present in all power law fits in Figure \ref{flux_spectra_fits} and noted in Tables \ref{tab:thermal_fits}, \ref{tab:quad_fits}, \ref{tab:q+thermal_fits}, \ref{tab:quint_fits}, and \ref{tab:2t3p fits} takes very similar values, at \(32.29 - 33.16\) keV. The final spectral index in those functions takes the value of \(-3.58\) no matter what function form is used, agreeing with prompt event observations from other work \citep[e.g., ][]{2007-Krucker, 2013-reid, Jebaraj_2023},  and consistent with solar flare models. These notably do not include the worst fitting triple thermal function. When we compare the residuals at the highest energies between the thermal and non-thermal fits, we see that the best fits are the non-thermal power laws. This provides strong evidence for the non-thermal component we expected to see in this energy range as described in other studies \citep[e.g., ][]{2007-Krucker}. Because the fit statistics strongly favour the STEP data, which has smaller uncertainties, this masks the fact that the lowest energies are best fitted by thermal curves. When we review the residuals shown in Figure \ref{flux_spectra_fits}, it is not any clearer which function form fits best, with each form performing similarly for the EAS data with the exception of the quadruple power law. This can also be visually assessed, as it is clear from fit \ref{5min flux 4pl spec fit fig} that a power law struggles to match the shape created by the data points when compared to thermal fits or the spectral break at low energies present in the quintuple power law. The lowest temperature thermal curves fitted in fits \ref{5min flux Thermal and 4pl spec fit fig}, \ref{5min flux 3 thermal spec fit fig}, and \ref{5min flux 2 therm 2 power spec fit fig} of Figure \ref{flux_spectra_fits} take similar values in the region of 2.3 to 2.7 MK with overlapping error bars, as seen in Tables \ref{tab:thermal_fits}, \ref{tab:3_thermal_fits}, and \ref{tab:q+thermal_fits}. This temperature matches hot corona/active region temperature population \citep{2013-DelZanna,2021-DelZanna}, which we may expect to accompany a solar flare. This population accompanies the background solar wind and flare accelerated electron populations. The values are most similar in \ref{5min flux 3 thermal spec fit fig} and \ref{5min flux 2 therm 2 power spec fit fig}, where they are followed by a second thermal curve which also matches closely in the 17 to 20 MK region (see Tables \ref{tab:thermal_fits} and \ref{tab:3_thermal_fits}), similar to flaring plasma temperatures measured by X-ray observations \citep[e.g.,][]{2014-Jeffrey, 2015-Kontar}. 

In Figure \ref{fluence_spectra_fits}, we see that according to the reduced chi-squared, the quintuple power law best fits our fluence spectrum, while a triple thermal function fits worst. However, the BIC shows that the single thermal and quadruple power law is the best fit. It is worth noting that the single thermal/quadruple power law and quintuple power law fits of Figure \ref{fluence_spectra_fits} show BIC within 10 of each other, and therefore neither fit can be picked out as the best \citep{kass-1995}. However, the triple thermal law is clearly the worst by both BIC and chi-squared. When we look at the residuals in Figure \ref{fluence_spectra_fits}, we see that between 0.5 and 3 keV the triple thermal function fits as well or better than the other fits. The major difference comes between 10 and 80 keV, where the triple thermal law is significantly worse than any other. This suggests that a non-thermal power law is required to achieve a good fit to the data above 10 keV, as we would expect, and provides evidence that a thermal mechanism is at least as good as a non-thermal mechanism for explaining the spectral shape at lower energies, below 3 keV. We see that in fits \ref{5min flue 2 therm 2 power spec fit fig}, \ref{5min flue  therm 4 power spec fit fig}, \ref{5min flue 3 thermal spec fit fig}, and \ref{5min flue 2 therm 3 power spec fit fig} of Figure \ref{fluence_spectra_fits} (and their corresponding Tables \ref{tab:thermal_fits}, \ref{tab:3_thermal_fits}, \ref{tab:q+thermal_fits} and \ref{tab:2t3p fits}), the four \(T_1\) thermal curves fitted  take different values in the region of \(2.4 - 4\) MK. These are grouped much less closely than we saw in the peak flux spectra, and, as occurred with flux, none lie within each other's error bars. In addition, the difference in \(T_2\) from fits \ref{5min flue 3 thermal spec fit fig} and \ref{5min flue 2 therm 2 power spec fit fig} is significantly higher. In the 3 to 20 keV range these second thermal functions try to fit, the higher temperature of 23.07 MK (fit \ref{5min flue 3 thermal spec fit fig}) seems to fit more poorly than the lower value of 18.33 MK (fit \ref{5min flue 2 therm 2 power spec fit fig}), though both are within the expected range for flaring plasma temperatures. It is worth noting that due to low measurement uncertainties we see very high values of residuals, reduced chi-squared, and BIC for fluence, so they are best used here comparatively rather than a direct measures of fit quality. This is true for peak flux too, though not as extremely as in the fluence case.

\begin{figure*}[hbpt]
    \centering

    \begin{subfigure}{0.49\textwidth}
        \centering
        \includegraphics[width=\textwidth]{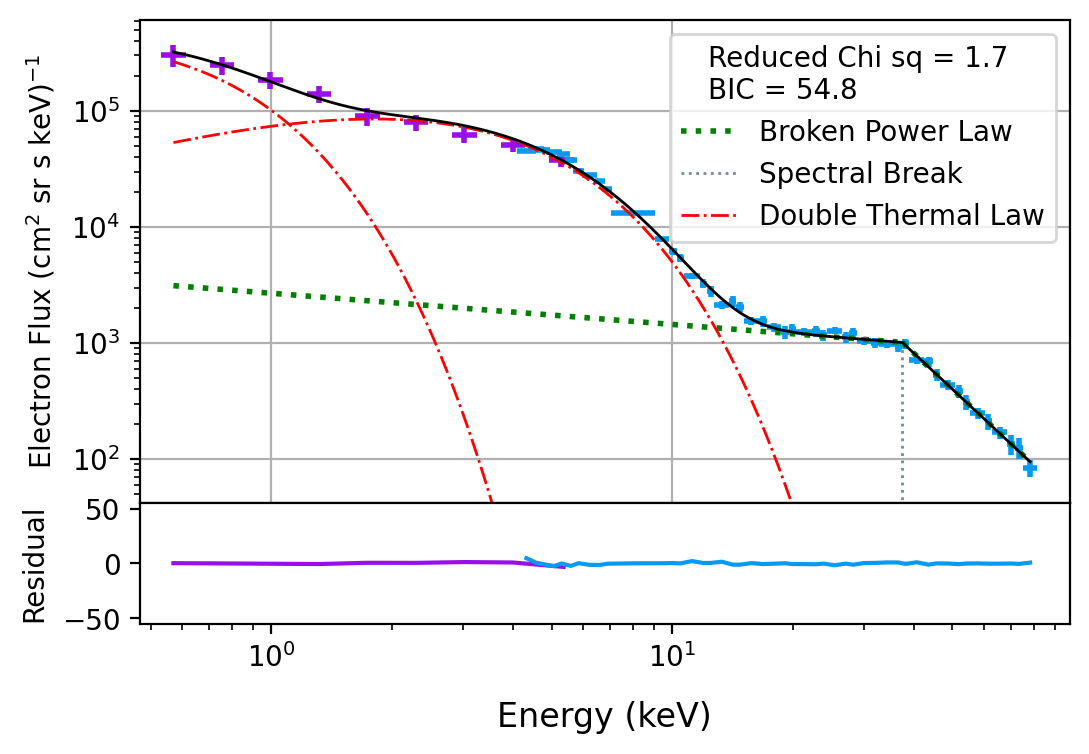 }
        \caption{Raw cadence}
        \label{raw}
    \end{subfigure}
    \begin{subfigure}{0.49\textwidth}
        \centering
        \includegraphics[width=\textwidth]{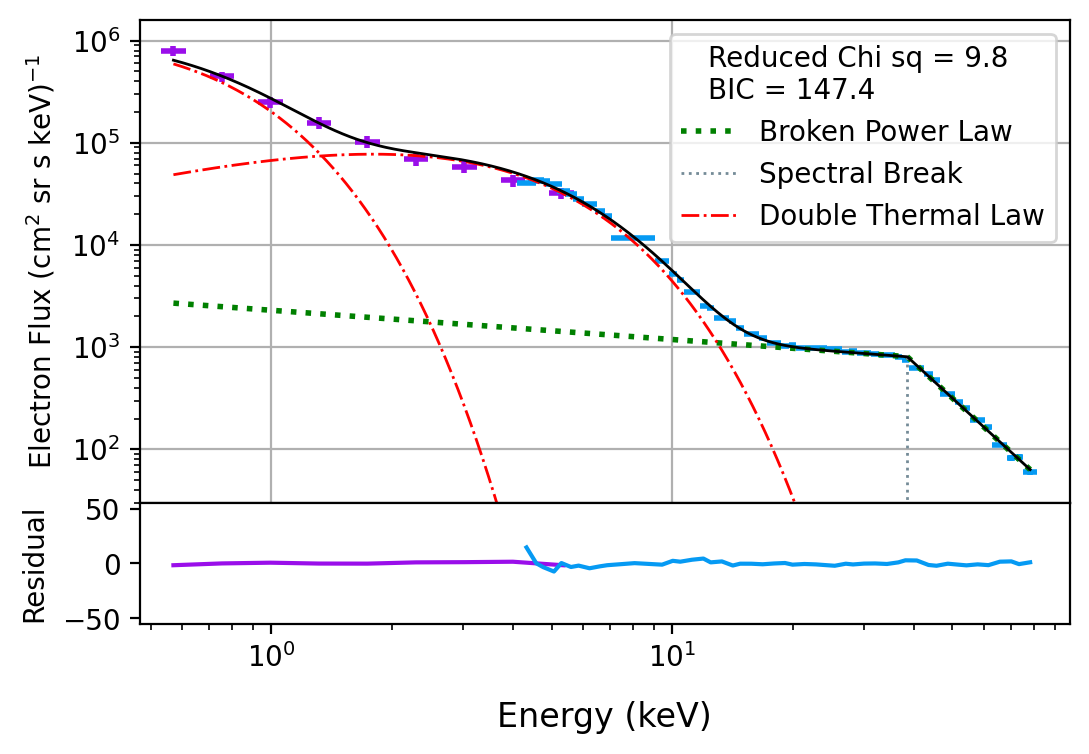}
        \caption{2 minute cadence}
        \label{2min flux 2 therm 2 power}
    \end{subfigure} 
    
    \begin{subfigure}{0.49\textwidth}
        \centering
        \includegraphics[width=\textwidth]{5min_flux_FAF_2p2t.png}
        \caption{5 minute cadence}
        \label{5min flux 2 therm 2 power}
    \end{subfigure}
    \begin{subfigure}{0.49\textwidth}
        \centering
        \includegraphics[width=\textwidth]{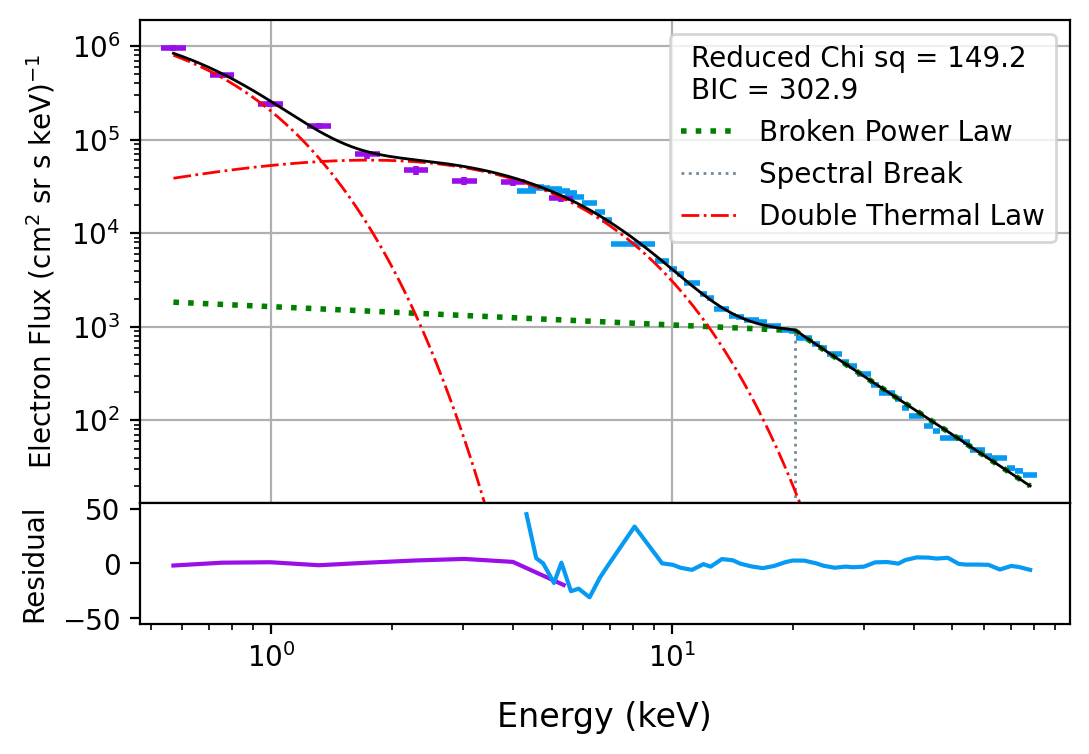}
        \caption{ 10 minute cadence}
        \label{10min flux 2 therm 2 power}
    \end{subfigure}
    %\vspace{0.5cm} % Adds some vertical space

    \begin{subfigure}{0.49\textwidth}
        \centering
        \includegraphics[width=\textwidth]{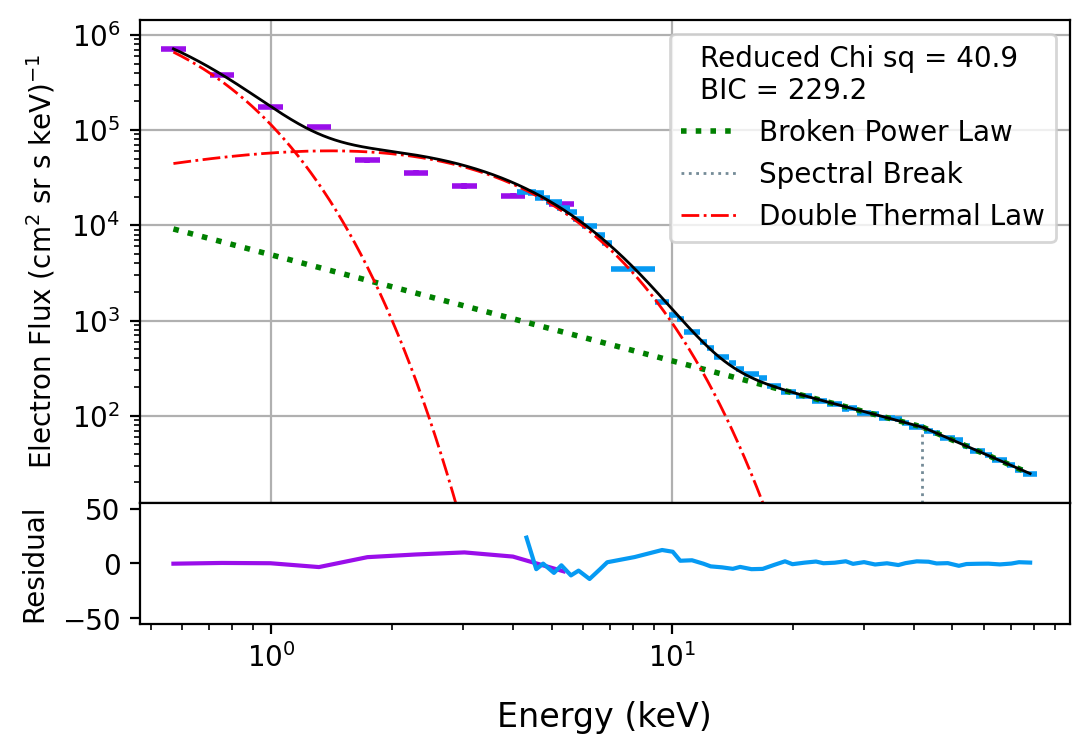}
        \caption{30 minute cadence}
        \label{30min flux 2 therm 2 power}
    \end{subfigure}
    \begin{subfigure}{0.49\textwidth}
        \centering
        \includegraphics[width=\textwidth]{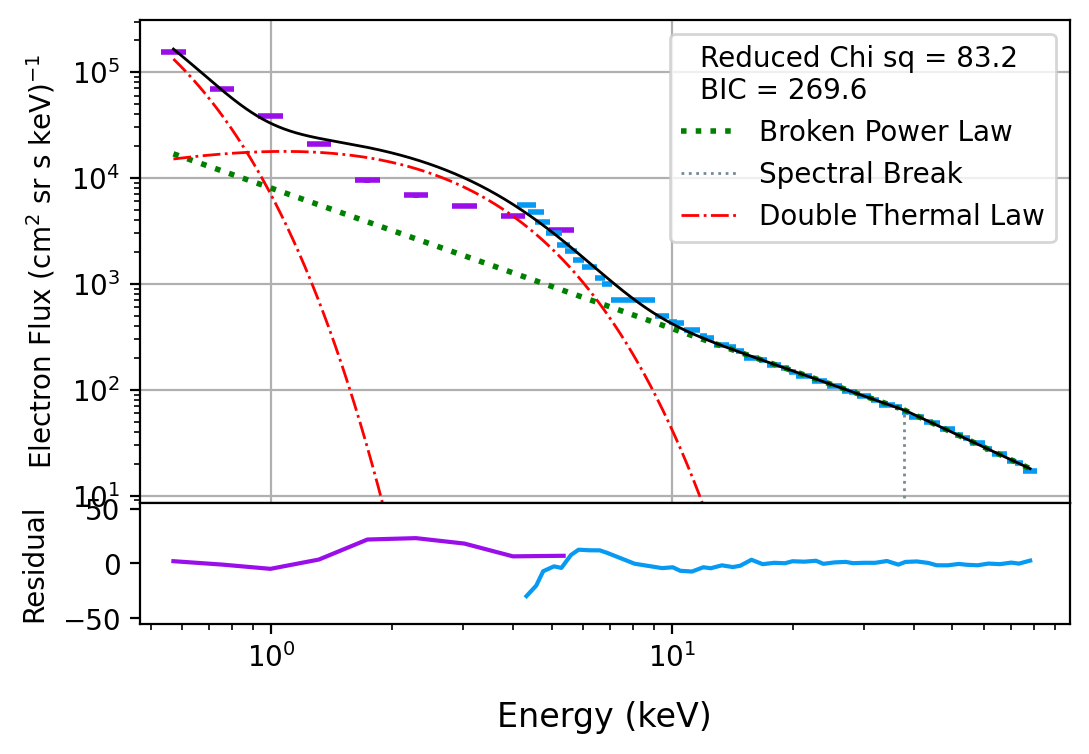}
        \caption{1 hour cadence}
        \label{1H flux 2 therm 2 power}
    \end{subfigure}

    \caption{Showing the effect of resampling on the fit of the double thermal function and double power law function. Each panel shows a peak flux spectrum generated from time series of a different cadence. The total fit is shown in black, with the spectral components added as shown in the legend. These fits have their respective residuals shown below to further illustrate fit quality. The parameters of these fits are given in Table \ref{tab:resample_fits}.}
\label{resample_fits}
\end{figure*}

In both peak flux and fluence spectra shown in Figures \ref{flux_spectra_fits} and \ref{fluence_spectra_fits}, a spectral break is visible. This spectral break is described in the literature \citep[e.g., ][]{2007-Krucker}. The break in \cite{2007-Krucker} is at around \(50\) keV, which is about 10 keV higher than we find, with our values ranging between 31 and 38 keV. For both spectra, there is a softening of the index at this point, with the index getting significantly greater in magnitude after that point. It could be related to wave-particle interactions which modify the spectrum and cause this flattening at lower energies \citep{2013-reid}, or possibly due to the transition between the thermal and non-thermal acceleration regimes \citep{2023-Pallister,2025-Pallister}. We see this first spectral component disappear as we increase resampling time (see Figure \ref{resample_fits}) and in our fluence spectra, causing the spectrum to tend towards a single power law rather than a broken one. However, the widths of the peaks are energy dependent, as seen in figure \ref{time_series_figure}. Therefore, as we change the time averaging, each peak will be affected differently and the spectral shape will change artificially. It is therefore important to consider that as the data is averaged to improve counting statistics, the shape of the resultant spectrum is changed in turn.

In most cases, parameters of the fits do not agree within error between peak flux and fluence. We expect the peak flux and fluence to take differing shapes, as they represent differing samples of the population, and therefore this result is as we expected, and as described in literature \citep[e.g., ][]{2013-reid}.

We also test how resampling alters the spectra and the fits in Figure \ref{resample_fits} and Table \ref{tab:resample_fits}. As we increase the resampling time, we can see that the first power law component steadily decreases in energy range, and is barely present at 1 hour. This is possibly a merging of the two spectral components as resampling time is increased, as we see the two spectral indices tending towards each other as resampling time increases. It is also possible, supposing that the thermal components are present, that the first power law disappears due to the lessening break energy moving that spectral component into a region where it would be hidden by the thermal curves. In this way, the  spectra at longer resample times appear similar to the fluence spectra we generated. Additionally, as resampling time is raised, the temperatures of the thermal components also decreases from 5.04 MK to 2.24 MK for \(T_1\) and 19.54 MK to 11.49 MK for \(T_2\), though these are all within typical active region and flaring material temperatures. The spectral components are made less distinct when resampling is performed, with the 1 hour resampling appearing similar to the fluence spectra. This is best illustrated by the difference between \(\delta_1\) and \(\delta_2\), which goes from a difference of 3.01 at raw cadence to 0.52 at 1 hour resampling. All FAFs are below 1, and the closer the FAF is to 1 the less change is required to align the data sets. The FAF increases with resampling, peaks at 0.39 at 10 minutes and then decreases again. The same trend is shown with the fit quality measures, suggesting that when the spectrum must be manipulated more to align the instruments, the fit quality improves. The fit is best when the raw binning is used, both in terms of chi-squared and BIC.

\begin{table*}[t]
    \centering
    \renewcommand{\arraystretch}{1.2}  % Adjust row spacing
    \setlength{\tabcolsep}{5pt}       % Adjust column spacing
    \begin{tabular}{c c c c c c c }
        \toprule
                & \(T_1\) (MK)& \(T_2\) (MK)& \(\delta_1\) & \(\delta_2\) & \(E_{b}\) (keV) & FAF\\ 
        \midrule
        Raw   & 3.27 \(\pm\) 0.50 & 20.91 \(\pm\) 0.34 & -0.27 \(\pm\) 0.10 & -3.23 \(\pm\) 0.16 & 37.45 \(\pm\) 0.95 &0.04\\ 
        2 minute   & 3.04 \(\pm\) 0.24 & 20.79 \(\pm\) 0.09 & -0.29 \(\pm\) 0.04 & -3.61 \(\pm\) 0.07 & 38.55\(\pm\) 0.47&0.27 \\ 
        5 minute   & 2.51 \(\pm\) 0.15 & 20.89 \(\pm\) 0.29 & -0.23 \(\pm\) 0.13 & -3.58 \(\pm\) 0.05 & 32.29 \(\pm\) 0.62&0.35\\
        10 minute   & 2.57 \(\pm\) 0.11 & 20.27 \(\pm\) 0.04 & -0.20 \(\pm\) 0.01 & -2.83 \(\pm\) 0.02 & 20.21 \(\pm\) 0.14 &0.39\\
        30 minute   &  2.14 \(\pm\) 0.06 & 16.28 \(\pm\) 0.03 &-1.11 \(\pm\) 0.02 & -1.82 \(\pm\) 0.06 & 42.08 \(\pm\) 1.81 &0.32\\ 
        1 hour  & 1.42 \(\pm\) 0.05 & 12.49 \(\pm\) 0.05 & -1.33 \(\pm\) 0.01 & -1.76 \(\pm\) 0.04 & 37.77 \(\pm\) 2.06 & 0.07\\ 
        \bottomrule
    \end{tabular}
    \caption{The parameters of the double thermal, double power law fits shown in Figure \ref{resample_fits} for different resampling times. \(T_1\) and \(T_2\) are the temperatures of the two thermal distributions, \(\delta_1\) and \(\delta_2\) are the two spectral indices, and \(E_{b_1}\) is the break energy of the double power law.}
    \label{tab:resample_fits}
\end{table*}

\subsection{Potential Issues and Event Re-selection}

\begin{figure*}[hbtp]
 \centering
    \begin{subfigure}{0.49\textwidth}
        \centering
        \includegraphics[width=\textwidth]{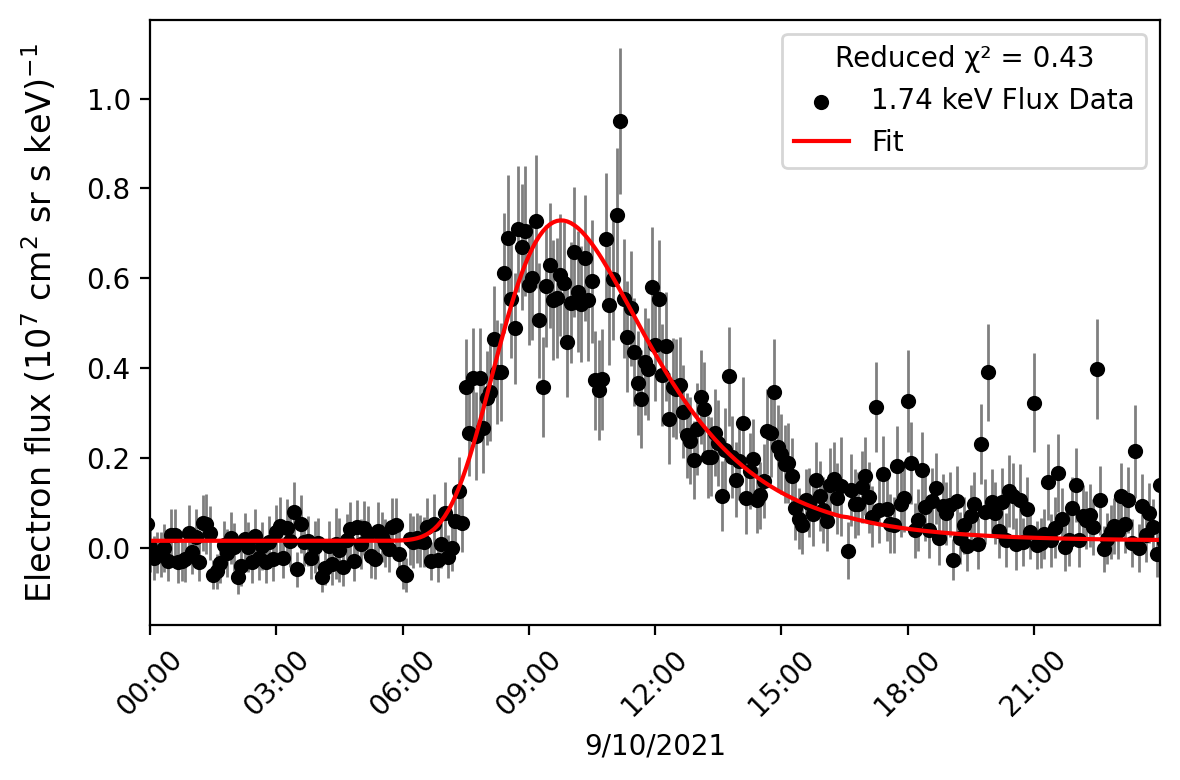}
        \caption{}
        \label{singlets}
    \end{subfigure}    
    \begin{subfigure}{0.49\textwidth}
        \centering
        \includegraphics[width=\textwidth]{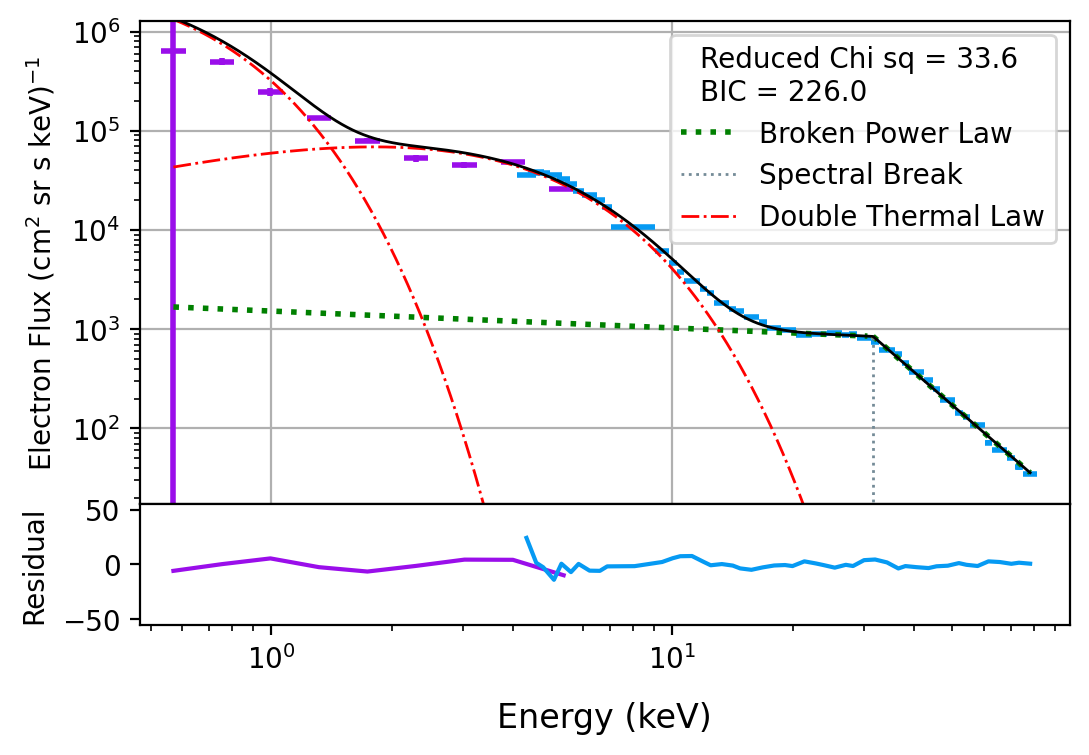}
        \caption{}
       \label{tsspec} 
    \end{subfigure}
    
\caption{By taking fits across the time series of each energy, as shown in \ref{singlets} for 1.74 keV, we generate a spectrum as shown in \ref{tsspec} by extracting the peak flux from the fit.  The total fit is shown in black, with the spectral components added as shown in the legend. The fit has residuals shown below to further illustrate fit quality.}
\label{Spectrum generation by fitting}
\end{figure*}

As the uncertainties in the data flux values are unexpectedly low (as discussed in \ref{Instrumental, Data and Fitting Issues}), these must be checked to ensure issues with them are not causing problems when chi-squared is calculated. The uncertainties on the STEP data are much smaller than those on EAS, and Figures \ref{flux_spectra_fits} and \ref{fluence_spectra_fits} show clearly that the minimiser favours the STEP data points heavily when doing the fit. Future work will include an expanded analysis of the uncertainties, including how the fit changes if the STEP uncertainty values are artificially increased to levels more consistent with our expectations. This will change the final fit, and tell us about the energy ranges that determine the parameter values most strongly when other constraints are not being imposed.

It is of note that SolO STEP is only able to observe 30 degrees of sky, which limits the usable pitch angle distribution data \citep{Lorfing-2023}. This means that a potentially significant portion of the electron population was not counted towards the generated spectrum. This unfortunately limits the reliability of the results deduced for this event from the SolO STEP observations. It is possible that the flattening seen in the peak flux spectrum between approximately 20 and 40 keV is due to the lack of pitch angle coverage.

There are two concerns we have identified with using this event in our analysis: the pitch angle issue with SolO STEP and the possible interference from the CME that occurs in the same active region that day. 

In Figure \ref{Spectrum generation by fitting}, we test fitting a curve to the EAS time series data following resampling to better eliminate the effect of the spikiness on the extracted peak value. The time series of each energy bin for EAS shown in Figure \ref{time_series_figure} was fitted with a simple straight line background and a Weibull  \citep{kahler_2017} to simulate the asymmetrical shape of the peak. The peak of that curve is taken as the peak flux, rather than directly extracting the maximum flux value from the resampled data. Additionally, we used a square-root Cauchy robust residual calculation \citep{zahra-2014,Mlotshwa-2022} rather than the standard method when fitting, as this method is more resilient to the outliers caused by the ``spikiness" we see in the time series data. The STEP data points are generated as before, because the time series data is not as ``spiky" for that instrument. This produces a spectrum similar to our previous method, requiring a FAF of 0.02 to align the EAS and STEP spectra. This produces a two thermal, two power law fit with temperatures of 2.51 and 20.94 MK, spectral indices of -0.17 and -3.51, and the spectral break at 31.68 keV. These values largely agree with previous fits. We expected to see a lower value, as the peak values taken from the time series are lower than when the direct extraction is used. This is because the time series fitting method is being used to eliminate the spikiness, so will naturally smooth the curve to a lower value. As is shown in the example time series in the first panel, this prevents any potential outliers from affecting the determined peak value. It is clear from the time series in Figure \ref{time_series_figure} that there is a lot of structure present in the time series data, and future studies into the rise times, peak structure, and decay could yield very interesting results.

\section{Discussion and Summary}

In this study, we fit curves of various forms to the in situ electron spectra observed by SolO EPD STEP and SWA EAS, using a new fitting tool we have named INSPEX. The spectra, which appear to consist of multiple components over our studied energy range, usually require a combination of several functions to fit. We were able to fit functions of varying forms, and found that, although not a unique fit to the spectrum, a function composed of two thermal curves and a double/triple power law fits the data well, both visually and in terms of fit statistics. When we perform the fitting, we find a shape and temperature close to that observed with X-ray methods in previous studies \citep[e.g.,][]{2014-caspi,Aschwanden_2015,2015-jeffrey}, which suggests that the features we see are originating in the flare, and that we are retrieving physically meaningful parameters from our fitting. These fits suggest that electron acceleration into the heliosphere occurs in a region that may have a temperature of the order of \(10-20\) MK during the event on 09/10/2021, corresponding to flaring material temperatures expected from established literature, possibly suggesting in situ electrons may be produced in a similar region to their X-ray emitting counterparts but are somehow escaping via a complicated magnetic topology. 

We also find a similar form at the high energy end, in the broken power law components. Most of the fits performed on 5 minute resampled peak flux spectra agree on a final break energy of 33 keV and a final spectral index of 3.58. While this break energy is lower than that found by \cite{2007-Krucker} in both their X-ray and in situ peak electron spectra, the spectral index found lies between their values, and well within the range of value identified in their Table 1. Additionally, we are within the range of values found in \cite{2009-Krucker}, suggesting that we have successfully interrogated the parameter space. 

\cite{fedeli_2026} analyse similar events using STEP and EPT, considering the peak intensity spectra of 44 events. The average of the lower energy break energies they find is \(38.0 \pm 6.8\) keV, which agrees well with the values of between 31 and 38 keV we find in both our peak flux and fluence fitting. Additionally, our spectral indices agree well with the spectral indices found by this study for knee-knee triple power laws.

Other fitted spectral shapes perform well, including those representing the case where no thermal acceleration in present. While our isothermal distribution fitting returns results similar to those previously found through remote sensing, further work is needed to determine whether an isothermal distribution is truly the underlying distribution present.

 We test several variations on our methodology, including changing the resampling period and performing time series fitting. All the curves fitted to these spectra produce possible parameters with plausible values within current literature; corona/active region temperatures between 2.24 and 5.89 MK and flaring material temperatures between 11.49 and 27.17 MK. While the temperatures are in a plausible range, the best fitting curves are composed of two thermal laws and a double power law.

Our promising new software package, INSPEX, allows the determination of possible properties from the acceleration region of the electrons detected in situ. It can be used in parallel with the established method for doing so through remote observations to create a well developed picture of the flaring corona and heliosphere. Using this new flexible methodology, we may have found signatures of hot flaring plasma in in situ electron spectra, indicating that in situ electrons may have passed through or been accelerated in such regions, close to the flare. Though still with some caveats including accounting for transport effects which may have occurred between the original acceleration and detection by in situ instruments, the determined properties are consistent across multiple spectral forms, allowing a good degree of confidence.

Due to possible interference from a CME and issues with the SolO STEP view angle during this flare, which complicate the interpretation of results for this event, we are considering other flares in future work. By using our developed methodology and its associated routines we will be able to bring the analysis up to the same level far faster than was possible thus far. This will also allow us to further refine our fitting methods and routines to obtain results with greater physical meaning. The convoluted nature of the in situ spectra's multiple components has made determining the correct functions difficult, so this is likely to form a core part of further work. These events will be analysed through the viewpoint of multiple spacecraft as we investigate heliospheric transport alterations.

\section{Acknowledgements}
NLSJ gratefully acknowledges financial support from the Science and Technology Facilities Council (STFC) Grant ST/X001008/1. The authors acknowledge IDL support provided by STFC.

Solar Orbiter is a mission of international cooperation between ESA and NASA, operated by ESA. Solar Orbiter Solar Wind Analyser (SWA) data are derived from scientific sensors that have been designed and created, and are operated, under funding provided in numerous contracts from the UK Space Agency (UKSA), the UK Science and Technology Facilities Council (STFC), the Agenzia Spaziale Italiana (ASI), the Centre National d’Etudes Spatiales (CNES; France), the Centre National de la Recherche Scientifique (CNRS; France), the Czech contribution to the ESA-PRODEX program, and NASA. Solar Orbiter SWA work at UCL/MSSL was funded under STFC grants ST/T001356/1, ST/S000240/1, ST/X002152/1, and ST/W001004/1. 

This research used version 6.0.4 \citep{sunpy-6.0.4}) of the SunPy open source software package \citep{sunpy_community2020}.

The authors gratefully acknowledge the insightful comments of the reviewers, which improved this work significantly.

\begin{appendix}

\section{Spectral Generation Methodology}\label{Spectral Generation Methodology}\label{app_method}
\subsection{Loading the Data}
Once downloaded and parsed, different processes are required to get the data from EAS and STEP ready for spectral analysis. 

\begin{figure*}[!hbtp]
 \centering
    \begin{subfigure}{0.49\textwidth}
        \centering
       \includegraphics[width=\textwidth]{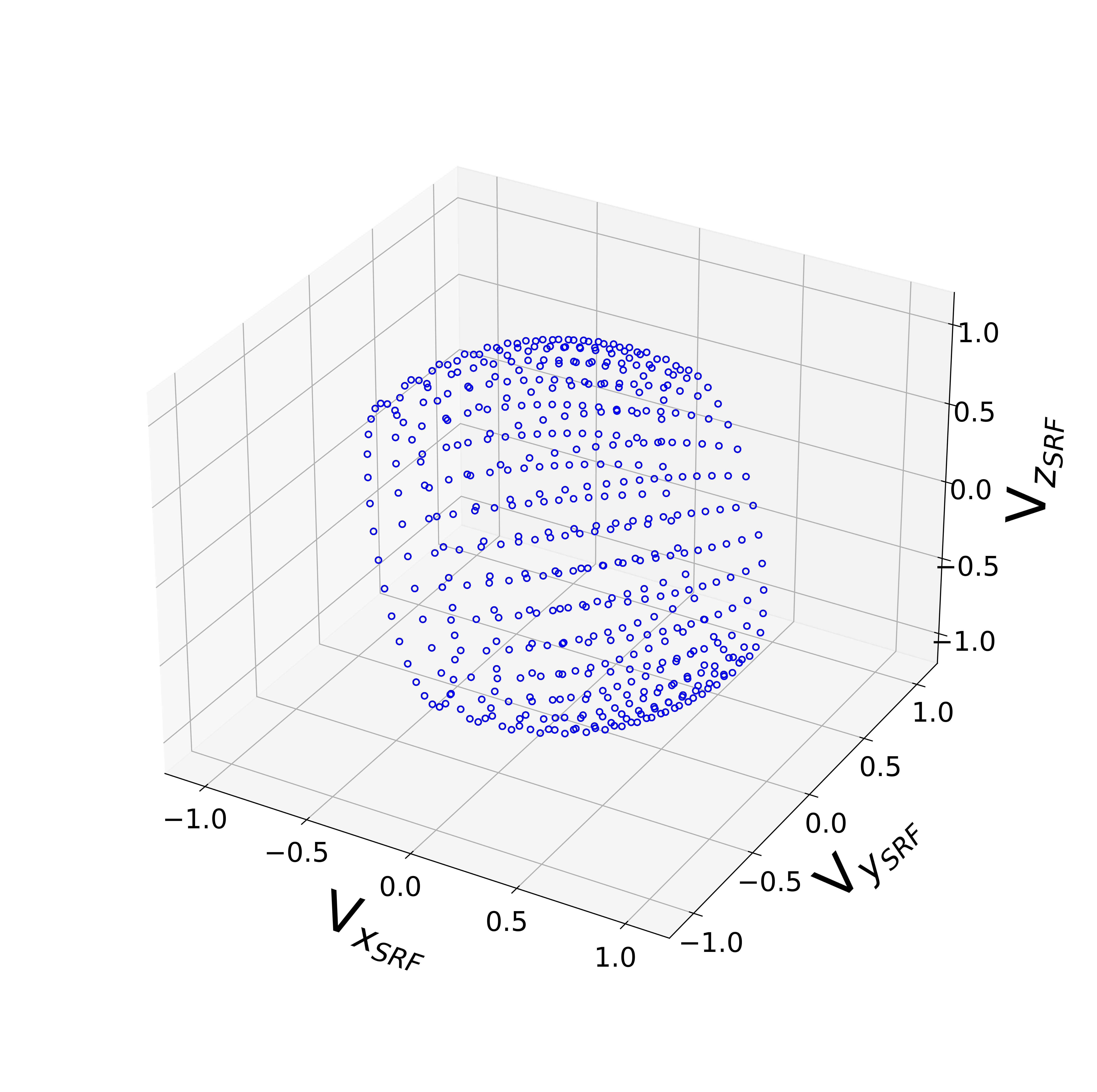}

    \end{subfigure}
    \begin{subfigure}{0.49\textwidth}
        \centering
       \includegraphics[width=\textwidth]{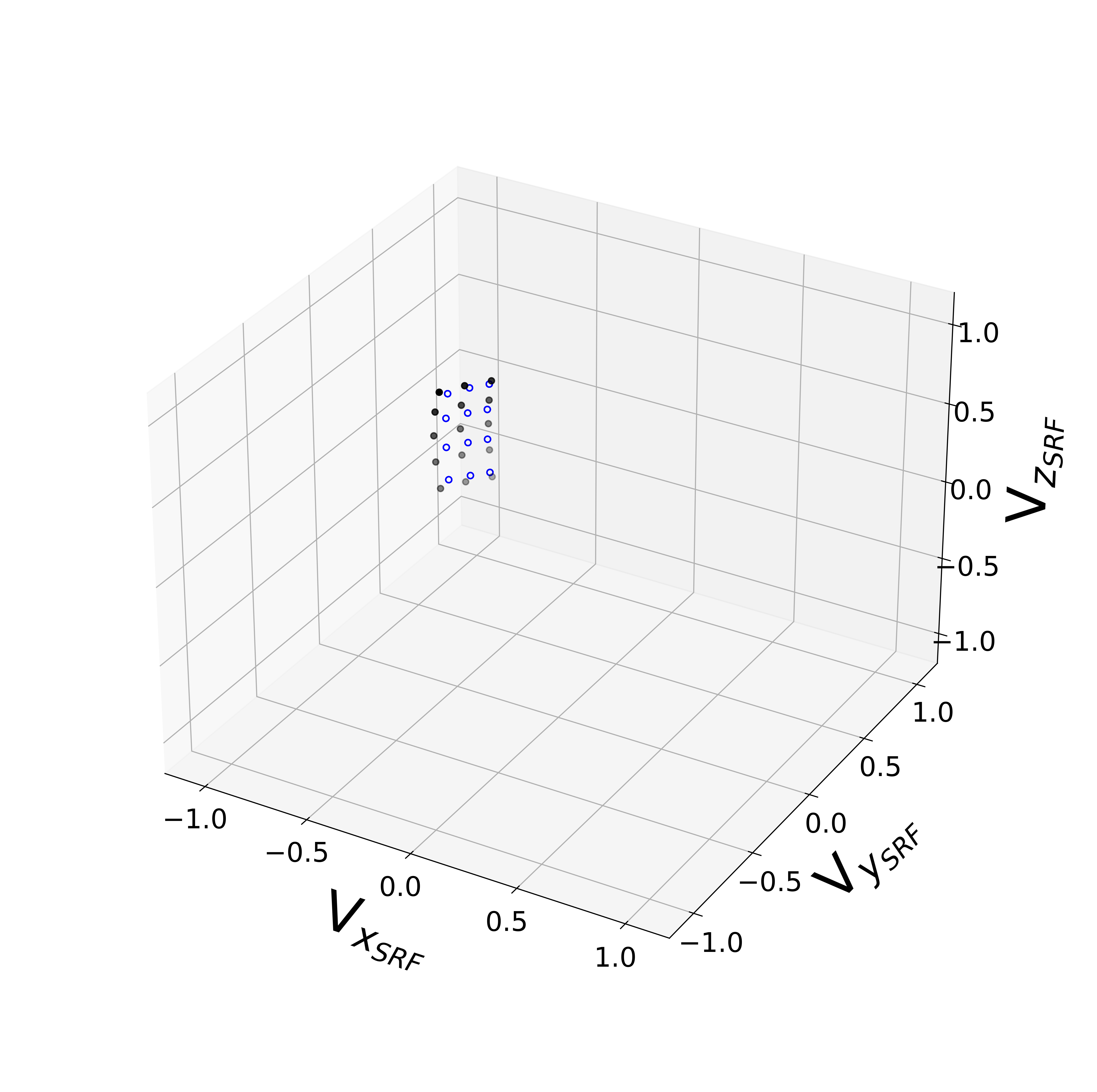}

    \end{subfigure}
\caption{On the left, the full view of EAS1 is shown, with each blue dot marking a single pixel. On the right, only the pixels which correspond to the STEP FOV in black are retained.}
\label{EAS1 3d map}
\end{figure*}

For EAS, the particle velocities must be calculated in the spacecraft frame so they can be filtered down to those which would also be seen by STEP, as EAS has a full sky FOV. The pixels left by this filtering are illustrated in Figure \ref{EAS1 3d map}. An average is then taken over the pixels to give a single value at each time in each energy channel. We then discard the odd indexed energy channels to correct for the sawtooth effect. The same analysis is done using the root count files to get series from which the uncertainties on the flux can be calculated \citep{Nicolaou_2018}.

For STEP, the first part of the parsing must always be determining whether the file is from before or after the instrument recalibration on 22/10/2021, as this determines the format/names of the variables and the required correction tables. Electron flux is found from the subtraction of the magnetic channels from the integral channels, using the pixel averaged data to match the process with EAS. This must be corrected for the difference in detector response to electrons and protons, and this table is different depending on whether the file is from before or after the instrument recalibration. Uncertainties on the electron are propagated quadratically from those on the integral and magnetic channels, and corrected using the same values in the electron correction tables.

\subsection{Processing the Data}
To account for poor counting statistics, we resample over a fixed time period, binning the time series data into the new cadence and taking a mean over the new bins. Each time series channel then has an average background taken for a period before the event, which is subtracted from the full time series to give a time series to be used in spectrum generation.

\end{appendix}

\bibliography{refs}{}

@ARTICLE{2017-Alaoui,
       author = {{Alaoui}, Meriem and {Holman}, Gordon D.},
        title = "{Understanding Breaks in Flare X-Ray Spectra: Evaluation of a Cospatial Collisional Return-current Model}",
      journal = {\apj},
         year = 2017,
        month = dec,
       volume = {851},
       number = {2},
          eid = {78},
        pages = {78},
          doi = {10.3847/1538-4357/aa98de},
archivePrefix = {arXiv},
       eprint = {1706.03897},
 primaryClass = {astro-ph.SR},
       adsurl = {https://ui.adsabs.harvard.edu/abs/2017ApJ...851...78A}
}

@ARTICLE{1978-Bai,
       author = {{Bai}, T. and {Ramaty}, R.},
        title = "{Backscatter, anisotropy, and polarization of solar hard X-rays.}",
      journal = {\apj},
         year = 1978,
        month = jan,
       volume = {219},
        pages = {705-726},
          doi = {10.1086/155830},
       adsurl = {https://ui.adsabs.harvard.edu/abs/1978ApJ...219..705B}
}

@article{2008-Benz,
   author = {Benz, A. O.},
   title = {Flare Observations},
   journal = {Living Rev Sol Phys},
   volume = {5},
   pages = {1},
   note = {Benz, Arnold O
eng
Review
Switzerland
2008/01/01
Living Rev Sol Phys. 2008;5:1. doi: 10.12942/lrsp-2008-1. Epub 2008 Feb 26.},
   ISSN = {1614-4961 (Print)
1614-4961 (Electronic)
1614-4961 (Linking)},
   DOI = {10.12942/lrsp-2008-1},
   url = {https://www.ncbi.nlm.nih.gov/pubmed/27194959},
   year = {2008},
   type = {Journal Article}
}

@ARTICLE{1971-Brown,
       author = {{Brown}, John C.},
        title = "{The Deduction of Energy Spectra of Non-Thermal Electrons in Flares from the Observed Dynamic Spectra of Hard X-Ray Bursts}",
      journal = {\solphys},
         year = 1971,
        month = jul,
       volume = {18},
       number = {3},
        pages = {489-502},
          doi = {10.1007/BF00149070},
       adsurl = {https://ui.adsabs.harvard.edu/abs/1971SoPh...18..489B}
}

@article{2020-Dressing,
   author = {Dresing, Nina and Effenberger, Frederic and Gómez-Herrero, Raúl and Heber, Bernd and Klassen, Andreas and Kollhoff, Alexander and Richardson, Ian and Theesen, Solveig},
   title = {Statistical Results for Solar Energetic Electron Spectra Observed over 12 yr with STEREO/SEPT},
   journal = {The Astrophysical Journal},
   volume = {889},
   number = {2},
   ISSN = {1538-4357},
   DOI = {10.3847/1538-4357/ab64e5},
   year = {2020},
   type = {Journal Article}
}

@article{2021-Dressing,
   author = {Dresing, N. and Warmuth, A. and Effenberger, F. and Klein, K. L. and Musset, S. and Glesener, L. and Brüdern, M.},
   title = {Connecting solar flare hard X-ray spectra to in situ electron spectra},
   journal = {Astronomy \& Astrophysics},
   volume = {654},
   ISSN = {0004-6361
1432-0746},
   DOI = {10.1051/0004-6361/202141365},
   year = {2021},
   type = {Journal Article}
}

@ARTICLE{2011-Fletcher,
       author = {{Fletcher}, L. and {Dennis}, B.~R. and {Hudson}, H.~S. and {Krucker}, S. and {Phillips}, K. and {Veronig}, A. and {Battaglia}, M. and {Bone}, L. and {Caspi}, A. and {Chen}, Q. and {Gallagher}, P. and {Grigis}, P.~T. and {Ji}, H. and {Liu}, W. and {Milligan}, R.~O. and {Temmer}, M.},
        title = "{An Observational Overview of Solar Flares}",
      journal = {\ssr},
         year = 2011,
        month = sep,
       volume = {159},
       number = {1-4},
        pages = {19-106},
          doi = {10.1007/s11214-010-9701-8},
archivePrefix = {arXiv},
       eprint = {1109.5932},
 primaryClass = {astro-ph.SR},
       adsurl = {https://ui.adsabs.harvard.edu/abs/2011SSRv..159...19F}
}

@article{2015-Fox,
   author = {Fox, N. J. and Velli, M. C. and Bale, S. D. and Decker, R. and Driesman, A. and Howard, R. A. and Kasper, J. C. and Kinnison, J. and Kusterer, M. and Lario, D. and Lockwood, M. K. and McComas, D. J. and Raouafi, N. E. and Szabo, A.},
   title = {The Solar Probe Plus Mission: Humanity’s First Visit to Our Star},
   journal = {Space Science Reviews},
   volume = {204},
   number = {1-4},
   pages = {7-48},
   ISSN = {0038-6308
1572-9672},
   DOI = {10.1007/s11214-015-0211-6},
   year = {2015},
   type = {Journal Article}
}

@ARTICLE{2020-French,
       author = {{French}, Ryan J. and {Matthews}, Sarah A. and {van Driel-Gesztelyi}, Lidia and {Long}, David M. and {Judge}, Philip G.},
        title = "{Dynamics of Late-stage Reconnection in the 2017 September 10 Solar Flare}",
      journal = {\apj},
         year = 2020,
        month = sep,
       volume = {900},
       number = {2},
          eid = {192},
        pages = {192},
          doi = {10.3847/1538-4357/aba94b},
archivePrefix = {arXiv},
       eprint = {2007.13377},
 primaryClass = {astro-ph.SR},
       adsurl = {https://ui.adsabs.harvard.edu/abs/2020ApJ...900..192F}
}

@ARTICLE{2011-Holman,
       author = {{Holman}, G.~D. and {Aschwanden}, M.~J. and {Aurass}, H. and {Battaglia}, M. and {Grigis}, P.~C. and {Kontar}, E.~P. and {Liu}, W. and {Saint-Hilaire}, P. and {Zharkova}, V.~V.},
        title = "{Implications of X-ray Observations for Electron Acceleration and Propagation in Solar Flares}",
      journal = {\ssr},
         year = 2011,
        month = sep,
       volume = {159},
       number = {1-4},
        pages = {107-166},
          doi = {10.1007/s11214-010-9680-9},
archivePrefix = {arXiv},
       eprint = {1109.6496},
 primaryClass = {astro-ph.SR},
       adsurl = {https://ui.adsabs.harvard.edu/abs/2011SSRv..159..107H}
}

@ARTICLE{2014-Jeffrey,
       author = {{Jeffrey}, Natasha L.~S. and {Kontar}, Eduard P. and {Bian}, Nicolas H. and {Emslie}, A. Gordon},
        title = "{On the Variation of Solar Flare Coronal X-Ray Source Sizes with Energy}",
      journal = {\apj},
         year = 2014,
        month = may,
       volume = {787},
       number = {1},
          eid = {86},
        pages = {86},
          doi = {10.1088/0004-637X/787/1/86},
archivePrefix = {arXiv},
       eprint = {1404.1962},
 primaryClass = {astro-ph.SR},
       adsurl = {https://ui.adsabs.harvard.edu/abs/2014ApJ...787...86J}
}

@article{2007-Kaiser,
   author = {Kaiser, M. L. and Kucera, T. A. and Davila, J. M. and St. Cyr, O. C. and Guhathakurta, M. and Christian, E.},
   title = {The STEREO Mission: An Introduction},
   journal = {Space Science Reviews},
   volume = {136},
   number = {1-4},
   pages = {5-16},
   ISSN = {0038-6308
1572-9672},
   DOI = {10.1007/s11214-007-9277-0},
   year = {2007},
   type = {Journal Article}
}

@ARTICLE{2006-Kontar,
       author = {{Kontar}, E.~P. and {MacKinnon}, A.~L. and {Schwartz}, R.~A. and {Brown}, J.~C.},
        title = "{Compton backscattered and primary X-rays from solar flares: angle dependent Green's function correction for photospheric albedo}",
      journal = {\aap},
         year = 2006,
        month = feb,
       volume = {446},
       number = {3},
        pages = {1157-1163},
          doi = {10.1051/0004-6361:20053672},
archivePrefix = {arXiv},
       eprint = {astro-ph/0510167},
 primaryClass = {astro-ph},
       adsurl = {https://ui.adsabs.harvard.edu/abs/2006A\&A...446.1157K}
}

@ARTICLE{2011-Kontar,
       author = {{Kontar}, E.~P. and {Brown}, J.~C. and {Emslie}, A.~G. and {Hajdas}, W. and {Holman}, G.~D. and {Hurford}, G.~J. and {Ka{\v{s}}parov{\'a}}, J. and {Mallik}, P.~C.~V. and {Massone}, A.~M. and {McConnell}, M.~L. and {Piana}, M. and {Prato}, M. and {Schmahl}, E.~J. and {Suarez-Garcia}, E.},
        title = "{Deducing Electron Properties from Hard X-ray Observations}",
      journal = {\ssr},
         year = 2011,
        month = sep,
       volume = {159},
       number = {1-4},
        pages = {301-355},
          doi = {10.1007/s11214-011-9804-x},
archivePrefix = {arXiv},
       eprint = {1110.1755},
 primaryClass = {astro-ph.SR},
       adsurl = {https://ui.adsabs.harvard.edu/abs/2011SSRv..159..301K}
}

@ARTICLE{2015-Kontar,
       author = {{Kontar}, Eduard P. and {Jeffrey}, Natasha L.~S. and {Emslie}, A. Gordon and {Bian}, N.~H.},
        title = "{Collisional Relaxation of Electrons in a Warm Plasma and Accelerated Nonthermal Electron Spectra in Solar Flares}",
      journal = {\apj},
         year = 2015,
        month = aug,
       volume = {809},
       number = {1},
          eid = {35},
        pages = {35},
          doi = {10.1088/0004-637X/809/1/35},
archivePrefix = {arXiv},
       eprint = {1505.03733},
 primaryClass = {astro-ph.SR},
       adsurl = {https://ui.adsabs.harvard.edu/abs/2015ApJ...809...35K}
}

@ARTICLE{2019-Kontar,
       author = {{Kontar}, Eduard P. and {Jeffrey}, Natasha L.~S. and {Emslie}, A. Gordon},
        title = "{Determination of the Total Accelerated Electron Rate and Power Using Solar Flare Hard X-Ray Spectra}",
      journal = {\apj},
         year = 2019,
        month = feb,
       volume = {871},
       number = {2},
          eid = {225},
        pages = {225},
          doi = {10.3847/1538-4357/aafad3},
archivePrefix = {arXiv},
       eprint = {1812.09474},
 primaryClass = {astro-ph.SR},
       adsurl = {https://ui.adsabs.harvard.edu/abs/2019ApJ...871..225K}
}

@article{2020-Krucker,
   author = {Krucker, Säm and Hurford, G. J. and Grimm, O. and Kögl, S. and Gröbelbauer, H. P. and Etesi, L. and Casadei, D. and Csillaghy, A. and Benz, A. O. and Arnold, N. G. and Molendini, F. and Orleanski, P. and Schori, D. and Xiao, H. and Kuhar, M. and Hochmuth, N. and Felix, S. and Schramka, F. and Marcin, S. and Kobler, S. and Iseli, L. and Dreier, M. and Wiehl, H. J. and Kleint, L. and Battaglia, M. and Lastufka, E. and Sathiapal, H. and Lapadula, K. and Bednarzik, M. and Birrer, G. and Stutz, St and Wild, Ch and Marone, F. and Skup, K. R. and Cichocki, A. and Ber, K. and Rutkowski, K. and Bujwan, W. and Juchnikowski, G. and Winkler, M. and Darmetko, M. and Michalska, M. and Seweryn, K. and Białek, A. and Osica, P. and Sylwester, J. and Kowalinski, M. and Ścisłowski, D. and Siarkowski, M. and Stęślicki, M. and Mrozek, T. and Podgórski, P. and Meuris, A. and Limousin, O. and Gevin, O. and Le Mer, I. and Brun, S. and Strugarek, A. and Vilmer, N. and Musset, S. and Maksimović, M. and Fárník, F. and Kozáček, Z. and Kašparová, J. and Mann, G. and Önel, H. and Warmuth, A. and Rendtel, J. and Anderson, J. and Bauer, S. and Dionies, F. and Paschke, J. and Plüschke, D. and Woche, M. and Schuller, F. and Veronig, A. M. and Dickson, E. C. M. and Gallagher, P. T. and Maloney, S. A. and Bloomfield, D. S. and Piana, M. and Massone, A. M. and Benvenuto, F. and Massa, P. and Schwartz, R. A. and Dennis, B. R. and van Beek, H. F. and Rodríguez-Pacheco, J. and Lin, R. P.},
   title = {The Spectrometer/Telescope for Imaging X-rays (STIX)},
   journal = {Astronomy \& Astrophysics},
   volume = {642},
   ISSN = {0004-6361
1432-0746},
   DOI = {10.1051/0004-6361/201937362},
   year = {2020},
   type = {Journal Article}
}

@article{2007-Krucker,
   author = {Krucker, Säm and Kontar, E. P. and Christe, S. and Lin, R. P.},
   title = {Solar Flare Electron Spectra at the Sun and near the Earth},
   journal = {The Astrophysical Journal},
   volume = {663},
   number = {2},
   pages = {L109-L112},
   ISSN = {0004-637X
1538-4357},
   DOI = {10.1086/519373},
   year = {2007},
   type = {Journal Article}
}

@ARTICLE{1985-Lin,
       author = {{Lin}, R.~P.},
        title = "{Energetic Solar Electrons in the Interplanetary Medium}",
      journal = {\solphys},
         year = 1985,
        month = oct,
       volume = {100},
        pages = {537},
          doi = {10.1007/BF00158444},
       adsurl = {https://ui.adsabs.harvard.edu/abs/1985SoPh..100..537L}
}

@article{1995-Lin,
   author = {Lin, R. P. and Anderson, K. A. and Ashford, S. and Carlson, C. and Curtis, D. and Ergun, R. and Larson, D. and McFadden, J. and McCarthy, M. and Parks, G. K. and Rème, H. and Bosqued, J. M. and Coutelier, J. and Cotin, F. and D'Uston, C. and Wenzel, K. -P. and Sanderson, T. R. and Henrion, J. and Ronnet, J. C. and Paschmann, G.},
   title = {A Three-Dimensional Plasma and Energetic Particle Investigation for the Wind Spacecraft},
   journal = {Space Science Reviews},
   volume = {71},
   pages = {125-153},
   ISSN = {0038-6308},
   DOI = {10.1007/bf00751328},
   url = {https://ui.adsabs.harvard.edu/abs/1995SSRv...71..125L},
   year = {1995},
   type = {Journal Article}
}

@article{2003-Lin,
   author = {Lin, Robert P. and Dennis, Brian R. and Benz, Arnold O. and Smith, D. M. and Lin, R. P. and Turin, P. and Curtis, D. W. and Primbsch, J. H. and Campbell, R. D. and Abiad, R. and Schroeder, P. and Cork, C. P. and Hull, E. L. and Landis, D. A. and Madden, N. W. and Malone, D. and Pehl, R. H. and Raudorf, T. and Sangsingkeow, P. and Boyle, R.},
   title = {The RHESSI Spectrometer},
   journal = {The Reuven Ramaty High-Energy Solar Spectroscopic Imager (RHESSI)},
   pages = {33-60},
   DOI = {info:doi/10.1007/978-94-017-3452-3_2},
   year = {2003},
   type = {Journal Article}
}

@article{2007-Luhmann,
   author = {Luhmann, J. G. and Curtis, D. W. and Schroeder, P. and McCauley, J. and Lin, R. P. and Larson, D. E. and Bale, S. D. and Sauvaud, J. A. and Aoustin, C. and Mewaldt, R. A. and Cummings, A. C. and Stone, E. C. and Davis, A. J. and Cook, W. R. and Kecman, B. and Wiedenbeck, M. E. and von Rosenvinge, T. and Acuna, M. H. and Reichenthal, L. S. and Shuman, S. and Wortman, K. A. and Reames, D. V. and Mueller-Mellin, R. and Kunow, H. and Mason, G. M. and Walpole, P. and Korth, A. and Sanderson, T. R. and Russell, C. T. and Gosling, J. T.},
   title = {STEREO IMPACT Investigation Goals, Measurements, and Data Products Overview},
   journal = {Space Science Reviews},
   volume = {136},
   number = {1-4},
   pages = {117-184},
   ISSN = {0038-6308
1572-9672},
   DOI = {10.1007/s11214-007-9170-x},
   year = {2007},
   type = {Journal Article}
}

@article{2023-Pallister,
   author = {Pallister, Ross and Jeffrey, Natasha L. S.},
   title = {Exploring the Origin of Solar Energetic Electrons. I. Constraining the Properties of the Acceleration Region Plasma Environment},
   journal = {The Astrophysical Journal},
   volume = {958},
   number = {1},
   pages = {18},
   ISSN = {0004-637X},
   DOI = {10.3847/1538-4357/ad0035},
   year = {2023},
   type = {Journal Article}
}

@ARTICLE{2025-Pallister,
       author = {{Pallister}, Ross and {Jeffrey}, Natasha L.~S. and {Stores}, Morgan},
        title = "{Exploring the Origin of Solar Energetic Electrons. II. Investigating Turbulent Coronal Acceleration}",
      journal = {\apj},
         year = 2025,
        month = apr,
       volume = {983},
       number = {1},
          eid = {58},
        pages = {58},
          doi = {10.3847/1538-4357/adbaf4},
archivePrefix = {arXiv},
       eprint = {2502.19105},
 primaryClass = {astro-ph.SR},
       adsurl = {https://ui.adsabs.harvard.edu/abs/2025ApJ...983...58P}
}

@ARTICLE{2008-Pick,
       author = {{Pick}, Monique and {Vilmer}, Nicole},
        title = "{Sixty-five years of solar radioastronomy: flares, coronal mass ejections and Sun Earth connection}",
      journal = {\aapr},
         year = 2008,
        month = oct,
       volume = {16},
        pages = {1-153},
          doi = {10.1007/s00159-008-0013-x},
       adsurl = {https://ui.adsabs.harvard.edu/abs/2008A\&ARv..16....1P}
}

@ARTICLE{2020-Rodriguez-Pacheco,
       author = {{Rodr{\'\i}guez-Pacheco}, J. and {Wimmer-Schweingruber}, R.~F. and {Mason}, G.~M. and {Ho}, G.~C. and {S{\'a}nchez-Prieto}, S. and {Prieto}, M. and {Mart{\'\i}n}, C. and {Seifert}, H. and {Andrews}, G.~B. and {Kulkarni}, S.~R. and {Panitzsch}, L. and {Boden}, S. and {B{\"o}ttcher}, S.~I. and {Cernuda}, I. and {Elftmann}, R. and {Espinosa Lara}, F. and {G{\'o}mez-Herrero}, R. and {Terasa}, C. and {Almena}, J. and {Begley}, S. and {B{\"o}hm}, E. and {Blanco}, J.~J. and {Boogaerts}, W. and {Carrasco}, A. and {Castillo}, R. and {da Silva Fari{\~n}a}, A. and {de Manuel Gonz{\'a}lez}, V. and {Drews}, C. and {Dupont}, A.~R. and {Eldrum}, S. and {Gordillo}, C. and {Guti{\'e}rrez}, O. and {Haggerty}, D.~K. and {Hayes}, J.~R. and {Heber}, B. and {Hill}, M.~E. and {J{\"u}ngling}, M. and {Kerem}, S. and {Knierim}, V. and {K{\"o}hler}, J. and {Kolbe}, S. and {Kulemzin}, A. and {Lario}, D. and {Lees}, W.~J. and {Liang}, S. and {Mart{\'\i}nez Hell{\'\i}n}, A. and {Meziat}, D. and {Montalvo}, A. and {Nelson}, K.~S. and {Parra}, P. and {Paspirgilis}, R. and {Ravanbakhsh}, A. and {Richards}, M. and {Rodr{\'\i}guez-Polo}, O. and {Russu}, A. and {S{\'a}nchez}, I. and {Schlemm}, C.~E. and {Schuster}, B. and {Seimetz}, L. and {Steinhagen}, J. and {Tammen}, J. and {Tyagi}, K. and {Varela}, T. and {Yedla}, M. and {Yu}, J. and {Agueda}, N. and {Aran}, A. and {Horbury}, T.~S. and {Klecker}, B. and {Klein}, K. -L. and {Kontar}, E. and {Krucker}, S. and {Maksimovic}, M. and {Malandraki}, O. and {Owen}, C.~J. and {Pacheco}, D. and {Sanahuja}, B. and {Vainio}, R. and {Connell}, J.~J. and {Dalla}, S. and {Dr{\"o}ge}, W. and {Gevin}, O. and {Gopalswamy}, N. and {Kartavykh}, Y.~Y. and {Kudela}, K. and {Limousin}, O. and {Makela}, P. and {Mann}, G. and {{\"O}nel}, H. and {Posner}, A. and {Ryan}, J.~M. and {Soucek}, J. and {Hofmeister}, S. and {Vilmer}, N. and {Walsh}, A.~P. and {Wang}, L. and {Wiedenbeck}, M.~E. and {Wirth}, K. and {Zong}, Q.},
        title = "{The Energetic Particle Detector. Energetic particle instrument suite for the Solar Orbiter mission}",
      journal = {\aap},
         year = 2020,
        month = oct,
       volume = {642},
          eid = {A7},
        pages = {A7},
          doi = {10.1051/0004-6361/201935287},
       adsurl = {https://ui.adsabs.harvard.edu/abs/2020A\&A...642A...7R}
}

@article{2016-McComas,
	title = {Integrated {Science} {Investigation} of the {Sun} ({ISIS}): {Design} of the {Energetic} {Particle} {Investigation}},
	volume = {204},
	issn = {1572-9672},
	url = {https://doi.org/10.1007/s11214-014-0059-1},
	doi = {10.1007/s11214-014-0059-1},
	number = {1},
	journal = {Space Science Reviews},
	author = {McComas, D. J. and Alexander, N. and Angold, N. and Bale, S. and Beebe, C. and Birdwell, B. and Boyle, M. and Burgum, J. M. and Burnham, J. A. and Christian, E. R. and Cook, W. R. and Cooper, S. A. and Cummings, A. C. and Davis, A. J. and Desai, M. I. and Dickinson, J. and Dirks, G. and Do, D. H. and Fox, N. and Giacalone, J. and Gold, R. E. and Gurnee, R. S. and Hayes, J. R. and Hill, M. E. and Kasper, J. C. and Kecman, B. and Klemic, J. and Krimigis, S. M. and Labrador, A. W. and Layman, R. S. and Leske, R. A. and Livi, S. and Matthaeus, W. H. and McNutt, R. L. and Mewaldt, R. A. and Mitchell, D. G. and Nelson, K. S. and Parker, C. and Rankin, J. S. and Roelof, E. C. and Schwadron, N. A. and Seifert, H. and Shuman, S. and Stokes, M. R. and Stone, E. C. and Vandegriff, J. D. and Velli, M. and von Rosenvinge, T. T. and Weidner, S. E. and Wiedenbeck, M. E. and Wilson, P.},
	month = dec,
	year = {2016},
	pages = {187--256},
}

@article{2020-Muller,
	title = {The {Solar} {Orbiter} mission - {Science} overview},
	volume = {642},
	url = {https://doi.org/10.1051/0004-6361/202038467},
	doi = {10.1051/0004-6361/202038467},
	journal = {A\&A},
	author = {{Müller, D.} and {St. Cyr, O. C.} and {Zouganelis, I.} and {Gilbert, H. R.} and {Marsden, R.} and {Nieves-Chinchilla, T.} and {Antonucci, E.} and {Auchère, F.} and {Berghmans, D.} and {Horbury, T. S.} and {Howard, R. A.} and {Krucker, S.} and {Maksimovic, M.} and {Owen, C. J.} and {Rochus, P.} and {Rodriguez-Pacheco, J.} and {Romoli, M.} and {Solanki, S. K.} and {Bruno, R.} and {Carlsson, M.} and {Fludra, A.} and {Harra, L.} and {Hassler, D. M.} and {Livi, S.} and {Louarn, P.} and {Peter, H.} and {Schühle, U.} and {Teriaca, L.} and {del Toro Iniesta, J. C.} and {Wimmer-Schweingruber, R. F.} and {Marsch, E.} and {Velli, M.} and {De Groof, A.} and {Walsh, A.} and {Williams, D.}},
	year = {2020},
	pages = {A1},
}

@ARTICLE{2010-Benkhoff,
       author = {{Benkhoff}, Johannes and {van Casteren}, Jan and {Hayakawa}, Hajime and {Fujimoto}, Masaki and {Laakso}, Harri and {Novara}, Mauro and {Ferri}, Paolo and {Middleton}, Helen R. and {Ziethe}, Ruth},
        title = "{BepiColombo{\textemdash}Comprehensive exploration of Mercury: Mission overview and science goals}",
      journal = {\planss},
         year = 2010,
        month = jan,
       volume = {58},
       number = {1-2},
        pages = {2-20},
          doi = {10.1016/j.pss.2009.09.020},
       adsurl = {https://ui.adsabs.harvard.edu/abs/2010P&SS...58....2B}
}

@article{2012-Lemen,
	title = {The {Atmospheric} {Imaging} {Assembly} ({AIA}) on the {Solar} {Dynamics} {Observatory} ({SDO})},
	volume = {275},
	issn = {1573-093X},
	url = {https://doi.org/10.1007/s11207-011-9776-8},
	doi = {10.1007/s11207-011-9776-8},
	number = {1},
	journal = {Solar Physics},
	author = {Lemen, James R. and Title, Alan M. and Akin, David J. and Boerner, Paul F. and Chou, Catherine and Drake, Jerry F. and Duncan, Dexter W. and Edwards, Christopher G. and Friedlaender, Frank M. and Heyman, Gary F. and Hurlburt, Neal E. and Katz, Noah L. and Kushner, Gary D. and Levay, Michael and Lindgren, Russell W. and Mathur, Dnyanesh P. and McFeaters, Edward L. and Mitchell, Sarah and Rehse, Roger A. and Schrijver, Carolus J. and Springer, Larry A. and Stern, Robert A. and Tarbell, Theodore D. and Wuelser, Jean-Pierre and Wolfson, C. Jacob and Yanari, Carl and Bookbinder, Jay A. and Cheimets, Peter N. and Caldwell, David and Deluca, Edward E. and Gates, Richard and Golub, Leon and Park, Sang and Podgorski, William A. and Bush, Rock I. and Scherrer, Philip H. and Gummin, Mark A. and Smith, Peter and Auker, Gary and Jerram, Paul and Pool, Peter and Soufli, Regina and Windt, David L. and Beardsley, Sarah and Clapp, Matthew and Lang, James and Waltham, Nicholas},
	month = jan,
	year = {2012},
	pages = {17--40},
}

@article{gieseler_solar-mach_2023,
	title = {Solar-{MACH}: {An} open-source tool to analyze solar magnetic connection configurations},
	volume = {9},
	issn = {2296-987X},
	url = {https://www.frontiersin.org/articles/10.3389/fspas.2022.1058810},
	doi = {10.3389/fspas.2022.1058810},
	journal = {Frontiers in Astronomy and Space Sciences},
	author = {Gieseler, Jan and Dresing, Nina and Palmroos, Christian and Freiherr von Forstner, Johan L. and Price, Daniel J. and Vainio, Rami and Kouloumvakos, Athanasios and Rodríguez-García, Laura and Trotta, Domenico and Génot, Vincent and Masson, Arnaud and Roth, Markus and Veronig, Astrid},
	year = {2023},
}

@misc{ospex,
       author = {{Tolbert}, Kim and {Schwartz}, Richard},
        title = "{OSPEX: Object Spectral Executive}",
 howpublished = {Astrophysics Source Code Library, record ascl:2007.018},
         year = 2020,
        month = jul,
          eid = {ascl:2007.018},
       adsurl = {https://ui.adsabs.harvard.edu/abs/2020ascl.soft07018T}
}

@article{pierrard_2010,
	title = {Kappa {Distributions}: {Theory} and {Applications} in {Space} {Plasmas}},
	volume = {267},
	issn = {1573-093X},
	url = {https://doi.org/10.1007/s11207-010-9640-2},
	doi = {10.1007/s11207-010-9640-2},
	number = {1},
	journal = {Solar Physics},
	author = {Pierrard, V. and Lazar, M.},
	month = nov,
	year = {2010},
	pages = {153--174},
}

@misc{newville_2015,
  author       = {Newville, Matthew and
                  Stensitzki, Till and
                  Allen, Daniel B. and
                  Ingargiola,  Antonino},
  title        = {{LMFIT: Non-Linear Least-Square Minimization and 
                   Curve-Fitting for Python}},
  month        = oct,
  year         = 2015,
  publisher    = {Zenodo},
  version      = {0.8.0},
  doi          = {10.5281/zenodo.11813},
  url          = {https://doi.org/10.5281/zenodo.11813}
}

@article{jeffrey_non-gaussian_2017,
	title = {Non-{Gaussian} {Velocity} {Distributions} in {Solar} {Flares} from {Extreme} {Ultraviolet} {Lines}: {A} {Possible} {Diagnostic} of {Ion} {Acceleration}},
	volume = {836},
	url = {https://dx.doi.org/10.3847/1538-4357/836/1/35},
	doi = {10.3847/1538-4357/836/1/35},
	number = {1},
	journal = {The Astrophysical Journal},
	author = {Jeffrey, Natasha L. S. and Fletcher, Lyndsay and Labrosse, Nicolas},
	month = feb,
	year = {2017},
	note = {Publisher: The American Astronomical Society},
	pages = {35},
}

@article{Emslie_2012,
doi = {10.1088/0004-637X/759/1/71},
url = {https://dx.doi.org/10.1088/0004-637X/759/1/71},
year = {2012},
month = {oct},
publisher = {The American Astronomical Society},
volume = {759},
number = {1},
pages = {71},
author = {A. G. Emslie and B. R. Dennis and A. Y. Shih and P. C. Chamberlin and R. A. Mewaldt and C. S. Moore and G. H. Share and A. Vourlidas and B. T. Welsch},
title = {GLOBAL ENERGETICS OF THIRTY-EIGHT LARGE SOLAR ERUPTIVE EVENTS},
journal = {The Astrophysical Journal}
}

@INPROCEEDINGS{2013_oka,
       author = {{Oka}, M. and {Krucker}, S. and {Phan}, T.},
        title = "{Electron Kappa Distributions in Solar Flares and the Earth's Magnetotail}",
    booktitle = {AGU Fall Meeting Abstracts},
         year = 2013,
       volume = {2013},
        month = dec,
          eid = {SH41D-2211},
        pages = {SH41D-2211},
       adsurl = {https://ui.adsabs.harvard.edu/abs/2013AGUFMSH41D2211O}
}

@article{ 2020-Owen,
	author = {{Owen, C. J.} and {Bruno, R.} and {Livi, S.} and {Louarn, P.} and {Al Janabi, K.} and {Allegrini, F.} and {Amoros, C.} and {Baruah, R.} and {Barthe, A.} and {Berthomier, M.} and {Bordon, S.} and {Brockley-Blatt, C.} and {Brysbaert, C.} and {Capuano, G.} and {Collier, M.} and {DeMarco, R.} and {Fedorov, A.} and {Ford, J.} and {Fortunato, V.} and {Fratter, I.} and {Galvin, A. B.} and {Hancock, B.} and {Heirtzler, D.} and {Kataria, D.} and {Kistler, L.} and {Lepri, S. T.} and {Lewis, G.} and {Loeffler, C.} and {Marty, W.} and {Mathon, R.} and {Mayall, A.} and {Mele, G.} and {Ogasawara, K.} and {Orlandi, M.} and {Pacros, A.} and {Penou, E.} and {Persyn, S.} and {Petiot, M.} and {Phillips, M.} and {Přech, L.} and {Raines, J. M.} and {Reden, M.} and {Rouillard, A. P.} and {Rousseau, A.} and {Rubiella, J.} and {Seran, H.} and {Spencer, A.} and {Thomas, J. W.} and {Trevino, J.} and {Verscharen, D.} and {Wurz, P.} and {Alapide, A.} and {Amoruso, L.} and {André, N.} and {Anekallu, C.} and {Arciuli, V.} and {Arnett, K. L.} and {Ascolese, R.} and {Bancroft, C.} and {Bland, P.} and {Brysch, M.} and {Calvanese, R.} and {Castronuovo, M.} and {Čermák, I.} and {Chornay, D.} and {Clemens, S.} and {Coker, J.} and {Collinson, G.} and {D’Amicis, R.} and {Dandouras, I.} and {Darnley, R.} and {Davies, D.} and {Davison, G.} and {De Los Santos, A.} and {Devoto, P.} and {Dirks, G.} and {Edlund, E.} and {Fazakerley, A.} and {Ferris, M.} and {Frost, C.} and {Fruit, G.} and {Garat, C.} and {Génot, V.} and {Gibson, W.} and {Gilbert, J. A.} and {de Giosa, V.} and {Gradone, S.} and {Hailey, M.} and {Horbury, T. S.} and {Hunt, T.} and {Jacquey, C.} and {Johnson, M.} and {Lavraud, B.} and {Lawrenson, A.} and {Leblanc, F.} and {Lockhart, W.} and {Maksimovic, M.} and {Malpus, A.} and {Marcucci, F.} and {Mazelle, C.} and {Monti, F.} and {Myers, S.} and {Nguyen, T.} and {Rodriguez-Pacheco, J.} and {Phillips, I.} and {Popecki, M.} and {Rees, K.} and {Rogacki, S. A.} and {Ruane, K.} and {Rust, D.} and {Salatti, M.} and {Sauvaud, J. A.} and {Stakhiv, M. O.} and {Stange, J.} and {Stubbs, T.} and {Taylor, T.} and {Techer, J.-D.} and {Terrier, G.} and {Thibodeaux, R.} and {Urdiales, C.} and {Varsani, A.} and {Walsh, A. P.} and {Watson, G.} and {Wheeler, P.} and {Willis, G.} and {Wimmer-Schweingruber, R. F.} and {Winter, B.} and {Yardley, J.} and {Zouganelis, I.}},
	title = {The Solar Orbiter Solar Wind Analyser (SWA) suite},
	DOI= "10.1051/0004-6361/201937259",
	url= "https://doi.org/10.1051/0004-6361/201937259",
	journal = {A\&A},
	year = 2020,
	volume = 642,
	pages = "A16",
}

@article{Lorfing-2023,
doi = {10.3847/1538-4357/ad0be3},
url = {https://dx.doi.org/10.3847/1538-4357/ad0be3},
year = {2023},
month = {dec},
publisher = {The American Astronomical Society},
volume = {959},
number = {2},
pages = {128},
author = {Camille Y. Lorfing and Hamish A. S. Reid and Raúl Gómez-Herrero and Milan Maksimovic and Georgios Nicolaou and Christopher J. Owen and Javier Rodriguez-Pacheco and Daniel F. Ryan and Domenico Trotta and Daniel Verscharen},
title = {Solar Electron Beam—Langmuir Wave Interactions and How They Modify Solar Electron Beam Spectra: Solar Orbiter Observations of a Match Made in the Heliosphere},
journal = {The Astrophysical Journal}
}

@ARTICLE{Jebaraj_2023,
       author = {{Jebaraj}, Immanuel C. and {Kouloumvakos}, A. and {Dresing}, N. and {Warmuth}, A. and {Wijsen}, N. and {Palmroos}, C. and {Gieseler}, J. and {Marmyleva}, A. and {Vainio}, R. and {Krupar}, V. and {Wiegelmann}, T. and {Magdalenic}, J. and {Schuller}, F. and {Battaglia}, A.~F. and {Fedeli}, A.},
        title = "{Multiple injections of energetic electrons associated with the flare and CME event on 9 October 2021}",
      journal = {\aap},
         year = 2023,
        month = jul,
       volume = {675},
          eid = {A27},
        pages = {A27},
          doi = {10.1051/0004-6361/202245716},
archivePrefix = {arXiv},
       eprint = {2301.03650},
 primaryClass = {astro-ph.SR},
       adsurl = {https://ui.adsabs.harvard.edu/abs/2023A\&A...675A..27J}
}

@ARTICLE{2006-Zharkova,
       author = {{Zharkova}, Valentina V. and {Gordovskyy}, Mykola},
        title = "{The Effect of the Electric Field Induced by Precipitating Electron Beams on Hard X-Ray Photon and Mean Electron Spectra}",
      journal = {\apj},
         year = 2006,
        month = nov,
       volume = {651},
       number = {1},
        pages = {553-565},
          doi = {10.1086/506423},
       adsurl = {https://ui.adsabs.harvard.edu/abs/2006ApJ...651..553Z}
}

@article{2017-Graham,
author = {Graham, G. A. and Rae, I. J. and Owen, C. J. and Walsh, A. P. and Arridge, C. S. and Gilbert, L. and Lewis, G. R. and Jones, G. H. and Forsyth, C. and Coates, A. J. and Waite, J. H.},
title = {The evolution of solar wind strahl with heliospheric distance},
journal = {Journal of Geophysical Research: Space Physics},
volume = {122},
number = {4},
pages = {3858-3874},
doi = {https://doi.org/10.1002/2016JA023656},
url = {https://agupubs.onlinelibrary.wiley.com/doi/abs/10.1002/2016JA023656},
eprint = {https://agupubs.onlinelibrary.wiley.com/doi/pdf/10.1002/2016JA023656},
year = {2017}
}

@article{1994-menzel,
  title={Introducing GOES-I: The first of a new generation of geostationary operational environmental satellites},
  author={Menzel, W Paul and Purdom, James FW},
  journal={Bulletin of the American Meteorological Society},
  volume={75},
  number={5},
  pages={757--782},
  year={1994},
  publisher={American Meteorological Society}
}

@book{2019-goodman,
  title={The GOES-R series: a new generation of geostationary environmental satellites},
  author={Goodman, Steven J and Schmit, Timothy J and Daniels, Jaime and Redmon, Robert J},
  year={2019},
  publisher={Elsevier}}

@article{ 2021-klein, 
title={Radio Astronomical Tools for the Study of Solar Energetic Particles I. Correlations and Diagnostics of Impulsive Acceleration and Particle Propagation}, 
volume={7}, 
ISSN={2296-987X}, 
DOI={10.3389/fspas.2020.580436}, 
journal={Frontiers in Astronomy and Space Sciences}, 
publisher={Frontiers in Astronomy and Space Sciences}, 
author={Klein, Karl-Ludwig}, 
year={2021} }

@article{ 2004-klein,
	author = {{Klein, K.-L.} and {Krucker, S.} and {Trottet, G.} and {Hoang, S.}},
	title = {Coronal phenomena at the release  of solar energetic electron events*},
	DOI= "10.1051/0004-6361:20041258",
	url= "https://doi.org/10.1051/0004-6361:20041258",
	journal = {A\&A},
	year = 2005,
	volume = 431,
	number = 3,
	pages = "1047-1060",
}

@article{1994-garcia,
	title = {Temperature and emission measure from goes soft {X}-ray measurements},
	volume = {154},
	issn = {1573-093X},
	url = {https://doi.org/10.1007/BF00681100},
	doi = {10.1007/BF00681100},
	number = {2},
	journal = {Solar Physics},
	author = {Garcia, Howard A.},
	month = oct,
	year = {1994},
	pages = {275--308},
}

@article{2009-Liavdiotis,
author = {Livadiotis, G. and McComas, D. J.},
title = {Beyond kappa distributions: Exploiting Tsallis statistical mechanics in space plasmas},
journal = {Journal of Geophysical Research: Space Physics},
volume = {114},
number = {A11},
pages = {},
doi = {https://doi.org/10.1029/2009JA014352},
url = {https://agupubs.onlinelibrary.wiley.com/doi/abs/10.1029/2009JA014352},
eprint = {https://agupubs.onlinelibrary.wiley.com/doi/pdf/10.1029/2009JA014352},
year = {2009}
}

@article{2013-livadiotis,
	title = {Understanding {Kappa} {Distributions}: {A} {Toolbox} for {Space} {Science} and {Astrophysics}},
	volume = {175},
	issn = {1572-9672},
	url = {https://doi.org/10.1007/s11214-013-9982-9},
	doi = {10.1007/s11214-013-9982-9},
	number = {1},
	journal = {Space Science Reviews},
	author = {Livadiotis, G. and McComas, D. J.},
	month = jun,
	year = {2013},
	pages = {183--214},
}

@article{2016-nicolaou,
	title = {Misestimation of temperature when applying {Maxwellian} distributions to space plasmas described by kappa distributions},
	volume = {361},
	issn = {1572-946X},
	url = {https://doi.org/10.1007/s10509-016-2949-z},
	doi = {10.1007/s10509-016-2949-z},
	number = {11},
	journal = {Astrophysics and Space Science},
	author = {Nicolaou, Georgios and Livadiotis, George},
	month = oct,
	year = {2016},
	pages = {359},
}

@article{ 2021-Nicolaou,
	author = {{Nicolaou, G.} and {Wicks, R. T.} and {Owen, C. J.} and {Kataria, D. O.} and {Chandrasekhar, A.} and {Lewis, G. R.} and {Verscharen, D.} and {Fortunato, V.} and {Mele, G.} and {DeMarco, R.} and {Bruno, R.}},
	title = {Deriving the bulk properties of solar wind electrons observed by Solar Orbiter - A preliminary study of electron plasma thermodynamics},
	DOI= "10.1051/0004-6361/202140875",
	url= "https://doi.org/10.1051/0004-6361/202140875",
	journal = {A\&A},
	year = 2021,
	volume = 656,
	pages = "A10",
}

@article{Nicolaou_2018,
doi = {10.3847/1538-4357/aad45d},
url = {https://dx.doi.org/10.3847/1538-4357/aad45d},
year = {2018},
month = {aug},
publisher = {The American Astronomical Society},
volume = {864},
number = {1},
pages = {3},
author = {Nicolaou, G. and Livadiotis, G. and Owen, C. J. and Verscharen, D. and Wicks, R. T.},
title = {Determining the Kappa Distributions of Space Plasmas from Observations in a Limited Energy Range},
journal = {The Astrophysical Journal}
}

@article{ 2003-conway,
	author = {{Conway, A. J.} and {Brown, J. C.} and {Eves, B. A. C.} and {Kontar, E.}},
	title = {Implications of solar flare hard X-ray “knee” spectra observed
  by RHESSI },
	DOI= "10.1051/0004-6361:20030897",
	url= "https://doi.org/10.1051/0004-6361:20030897",
	journal = {A\&A},
	year = 2003,
	volume = 407,
	number = 2,
	pages = "725-734",
}

@article{2005-Zharkova,
	author = {{Zharkova, V. V.} and {Gordovskyy, M.}},
	title = {The kinetic effects of electron beam precipitation  and resulting hard X-ray intensity in solar flares*},
	DOI= "10.1051/0004-6361:20041102",
	url= "https://doi.org/10.1051/0004-6361:20041102",
	journal = {A\&A},
	year = 2005,
	volume = 432,
	number = 3,
	pages = "1033-1047",
}

@article{2005-Maksimovic,
author = {Maksimovic, M. and Zouganelis, I. and Chaufray, J.-Y. and Issautier, K. and Scime, E. E. and Littleton, J. E. and Marsch, E. and McComas, D. J. and Salem, C. and Lin, R. P. and Elliott, H.},
title = {Radial evolution of the electron distribution functions in the fast solar wind between 0.3 and 1.5 AU},
journal = {Journal of Geophysical Research: Space Physics},
volume = {110},
number = {A9},
pages = {},
doi = {https://doi.org/10.1029/2005JA011119},
url = {https://agupubs.onlinelibrary.wiley.com/doi/abs/10.1029/2005JA011119},
eprint = {https://agupubs.onlinelibrary.wiley.com/doi/pdf/10.1029/2005JA011119},
year = {2005}
}

@article{2024-Nicolaou,
    author = {Nicolaou, G and Livadiotis, G and Sarlis, N and Ioannou, C},
    title = {Resolving velocity distribution function parameters from observations with significant Poisson statistical uncertainty},
    journal = {RAS Techniques and Instruments},
    volume = {3},
    number = {1},
    pages = {874-878},
    year = {2024},
    month = {12},
    issn = {2752-8200},
    doi = {10.1093/rasti/rzae059},
    url = {https://doi.org/10.1093/rasti/rzae059},
    eprint = {https://academic.oup.com/rasti/article-pdf/3/1/874/61025691/rzae059.pdf},
}

@article{ STIX-datacentre,
	author = {{Xiao, Hualin} and {Maloney, Shane} and {Krucker, Säm} and {Dickson, Ewan} and {Massa, Paolo} and {Lastufka, Erica} and {Francesco Battaglia, Andrea} and {Etesi, László} and {Hochmuth, Nicky} and {Schuller, Frédéric} and {Ryan, Daniel F.} and {Limousin, Olivier} and {Collier, Hannah} and {Warmuth, Alexander} and {Piana, Michele}},
	title = {The data center for the Spectrometer and Telescope for Imaging X-rays (STIX) on board Solar Orbiter},
	DOI= "10.1051/0004-6361/202346031",
	url= "https://doi.org/10.1051/0004-6361/202346031",
	journal = {A\&A},
	year = 2023,
	volume = 673,
	pages = "A142",
}

@article{2013-reid,
	title = {Evolution of the {Solar} {Flare} {Energetic} {Electrons} in the {Inhomogeneous} {Inner} {Heliosphere}},
	volume = {285},
	issn = {1573-093X},
	url = {https://doi.org/10.1007/s11207-012-0013-x},
	doi = {10.1007/s11207-012-0013-x},
	number = {1},
	journal = {Solar Physics},
	author = {Reid, Hamish A. S. and Kontar, Eduard P.},
	month = jul,
	year = {2013},
	pages = {217--232},
}

@article{ 2025-stverak,
	author = {{Stverák, S.} and {Hercík, D.} and {Nicolaou, G.} and {Hellinger, P.} and {Popďakunik, M.} and {Khotyaintsev, Yu. V.} and {Kataria, D. O.} and {Owen, C. J.} and {Maksimovic, M.}},
	title = {Effects of cold electron emissions on thermal plasma measurements on board Solar Orbiter spacecraft},
	DOI= "10.1051/0004-6361/202452030",
	url= "https://doi.org/10.1051/0004-6361/202452030",
	journal = {A\&A},
	year = 2025,
	volume = 693,
	pages = "A185",
}

@article{2015-Battaglia,
doi = {10.1088/0004-637X/815/1/73},
url = {https://dx.doi.org/10.1088/0004-637X/815/1/73},
year = {2015},
month = {dec},
publisher = {The American Astronomical Society},
volume = {815},
number = {1},
pages = {73},
author = {Battaglia, Marina and Motorina, Galina and Kontar, Eduard P.},
title = {MULTITHERMAL REPRESENTATION OF THE KAPPA-DISTRIBUTION OF SOLAR FLARE ELECTRONS AND APPLICATION TO SIMULTANEOUS X-RAY AND EUV OBSERVATIONS},
journal = {The Astrophysical Journal}
}

@article{Aschwanden_2015,
doi = {10.1088/0004-637X/802/1/53},
url = {https://dx.doi.org/10.1088/0004-637X/802/1/53},
year = {2015},
month = {mar},
publisher = {The American Astronomical Society},
volume = {802},
number = {1},
pages = {53},
author = {Aschwanden, Markus J. and Boerner, Paul and Ryan, Daniel and Caspi, Amir and McTiernan, James M. and Warren, Harry P.},
title = {GLOBAL ENERGETICS OF SOLAR FLARES. II. THERMAL ENERGIES},
journal = {The Astrophysical Journal}
}

@ARTICLE{2015-jeffrey,
       author = {{Jeffrey}, Natasha L.~S. and {Kontar}, Eduard P. and {Dennis}, Brian R.},
        title = "{High-temperature differential emission measure and altitude variations in the temperature and density of solar flare coronal X-ray sources}",
      journal = {\aap},
         year = 2015,
        month = dec,
       volume = {584},
          eid = {A89},
        pages = {A89},
          doi = {10.1051/0004-6361/201526665},
archivePrefix = {arXiv},
       eprint = {1510.04095},
 primaryClass = {astro-ph.SR},
       adsurl = {https://ui.adsabs.harvard.edu/abs/2015A\&A...584A..89J}
}

@ARTICLE{2009-Krucker,
       author = {{Krucker}, S{\"a}m and {Oakley}, P.~H. and {Lin}, R.~P.},
        title = "{Spectra of Solar Impulsive Electron Events Observed Near Earth}",
      journal = {\apj},
         year = 2009,
        month = jan,
       volume = {691},
       number = {1},
        pages = {806-810},
          doi = {10.1088/0004-637X/691/1/806},
       adsurl = {https://ui.adsabs.harvard.edu/abs/2009ApJ...691..806K}
}

@ARTICLE{2014-caspi,
       author = {{Caspi}, Amir and {Krucker}, S{\"a}m and {Lin}, R.~P.},
        title = "{Statistical Properties of Super-hot Solar Flares}",
      journal = {\apj},
         year = 2014,
        month = jan,
       volume = {781},
       number = {1},
          eid = {43},
        pages = {43},
          doi = {10.1088/0004-637X/781/1/43},
archivePrefix = {arXiv},
       eprint = {1312.0371},
 primaryClass = {astro-ph.SR},
       adsurl = {https://ui.adsabs.harvard.edu/abs/2014ApJ...781...43C}
}

@ARTICLE{2013-DelZanna,
       author = {{Del Zanna}, G.},
        title = "{The multi-thermal emission in solar active regions}",
      journal = {\aap},
         year = 2013,
        month = oct,
       volume = {558},
          eid = {A73},
        pages = {A73},
          doi = {10.1051/0004-6361/201321653},
       adsurl = {https://ui.adsabs.harvard.edu/abs/2013A\&A...558A..73D}
}

@ARTICLE{2021-DelZanna,
       author = {{Del Zanna}, Giulio and {Andretta}, Vincenzo and {Cargill}, Peter J. and {Corso}, Alain J. and {Daw}, Adrian N. and {Golub}, Leon and {Klimchuk}, James A. and {Mason}, Helen E.},
        title = "{High resolution soft X-ray spectroscopy and the quest for the hot (5-10 MK) plasma in solar active regions}",
      journal = {Frontiers in Astronomy and Space Sciences},
         year = 2021,
        month = apr,
       volume = {8},
          eid = {33},
        pages = {33},
          doi = {10.3389/fspas.2021.638489},
archivePrefix = {arXiv},
       eprint = {2103.06156},
 primaryClass = {astro-ph.SR},
       adsurl = {https://ui.adsabs.harvard.edu/abs/2021FrASS...8...33D}
}

@ARTICLE{sunpy_community2020,
  doi = {10.3847/1538-4357/ab4f7a},
  url = {https://iopscience.iop.org/article/10.3847/1538-4357/ab4f7a},
  author = {{The SunPy Community} and Barnes, Will T. and Bobra, Monica G. and Christe, Steven D. and Freij, Nabil and Hayes, Laura A. and Ireland, Jack and Mumford, Stuart and Perez-Suarez, David and Ryan, Daniel F. and Shih, Albert Y. and Chanda, Prateek and Glogowski, Kolja and Hewett, Russell and Hughitt, V. Keith and Hill, Andrew and Hiware, Kaustubh and Inglis, Andrew and Kirk, Michael S. F. and Konge, Sudarshan and Mason, James Paul and Maloney, Shane Anthony and Murray, Sophie A. and Panda, Asish and Park, Jongyeob and Pereira, Tiago M. D. and Reardon, Kevin and Savage, Sabrina and Sipőcz, Brigitta M. and Stansby, David and Jain, Yash and Taylor, Garrison and Yadav, Tannmay and Rajul and Dang, Trung Kien},
  title = {The SunPy Project: Open Source Development and Status of the Version 1.0 Core Package},
  journal = {The Astrophysical Journal},
  volume = {890},
  issue = {1},
  pages = {68-},
  publisher = {American Astronomical Society},
  year = {2020}
}

@software{sunpy-6.0.4,
  author       = {Stuart J. Mumford and
                  Nabil Freij and
                  David Stansby and
                  Steven Christe and
                  Albert Y. Shih and
                  Jack Ireland and
                  Florian Mayer and
                  V. Keith Hughitt and
                  Daniel F. Ryan and
                  Simon Liedtke and
                  Will Barnes and
                  Laura Hayes and
                  David Pérez-Suárez and
                  Vishnunarayan K I. and
                  Pritish Chakraborty and
                  Andrew Inglis and
                  Punyaslok Pattnaik and
                  Brigitta Sipőcz and
                  Conor MacBride and
                  Rishabh Sharma and
                  Andrew Leonard and
                  Russell Hewett and
                  Alex Hamilton and
                  Abhijeet Manhas and
                  Asish Panda and
                  Matt Earnshaw and
                  Nitin Choudhary and
                  Ankit Kumar and
                  Raahul Singh and
                  Prateek Chanda and
                  Md Akramul Haque and
                  Alasdair Wilson and
                  Michael S Kirk and
                  Shane Maloney and
                  Michael Mueller and
                  Sudarshan Konge and
                  Matt Wentzel-Long and
                  Rajul Srivastava and
                  Samuel Bennett and
                  Yash Jain and
                  Lazar Zivadinovic and
                  Ankit Baruah and
                  Quinn Arbolante and
                  Trestan F. Simon and
                  Michael Charlton and
                  Sashank Mishra and
                  Jeffrey Aaron Paul and
                  Akash Verma and
                  Nicky Chorley and
                  Aryan Chouhan and
                  Chris R. Gilly and
                  Himanshu and
                  James Paul Mason and
                  Sanskar Modi and
                  Yash Sharma and
                  Naman9639 and
                  Monica Bobra and
                  Akshit Tyagi and
                  Jose Ivan Campos Rozo and
                  Larry Manley and
                  Kateryna Ivashkiv and
                  Timo Laitinen and
                  Agneet Chatterjee and
                  Ansh Dixit and
                  Jan Gieseler and
                  Jayraj Dulange and
                  Johan Freiherr von Forstner and
                  Juanjo Bazán and
                  Kris Akira Stern and
                  Aryan Shukla and
                  John Evans and
                  Sarthak Jain and
                  Michael Malocha and
                  Sourav Ghosh and
                  Airmansmith97 and
                  Dominik Stańczak and
                  Manit Singh and
                  Rajiv Ranjan Singh and
                  Ruben De Visscher and
                  Shresth Verma and
                  Sophie Lemos and
                  Ankit Agrawal and
                  Arib Alam and
                  Aritra Sinha and
                  Brett J Graham and
                  Dumindu Buddhika and
                  Hannah Collier and
                  Himanshu Pathak and
                  Jai Ram Rideout and
                  Swapnil Sharma and
                  Daniel Garcia Briseno and
                  Harsh Shah and
                  Jongyeob Park and
                  Matt Bates and
                  Devansh Shukla and
                  Marius Giger and
                  Pankaj Mishra and
                  Deepankar Sharma and
                  Dhruv Goel and
                  Garrison Taylor and
                  Goran Cetusic and
                  Guntbert Reiter and
                  Jacob and
                  Mateo Inchaurrandieta and
                  Piyush Sharma and
                  Sally Dacie and
                  Sanjeev Dubey and
                  Arthur Eigenbrot and
                  Benjamin Mampaey and
                  Erik Bray and
                  Nick Murphy and
                  Rutuja Surve and
                  Serge Zahniy and
                  Sudeep Sidhu and
                  Tomas Meszaros and
                  Utkarsh Parkhi and
                  William Russell and
                  Abhigyan Bose and
                  Abhishek Pandey and
                  Adrian Price-Whelan and
                  Ahmed Hossam and
                  Amogh Jahagirdar and
                  André Chicrala and
                  Aniket Mishra and
                  Ankit and
                  Chloé Guennou and
                  Daniel D'Avella and
                  Daniel Williams and
                  Dipanshu Verma and
                  Jordan Ballew and
                  Krish Agrawal and
                  Mubin Manasia and
                  Neeraj Kulkarni and
                  Nischal Singh and
                  Priyank Lodha and
                  Samuel J. Van Kooten and
                  Shivansh Mishra and
                  Thomas Robitaille and
                  Tom Augspurger and
                  Yash Krishan and
                  Abijith Bahuleyan and
                  Adwait Bhope and
                  Amarjit Singh Gaba and
                  Andrew Hill and
                  Bernhard M. Wiedemann and
                  Carlos Molina and
                  Duygu Keşkek and
                  Ishtyaq Habib and
                  Joseph Letts and
                  Karthikeyan Singaravelan and
                  Kritika Ranjan and
                  Mridul Pandey and
                  Noah Altunian and
                  Ole Streicher and
                  Reid Gomillion and
                  Samriddhi Agarwal and
                  Yash Kothari and
                  Yash Malik and
                  Yukie Nomiya and
                  Zach Burnett and
                  Abigail L. Stevens and
                  Akhoury Shauryam and
                  Alex Kaszynski and
                  Alex Wang and
                  Ambar Mehrotra and
                  Andy Tang and
                  Anubhav Sinha and
                  Arfon Smith and
                  Arseniy Kustov and
                  Ashish Bastola and
                  Brandon Stone and
                  Chris Bard and
                  Clément Robert and
                  Ed Behn and
                  Ed Mansky and
                  Emmanuel Arias and
                  Enrico Paganin and
                  Erik Tollerud and
                  Fionnlagh Mackenzie Dover and
                  Freek Verstringe and
                  Ghaith Kdimati and
                  Gulshan Kumar and
                  Harsh Mathur and
                  Igor Babuschkin and
                  James Calixto and
                  Jaylen Wimbish and
                  Jia Qing and
                  Juan Camilo Buitrago-Casas and
                  Kalpesh Krishna and
                  Kaustubh Chaudhari and
                  Kaustubh Hiware and
                  Koustav Ghosh and
                  Kurt McKee and
                  Manas Mangaonkar and
                  Mark Cheung and
                  Matthew Mendero and
                  Megh Dedhia and
                  Mickaël Schoentgen and
                  Mika and
                  Mouloudi Mohamed Lyes and
                  Nakul Shahdadpuri and
                  Naveen Srinivasan and
                  Norbert G Gyenge and
                  OussCHE and
                  Paul J. Wright and
                  Rajasekhar Reddy Mekala and
                  Ratul Das and
                  Rehan Chalana and
                  Rishabh Mishra and
                  Rohan Sharma and
                  Samuel T. Badman and
                  Shashank Srikanth and
                  Shubham Jain and
                  Sijie Yu and
                  Sirjan Hansda and
                  Suleiman Farah and
                  Swapnil Kannojia and
                  Syed Md Mihan Chistie and
                  Tan Jia Qing and
                  Tannmay Yadav and
                  Tathagata Paul and
                  Tessa D. Wilkinson and
                  Thomas A Caswell and
                  Thomas Braccia and
                  Tiago M. D. Pereira and
                  Tim Gates and
                  Trung Kien Dang and
                  Varun Bankar and
                  William Jamieson and
                  Yudhik Agrawal and
                  graham and
                  pradeep and
                  resakra and
                  yasintoda and
                  Raphael Attie and
                  Sophie A. Murray},
  title        = {sunpy: A Core Package for Solar Physics},
  month        = dec,
  year         = 2024,
  publisher    = {Zenodo},
  version      = {v6.0.4},
  doi          = {10.5281/zenodo.14292170},
  url          = {https://doi.org/10.5281/zenodo.14292170},
  swhid        = {swh:1:dir:846d4accd7bbb036dac6a7f4ddf94b2b09ef7a38
                   ;origin=https://doi.org/10.5281/zenodo.591887;visi
                   t=swh:1:snp:9adf20f615ddb0374d64693d09cdc0966c1c15
                   91;anchor=swh:1:rel:0e6b7fa3a1357e3306c84aa2c20c60
                   38b8ed1904;path=/
                  },
}

@misc{gieseler_solo-epd-loader_2025,
	title = {solo-epd-loader},
	copyright = {BSD 3-Clause "New" or "Revised" License},
	url = {https://zenodo.org/doi/10.5281/zenodo.15130823},
	urldate = {2025-08-29},
	publisher = {Zenodo},
	author = {Gieseler, Jan and Palmroos, Christian},
	month = apr,
	year = {2025},
	doi = {10.5281/ZENODO.15130823},
}

@ARTICLE{2022-Palmroos,
       author = {{Palmroos}, Christian and {Gieseler}, Jan and {Dresing}, Nina and {Morosan}, Diana E. and {Asvestari}, Eleanna and {Yli-Laurila}, Aleksi and {Price}, Daniel J. and {Valkila}, Saku and {Vainio}, Rami},
        title = "{Solar Energetic Particle Time Series Analysis with Python}",
      journal = {Frontiers in Astronomy and Space Sciences},
         year = 2022,
        month = dec,
       volume = {9},
          eid = {395},
        pages = {395},
          doi = {10.3389/fspas.2022.1073578},
archivePrefix = {arXiv},
       eprint = {2210.10432},
 primaryClass = {physics.space-ph},
       adsurl = {https://ui.adsabs.harvard.edu/abs/2022FrASS...973578P}
}

@article{1992-press-numerical,
  title={Numerical recipes in Fortran 77},
  author={Press, William H and Teukolsky, Saul A and Vetterling, William T and Flannery, Brian P},
  journal={The art of scientific computing},
  volume={1},
  year={1992}
}

@article{kass-1995,
  title={Bayes factors},
  author={Kass, Robert E and Raftery, Adrian E},
  journal={Journal of the American Statistical Association},
  volume={90},
  number={430},
  pages={773--795},
  year={1995},
  ISSN = {01621459, 1537274X},
  URL = {http://www.jstor.org/stable/2291091},
  publisher={Taylor \& Francis}
}

@article{Bell-1978,
    author = {Bell, A. R.},
    title = {The acceleration of cosmic rays in shock fronts – I},
    journal = {Monthly Notices of the Royal Astronomical Society},
    volume = {182},
    number = {2},
    pages = {147-156},
    year = {1978},
    month = {02},
    issn = {0035-8711},
    doi = {10.1093/mnras/182.2.147},
    url = {https://doi.org/10.1093/mnras/182.2.147},
    eprint = {https://academic.oup.com/mnras/article-pdf/182/2/147/3710138/mnras182-0147.pdf},
}

@Article{owen-2022,
AUTHOR = {Owen, Christopher J. and Abraham, Joel Baby and Nicolaou, Georgios and Verscharen, Daniel and Louarn, Philippe and Horbury, Timothy S.},
TITLE = {Solar Orbiter SWA Observations of Electron Strahl Properties Inside 1 AU},
JOURNAL = {Universe},
VOLUME = {8},
YEAR = {2022},
NUMBER = {10},
ARTICLE-NUMBER = {509},
URL = {https://www.mdpi.com/2218-1997/8/10/509},
ISSN = {2218-1997},
DOI = {10.3390/universe8100509}
}

@software{INSPEX_8_12_25_zenodo,
  author       = {Samuel Carter},
  title        = {SamuelCarter42/INSPEX: INSPEX for Carter et al
                   2026
                  },
  month        = may,
  year         = 2026,
  publisher    = {Zenodo},
  version      = {v1.8.0},
  doi          = {10.5281/zenodo.21276130},
  url          = {https://doi.org/10.5281/zenodo.21276130},
}

@article{zahra-2014,
  title={Performance functions alternatives of MSE for neural networks learning},
  author={Zahra, Mohamed M and Essai, Mohamed H and Abd Ellah, Ali R},
  journal={International Journal of Engineering Research \& Technology (IJERT)},
  volume={3},
  number={1},
  pages={967--970},
  year={2014}
}

@InProceedings{Mlotshwa-2022,
author="Mlotshwa, Thamsanqa
and van Deventer, Heinrich
and Bosman, Anna Sergeevna",
editor="Pillay, Anban
and Jembere, Edgar
and Gerber, Aurona",
title="Cauchy Loss Function: Robustness Under Gaussian and Cauchy Noise",
booktitle="Artificial Intelligence Research",
year="2022",
publisher="Springer Nature Switzerland",
address="Cham",
pages="123--138",
isbn="978-3-031-22321-1"
}

@article{kahler_2017,
	title = {Characterizing {Solar} {Energetic} {Particle} {Event} {Profiles} with {Two}-{Parameter} {Fits}},
	volume = {292},
	issn = {1573-093X},
	url = {https://doi.org/10.1007/s11207-017-1085-4},
	doi = {10.1007/s11207-017-1085-4},
	number = {4},
	journal = {Solar Physics},
	author = {Kahler, Stephen W. and Ling, Alan G.},
	month = apr,
	year = {2017},
	pages = {59},
}

@ARTICLE{2014-kontar,
       author = {{Kontar}, Eduard P. and {Bian}, Nicolas H. and {Emslie}, A. Gordon and {Vilmer}, Nicole},
        title = "{Turbulent Pitch-angle Scattering and Diffusive Transport of Hard X-Ray-producing Electrons in Flaring Coronal Loops}",
      journal = {\apj},
         year = 2014,
        month = jan,
       volume = {780},
       number = {2},
          eid = {176},
        pages = {176},
          doi = {10.1088/0004-637X/780/2/176},
archivePrefix = {arXiv},
       eprint = {1312.0266},
 primaryClass = {astro-ph.SR},
       adsurl = {https://ui.adsabs.harvard.edu/abs/2014ApJ...780..176K}
}

@ARTICLE{2016-laitinen,
       author = {{Laitinen}, T. and {Kopp}, A. and {Effenberger}, F. and {Dalla}, S. and {Marsh}, M.~S.},
        title = "{Solar energetic particle access to distant longitudes through turbulent field-line meandering}",
      journal = {\aap},
         year = 2016,
        month = jun,
       volume = {591},
          eid = {A18},
        pages = {A18},
          doi = {10.1051/0004-6361/201527801},
archivePrefix = {arXiv},
       eprint = {1508.03164},
 primaryClass = {astro-ph.SR},
       adsurl = {https://ui.adsabs.harvard.edu/abs/2016A\&A...591A..18L}
}

@article{ fedeli_2026,
	author = {{Fedeli, Annamaria} and {Dresing, Nina} and {Gieseler, Jan} and {Warmuth, Alexander} and {Schuller, Frederic} and {G\'omez-Herrero, Ra\'ul} and {Jebaraj, Immanuel Christopher} and {Espinosa, Francisco} and {Vainio, Rami}},
	title = {Peak-intensity energy spectra of intense solar energetic electron events measured with Solar Orbiter in 2020-2022},
	DOI= "10.1051/0004-6361/202555915",
	url= "https://doi.org/10.1051/0004-6361/202555915",
	journal = {A\&A},
	year = 2026,
	volume = 706,
	pages = "A107",
}

@article{boggs_1992,
  title={User's reference guide for odrpack version 2.01: Software for weighted orthogonal distance regression},
  author={Boggs, Paul T and Boggs, Paul T and Rogers, Janet E and Schnabel, Robert B},
  year={1992},
  publisher={US Department of Commerce, National Institute of Standards and Technology}
}
\bibliographystyle{aasjournal}

\end{document}